\documentclass[iop,numberedappendix,appendixfloats]{emulateapj}
\usepackage[colorlinks,urlcolor=blue,citecolor=blue,linkcolor=blue]{hyperref} 
\slugcomment{{\sc Accepted for publication in the Astrophysical Journal;} to appear May 2013}
\usepackage{array, amsmath}

\newcolumntype{x}[1]{>{\raggedleft\hspace{0pt}}p{#1}}
\newcolumntype{y}[1]{>{\centering\hspace{0pt}}p{#1}}
\newlength{\Oldarrayrulewidth}
\newcommand{\tn}{\tabularnewline}
\newcommand{\Cline}[2]{%
  \noalign{\global\setlength{\Oldarrayrulewidth}{\arrayrulewidth}}%
  \noalign{\global\setlength{\arrayrulewidth}{#1}}\cline{#2}%
  \noalign{\global\setlength{\arrayrulewidth}{\Oldarrayrulewidth}}}

\newcommand{\jone}{($J=1\rightarrow0$) }
\newcommand{\jtwo}{($J=2\rightarrow1$) }
\newcommand{\jthree}{($J=3\rightarrow2$) }
\newcommand{\jfour}{($J=4\rightarrow3$) }

\newcommand{\kms}{km~s$^{-1}$}
\newcommand{\kmssp}{km~s$^{-1}$~}

\newcommand{\Kkmspc}{K~km~s$^{-1}$~pc$^2$}

\def\h2{H$_2$}

\def\Mmgas{M_\mathrm{mgas}}

\shorttitle{The EGNoG Survey}
\shortauthors{Bauermeister et al.}
\begin{document}

\title{The EGNoG Survey: Molecular Gas in Intermediate-Redshift Star-Forming Galaxies}
\author{A. Bauermeister$^1$, L. Blitz$^1$, A. Bolatto$^2$, M. Bureau$^3$, A. Leroy$^4$, E. Ostriker$^5$, \\
P. Teuben$^2$, T. Wong$^6$, M. Wright$^1$}
\affil{$^1$Department of Astronomy and Radio Astronomy Laboratory, University of California at Berkeley, \\B-20 Hearst Field Annex, Berkeley, CA 94720, USA\\
$^2$Department of Astronomy and Laboratory for Millimeter-wave Astronomy, University of Maryland, College Park, MD 20742, USA\\
$^3$Sub-department of Astrophysics, Department of Physics, University of Oxford, \\Denys Wilkinson Building, Keble Road, Oxford OX1 3RH, UK\\ 
$^4$National Radio Astronomy Observatory, 520 Edgemont Road, Charlottesville, VA 22903, USA\\
$^5$Department of Astrophysical Sciences, Princeton University, Princeton, NJ 08544, USA\\
$^6$Department of Astronomy, University of Illinois, MC-221, 1002 W. Green Street, Urbana, IL 61801, USA}
\email{amberb@astro.berkeley.edu, blitz@berkeley.edu}
%\keywords{Galaxies:ISM --- Galaxies:evolution ---Stars:Formation}

\begin{abstract}
We present the Evolution of molecular Gas in Normal Galaxies (EGNoG) survey, 
an observational study of molecular gas in 31 star-forming galaxies from $z=0.05$ to $z=0.5$, 
with stellar masses of $(4-30)\times10^{10}$ M$_\odot$ and star formation rates of $4-100$ M$_\odot$ yr$^{-1}$. 
This survey probes a relatively un-observed redshift range in which the molecular gas content of galaxies
is expected to have evolved significantly. 
To trace the molecular gas in the EGNoG galaxies, we observe the CO\jone and CO\jthree rotational lines  
using the Combined Array for Research in Millimeter-wave Astronomy (CARMA).
We detect 24 of 31 galaxies and present resolved maps of 10 galaxies in the lower redshift portion of the
survey. We use a bimodal prescription for the CO to molecular gas conversion factor, based on specific
star formation rate, and compare the EGNoG galaxies to a large sample of galaxies assembled from the literature.
We find an  average molecular gas depletion time of $0.76\pm0.54$ Gyr
for normal galaxies and $0.06\pm0.04$ Gyr for starburst galaxies. We calculate an average
molecular gas fraction of 7-20\% at the intermediate redshifts probed by the EGNoG survey.
By expressing the molecular gas fraction in terms of the specific star formation rate and molecular gas depletion time
(using typical values), we also calculate the expected evolution of the molecular gas fraction with redshift. 
The predicted behavior agrees well with the significant evolution observed from $z\sim2.5$ to today.

\end{abstract}

\section{Introduction} 

In the past decade, molecular gas observations have begun probing the
high redshift universe in a systematic way  using increasingly powerful
millimeter instruments.
The picture that is emerging at redshifts $1-2$ is similar in some respects
to what we see in the local universe.  Sub-mm galaxies (SMGs) are observed to be
undergoing extreme starbursts as a result of major mergers
\citep[e.g.][]{Tacconi2008,Engel2010}, equivalent to local ultra-luminous infrared galaxies (ULIRGs)
\citep[e.g.][]{Sanders1986,Solomon1997,DownSol1998}. 
Normal star-forming galaxies (SFGs) at high redshifts, akin to local spirals, are becoming accessible as well. 
Recent works \citep{Tacconi2010,Daddi2010a} suggest that $z\sim1-2$ star-forming galaxies
(with star formation rates (SFRs) of $\approx50-200$ M$_\odot$ yr$^{-1}$)
are scaled up versions of local spirals, forming stars in a steady mode (not triggered by interaction), 
despite hosting star formation rates at the level of 
typical local starburst systems like luminous infrared galaxies (LIRGs) and ULIRGS.

%This comparison brings into focus the evolution of normal star-forming galaxies (SFGs) with redshift. }
While galaxies classified as LIRGs or ULIRGs (by their infrared luminosities only)
have typically been associated with starbursting and merging systems by analogy to galaxies in the local 
universe, it is becoming clear that this connection does not hold at high redshifts.
Morphological studies find that while 50\% of local LIRGs show evidence of major mergers \citep{wang2006}, 
that fraction appears to decrease toward high redshifts: \citet{Bell2005} find that more than half of intensely star-forming galaxies
 at $z\approx0.7$ have spiral morphologies and fewer than 30\% show evidence of strong interaction.
Further, the typical star formation rate of SFGs increases toward higher redshift.
SFGs have been observed to obey a tight relation between stellar mass and star formation rate
out to $z\sim4$ \citep[e.g.][]{Brinchmann2004, Noeske2007, Elbaz2007, Daddi2007, Pannella2009, Daddi2009, Magdis2010a}.
This `main sequence' of SFGs evolves with redshift, shifting to higher star formation rates at higher redshifts.

The increase in the star formation rate is mirrored in the molecular gas fraction 
($f_\mathrm{mgas} \equiv M_\mathrm{mgas}/(M_*+M_\mathrm{mgas}$), 
where $M_\mathrm{mgas}$ is the molecular gas mass (including He) and $M_*$ is the stellar mass)
of these systems. 
While studies of local spirals (e.g. FCRAO survey, \citealt{Young1995}; 
BIMA SONG, \citealt{Regan2001}, \citealt{Helfer2003}; HERACLES, \citealt{Leroy2009}) 
find average molecular gas fractions of $\sim$5\%,
observations of high redshift SFGs suggest molecular gas fractions of 20-80\%, an order
of magnitude higher than local spirals. Unfortunately, the past 8 Gyr of the evolution of $f_\mathrm{mgas}$
remains relatively unprobed, with only one study of normal SFGs between $z=1$ and $z=0.05$
\citep[at $z=0.4$]{Geach2011}.

\begin{table*}[t]
\centering
\begin{tabular}{ | c | c | c | c | c | c | c | c | c |}
\hline
Redshift & Redshift & Sample & Parent & M$_*$ Range & SFR Range & Observed & $\nu_\mathrm{obs}$ & CARMA \\
Bin & Range & Size & Sample & (M$_\odot$) & (M$_\odot$ yr$^{-1}$) & Transition & (GHz) & Configuration \\
\hline
A & 0.05 - 0.10 & 13 & SDSS & (4.0 - 16)$\times 10^{10}$ & 3.4 - 41 & CO\jone & 107 & C \\
B & 0.16 - 0.20 & 10 & SDSS & (6.3 - 32)$\times 10^{10}$ & 47 - 106  & CO\jone & 98 & D, some C \\
C & 0.28 - 0.32 & 4 & SDSS & (16 - 32)$\times 10^{10}$ &  39 - 65  & CO\jone & 88 & D \\
 & & & & & & CO\jthree & 266 & D, E \\
D & 0.47 - 0.53 & 4 & zCOSMOS & (4.0 - 5.5)$\times 10^{10}$ & 62 - 74  & CO\jthree & 230 & D, E \\
\hline
\end{tabular}
\caption{Summary of the EGNoG survey.}
\label{sampletable}
\end{table*}

To fill in this observational gap in redshift, we have carried out the Evolution of molecular Gas in Normal
Galaxies (EGNoG) survey\footnote{More information and data products of the EGNoG survey
can be found at \url{http://carma.astro.umd.edu/wiki/index.php/EGNoG}}, a key project at the 
Combined Array for Research in Millimeter-wave Astronomy (CARMA).\footnote{CARMA is a 3-band, 
23-element millimeter interferometer jointly operated by the
California Institute of Technology, University of California Berkeley, 
University of Chicago, University of Illinois at Urbana-Champaign, and University of Maryland.}
The EGNoG survey uses rotational transitions of the carbon monoxide (CO) molecule to 
trace the molecular gas content of 31 star-forming galaxies from $z=0.05$ to 0.5
(the CO\jone line for galaxies at $z=0-0.3$ and the CO\jthree line for galaxies at $z=0.3-0.5$). 
The survey includes four galaxies at $z\approx0.3$ (the gas excitation sample) observed in
both the CO\jone and CO\jthree lines, more than doubling the number of normal SFGs at $z>0.1$ in
which CO line ratios have been measured. The results from the gas excitation sample are presented
separately, in \citet{Bauermeister2013a}. 

In this paper, we present the data for the entire EGNoG sample and use these data to constrain
the evolution of the molecular gas fraction in galaxies at intermediate redshift. 
The paper is organized as follows. In Section \ref{sec:sample} we describe
the selection of the EGNoG survey galaxies and the properties of the sample.
In Section \ref{sec:data} we describe the observations and data reduction (see Appedix \ref{sec:datreducandflux}
for a detailed description of the data reduction and Appendix \ref{sec:polcal} for polarization measurements of calibrators
0854+201 and 1058+015). We present the CO maps, total CO luminosities,
and derived molecular gas masses in Section \ref{sec:results} and discuss the morphologies of the low-redshift
portion of the sample as well as the non-detections of the high-redshift portion.
In Section \ref{sec:discussion}, we present a compilation of data from the literature and analyze the behavior
of the molecular gas depletion time and molecular gas fraction in starburst and normal galaxies at low and high redshift.
We provide some concluding remarks in Section \ref{sec:conclusions}.

Throughout this work, we use a $\Lambda$CDM cosmology with 
(h, $\Omega_\mathrm{M}$, $\Omega_{\Lambda}$) =  (0.7, 0.3, 0.7). 
We use the following terms for star-forming galaxies at low and high redshift:
\begin{itemize} \itemsep -2pt
\item {\bf Star-forming galaxy} refers to any galaxy which is actively forming stars. Quantitatively, these galaxies
lie on or above the main sequence of star-forming galaxies. This term encompasses both normal star-forming and starburst galaxies.  
\item {\bf Normal star-forming galaxy or SFG} refers to star-forming galaxies that lie on the main sequence (within some scatter). 
While the acronym SFG stands for star-forming galaxy, we use it (in keeping with other authors)
to refer to main sequence galaxies only, which comprise the bulk of the star-forming galaxy population.
\item {\bf Starburst galaxy} refers to a galaxy undergoing a period of enhanced star formation. We give a quantitative definition
in Section \ref{sec:sampselect}.
\item {\bf LIRG, ULIRG} (luminous lnfrared galaxy, ultra-luminous lnfrared galaxy) denotes a galaxy with a total infrared luminosity 
between $10^{11}$ and $10^{12}$ L$_\odot$ (LIRG) or $> 10^{12}$ L$_\odot$ (ULIRG).
Local LIRGs and ULIRGs tend to be starbursts, but at high-redshift, normal star-forming galaxies would be classified
as LIRGs or ULIRGs based on their infrared luminosities alone.
\item {\bf SMG} (sub-mm galaxy) refers to the population of galaxies at $z \gtrsim 1$ which have 850-$\mu$m fluxes $\gtrsim 4$ mJy. SMGs
tend to be starbursts (often as a result of a merger, akin to local ULIRGs).
\end{itemize}

\section{The EGNoG Sample}
\label{sec:sample}

Utilizing the CO\jone and CO\jthree in the 3 mm and 1 mm bands of CARMA, 
it is possible to achieve continuous redshift coverage from $z=0$ to $z\sim 0.5$.
The EGNoG survey observed 31 galaxies spanning the accessible redshift range in 4 
redshift bins, A-D. 
The redshift range, sample size, parent sample and range of 
M$_*$ and SFR are given for each redshift bin in Table \ref{sampletable}.

\begin{table*}[!ht]
\centering
\renewcommand{\arraystretch}{1.4}
\begin{tabular}{|c|c|c|c|c|c|c|c|}
\hline
EGNoG & SDSS &  &  & &  & SFR  & sSFR \\
Name & Identification & RA & Dec & $z$ & $\log(M_*/M_\odot)$ & (M$_\odot$ yr$^{-1}$) & (Gyr$^{-1}$) \\
\hline
%DUPLICATES: A1, B4 and C1
 A1$^a$ & SDSS J234311.26+000524.3 & 23:43:11.257 & +00:05:24.319 & $0.096941 \pm 0.000007$ & $10.74^{+0.10}_{-0.09}$ & $30.0^{+10.5}_{-8.1}$ & 0.54 \\
 A2 & SDSS J231332.46+133845.3 & 23:13:32.468 & +13:38:45.301 & $0.081196 \pm 0.000008$ & $10.99^{+0.10}_{-0.10}$ & $28.4^{+5.9}_{-5.2}$ & 0.29 \\
 A3 & SDSS J233455.23+141731.0 & 23:34:55.239 & +14:17:31.093 & $0.062168 \pm 0.000010$ & $11.03^{+0.10}_{-0.09}$ & $21.5^{+9.0}_{-6.4}$ & 0.20 \\
 A4 & SDSS J085504.16+525248.4 & 08:55:04.171 & +52:52:48.320 & $0.089592 \pm 0.000028$ & $10.92^{+0.06}_{-0.10}$ & $49.9^{+5.9}_{-4.6}$ & 0.60 \\
 A5 & SDSS J085307.26+121900.8 & 08:53:07.258 & +12:19:00.827 & $0.080978 \pm 0.000011$ & $11.01^{+0.13}_{-0.11}$ & $42.3^{+6.2}_{-6.3}$ & 0.42 \\
 A6 & SDSS J084549.66+573239.3 & 08:45:49.662 & +57:32:39.329 & $0.085259 \pm 0.000005$ & $10.80^{+0.01}_{-0.01}$ & $35.4^{+26.5}_{-9.0}$ & 0.56 \\
 A7 & SDSS J211527.81-081234.4 & 21:15:27.817 & -08:12:34.441 & $0.090469 \pm 0.000006$ & $10.86^{+0.10}_{-0.09}$ & $40.9^{+12.1}_{-8.7}$ & 0.57 \\
 A8 & SDSS J135751.77+140527.3 & 13:57:51.775 & +14:05:27.317 & $0.099188 \pm 0.000027$ & $11.21^{+0.11}_{-0.10}$ & $ 7.5^{+5.6}_{-2.8}$ & 0.05 \\
 A9 & SDSS J105733.59+195154.2 & 10:57:33.589 & +19:51:54.274 & $0.077338 \pm 0.000007$ & $10.55^{+0.10}_{-0.09}$ & $ 3.7^{+1.8}_{-1.1}$ & 0.10 \\
A10 & SDSS J141601.21+183434.1 & 14:16:01.216 & +18:34:34.171 & $0.055121 \pm 0.000008$ & $10.75^{+0.09}_{-0.09}$ & $ 5.1^{+2.2}_{-1.4}$ & 0.09 \\
A11 & SDSS J100559.89+110919.6 & 10:05:59.890 & +11:09:19.688 & $0.076055 \pm 0.000006$ & $10.72^{+0.09}_{-0.08}$ & $ 8.9^{+5.0}_{-2.4}$ & 0.17 \\
A12 & SDSS J111150.65+281147.7 & 11:11:50.654 & +28:11:47.767 & $0.098171 \pm 0.000012$ & $10.55^{+0.11}_{-0.09}$ & $ 3.4^{+6.1}_{-1.9}$ & 0.10 \\
A13 & SDSS J221938.11+134213.9 & 22:19:38.115 & +13:42:13.890 & $0.083503 \pm 0.000012$ & $11.06^{+0.11}_{-0.10}$ & $11.7^{+12.8}_{-5.8}$ & 0.10 \\
\hline
 B1 & SDSS J223528.63+135812.6 & 22:35:28.638 & +13:58:12.619 & $0.183193 \pm 0.000017$ & $11.43^{+0.12}_{-0.13}$ & $88.2^{+129.9}_{-50.5}$ & 0.33 \\
 B2 & SDSS J002353.97+155947.8 & 00:23:53.973 & +15:59:47.757 & $0.191773 \pm 0.000020$ & $11.29^{+0.12}_{-0.11}$ & $54.6^{+30.1}_{-22.5}$ & 0.28 \\
 B3 & SDSS J100518.63+052544.2 & 10:05:18.640 & +05:25:44.225 & $0.165737 \pm 0.000012$ & $10.79^{+0.09}_{-0.09}$ & $47.1^{+55.9}_{-23.3}$ & 0.76 \\
 B4$^a$ & SDSS J105527.18+064015.0 & 10:55:27.188 & +06:40:15.025 & $0.173078 \pm 0.000013$ & $11.02^{+0.11}_{-0.10}$ & $49.7^{+82.8}_{-28.7}$ & 0.47 \\
 B5 & SDSS J115744.35+120750.8 & 11:57:44.348 & +12:07:50.792 & $0.182661 \pm 0.000010$ & $11.24^{+0.10}_{-0.09}$ & $51.1^{+19.3}_{-14.5}$ & 0.29 \\
 B6 & SDSS J124252.54+130944.2 & 12:42:52.548 & +13:09:44.215 & $0.174661 \pm 0.000013$ & $10.83^{+0.23}_{-0.12}$ & $59.2^{+15.9}_{-9.5}$ & 0.87 \\
 B7 & SDSS J091426.24+102409.6 & 09:14:26.239 & +10:24:09.649 & $0.176180 \pm 0.000018$ & $11.45^{+0.10}_{-0.10}$ & $61.5^{+83.6}_{-36.0}$ & 0.22 \\
 B8 & SDSS J114649.18+243647.7 & 11:46:49.182 & +24:36:47.703 & $0.176579 \pm 0.000021$ & $11.07^{+0.11}_{-0.09}$ & $106.1^{+149.5}_{-81.1}$ & 0.89 \\
 B9 & SDSS J134322.28+181114.1 & 13:43:22.288 & +18:11:14.135 & $0.178004 \pm 0.000019$ & $11.35^{+0.12}_{-0.12}$ & $67.7^{+84.8}_{-37.5}$ & 0.30 \\
B10 & SDSS J130529.30+222019.8 & 13:05:29.304 & +22:20:19.867 & $0.190227 \pm 0.000010$ & $10.96^{+0.09}_{-0.10}$ & $75.0^{+19.7}_{-15.8}$ & 0.82 \\
\hline
 C1$^a$ & SDSS J092831.94+252313.9 & 09:28:31.941 & +25:23:13.925 & $0.283020 \pm 0.000022$ & $11.24^{+0.10}_{-0.11}$ & $38.7^{+85.9}_{-25.6}$ & 0.22 \\
 C2 & SDSS J090636.69+162807.1 & 09:06:36.694 & +16:28:07.136 & $0.300622 \pm 0.000010$ & $11.20^{+0.29}_{-0.14}$ & $57.5^{+90.1}_{-21.9}$ & 0.37 \\
 C3 & SDSS J132047.13+160643.7 & 13:20:47.139 & +16:06:43.720 & $0.312361 \pm 0.000014$ & $11.46^{+0.25}_{-0.12}$ & $64.9^{+142.9}_{-28.0}$ & 0.23 \\
 C4 & SDSS J133849.18+403331.7 & 13:38:49.189 & +40:33:31.748 & $0.285380 \pm 0.000015$ & $11.26^{+0.18}_{-0.12}$ & $50.5^{+48.1}_{-15.4}$ & 0.28 \\
\hline
 D1$^b$ & SDSS J100055.81+015703.8 & 10:00:55.82 & +01:57:03.84 & $0.474100 \pm 0.000367$ & $10.60^{+0.15}_{-0.15}$ & $62.2^{+77.1}_{-34.4}$ & 1.55 \\
 D2$^b$ & SDSS J100052.41+014833.0 & 10:00:52.41 & +01:48:32.92 & $0.527900 \pm 0.000367$ & $10.69^{+0.15}_{-0.15}$ & $71.9^{+89.1}_{-39.8}$ & 1.47 \\
 D3$^b$ & SDSS J095939.07+022249.6 & 09:59:39.07 & +02:22:49.94 & $0.470800 \pm 0.000367$ & $10.73^{+0.15}_{-0.15}$ & $78.7^{+97.5}_{-43.5}$ & 1.47 \\
 D4$^b$ & SDSS J095900.61+022833.0 & 09:59:00.62 & +02:28:33.20 & $0.477800 \pm 0.000367$ & $10.74^{+0.15}_{-0.15}$ & $74.4^{+92.1}_{-41.2}$ & 1.36 \\
\hline
\end{tabular}
\caption{Basic information for EGNoG galaxies. \\
$^a$ Indicates duplicate source in SDSS. The average value is reported for $z$, $M_*$ and SFR. \\
$^b$ Bin D sources are selected from COSMOS. COSMOS identifications for galaxies D1-D4 are zCOSMOS 811469, zCOSMOS 811543, COSMOS 2019408
and zCOSMOS 840823, respectively.}
\label{tab:basicinfo}
\end{table*}

Sample galaxies in redshift bins A-C are drawn from the main spectroscopic sample of the 
Sloan Digital Sky Survey (SDSS), Data Release 7 \citep{York2000, Strauss2002, Abazajian2009}.
Spectroscopic redshifts are from David Schlegel's {\it spZbest} files produced by 
the Princeton-1D code, {\it specBS}\footnote{See \url{http://spectro.princeton.edu/} for more information}.
The stellar masses and SFRs of galaxies in 
the SDSS DR7 are provided by the Max-Planck-Institute for Astrophysics - John 
Hopkins University (MPA-JHU) group (\url{http://www.mpa-garching.mpg.de/SDSS}). 
Stellar masses are derived by fitting SDSS {\it ugriz} photometry to a grid of models
spanning a wide range of star formation histories. This method is found
to compare quite well with the \cite{Kauffmann2003} methodology
using spectral features (more detail on this comparison is found on the website above).
Star formation rates are derived by fitting the fluxes of no less than 5 emission lines
using the method described in \cite{Brinchmann2004}.
Both stellar masses and star formation rates are derived using a Bayesian analysis,
producing probability distributions of each quantity for each galaxy. The probability distributions
of the stellar mass and SFR for galaxies in redshift bins A-C
are given in Figure \ref{fig:masssfrPDFs} in Appendix \ref{sec:msfrPDFs}.
We take the median of 
the distribution, with errors indicated by the 16th and 84th percentile points ($\pm1\sigma$ for a Gaussian distribution). In cases
where a duplicate SDSS source exists (as a result of SDSS automated source-finding), 
we take the average of the two median values and use
the lowest(highest) 16th(84th) percentile value to indicate the negative(positive) error. 

The higher redshift portion of our sample (redshift bin D) is drawn from 
the Cosmic Evolution Survey 
(COSMOS; 2 square degrees at RA $\approx 10$ h, Dec $\approx 2^\circ$; \citealp{Scoville2007})
which has imaging and photometric redshifts for all the galaxies in the survey.
Spectroscopic redshifts are available for many sources (zCOSMOS; \citealp{Lilly2009}). 
We did not use the SDSS for this redshift bin due to the poor coverage of star-forming galaxies at this redshift.
\cite{Lilly2009} report an average accuracy of 110 \kmssp for the spectroscopic redshifts,
independent of redshift.
Stellar masses come from spectral energy distribution (SED) fitting by \cite{Bundy2010} and star formation rates are based on SED fits
provided by \cite{Ilbert2010}. Typical errors are 0.15 dex for the stellar masses and 0.3-0.4 dex for the
star formation rates.

The redshifts, stellar masses and star formation rates for the entire EGNoG sample are given in Table \ref{tab:basicinfo}.

\begin{figure}[t]
\centering
\includegraphics[width=\linewidth]{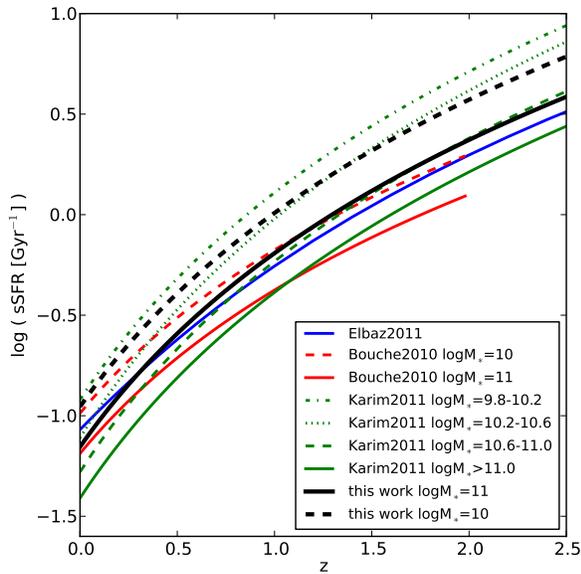}
\caption{Specific star formation rate (sSFR) versus redshift for different stellar masses. Three published forms are shown 
(\citealp{Elbaz2011} in blue, \citealp{Bouche2010} in red, and \citealp{Karim2011} in green) as well as the relation adopted
for this work in black.}
\label{fig:sSFRmodels}
\end{figure}

\subsection{Sample Selection}
\label{sec:sampselect}

The EGNoG galaxies were selected from the parent samples (SDSS and COSMOS)
to be as representative as possible of the `main sequence' (MS) of star-forming galaxies.
The main sequence is the tight correlation between $M_*$ and SFR that has been observed over a large range of redshifts:
e.g. $z\approx0$ \citep{Brinchmann2004}, $z\sim0.2-1$ \citep{Noeske2007}, $z\sim1-2$ \citep{Elbaz2007, Daddi2007, Pannella2009}, 
$z\sim3-4$ \citep{Daddi2009, Magdis2010a} (see also the summary of recent results in \citealp{Dutton2010}).
The sSFR increases with redshift, tracing the increase in the cosmological dark matter 
(and cold baryons, fuel for star formation) accretion rate onto halos \citep{Bouche2010}.
Building on these observations, a few authors have attempted to describe the main sequence relation at all redshifts
with one form. Figure \ref{fig:sSFRmodels} shows the functional forms of the specific star formation rate
(sSFR$ \equiv \mathrm{SFR} / M_*$) as a function of redshift
reported by three papers: \cite{Bouche2010}, \cite{Karim2011} and \cite{Elbaz2011}.
\cite{Bouche2010} suggest that existing literature data are well-fit by the form sSFR $\propto M_*^{-0.2} (1+z)^{2.7}$ for $z=0-2$.
\cite{Karim2011} use a deep 3.6 $\mu$m selected sample of galaxies from the COSMOS field, with average SFRs determined from
stacked 1.4 GHz radio continuum emission. Like \cite{Bouche2010}, they find the sSFR to be a function of stellar mass, with galaxies
with lower stellar masses having higher specific star formation rates. Specifically, the authors find sSFR $\propto M_*^{-0.4} (1+z)^{3.5}$ 
(on average) for star-forming galaxies at $z=0-3$ (in Figure \ref{fig:sSFRmodels}, we plot the sSFR fit for
each of the \cite{Karim2011} mass bins individually).
On the other hand, \cite{Elbaz2011} argue that observations at all redshifts are consistent with a sSFR independent of stellar mass. 
The authors use far-infrared observations from Herschel Space Observatory in conjunction with existing data on the GOODS-North and 
GOODS-South fields to fit the sSFR versus redshift, finding sSFR (Gyr$^{-1}$) $\propto t_\mathrm{cosmic}^{-2.2}$ for $z=0-3$
(where $t_\mathrm{cosmic}$ is the time since the Big Bang in Gyr). 
The sSFR forms from \cite{Bouche2010} (red) and \cite{Karim2011} (green) are shown for different stellar masses in 
Figure \ref{fig:sSFRmodels}. As \cite{Elbaz2011} argues that the sSFR is independent of sSFR, only one curve (blue) is plotted.

\begin{figure*}[t]
\centering
\includegraphics[width=0.24\linewidth]{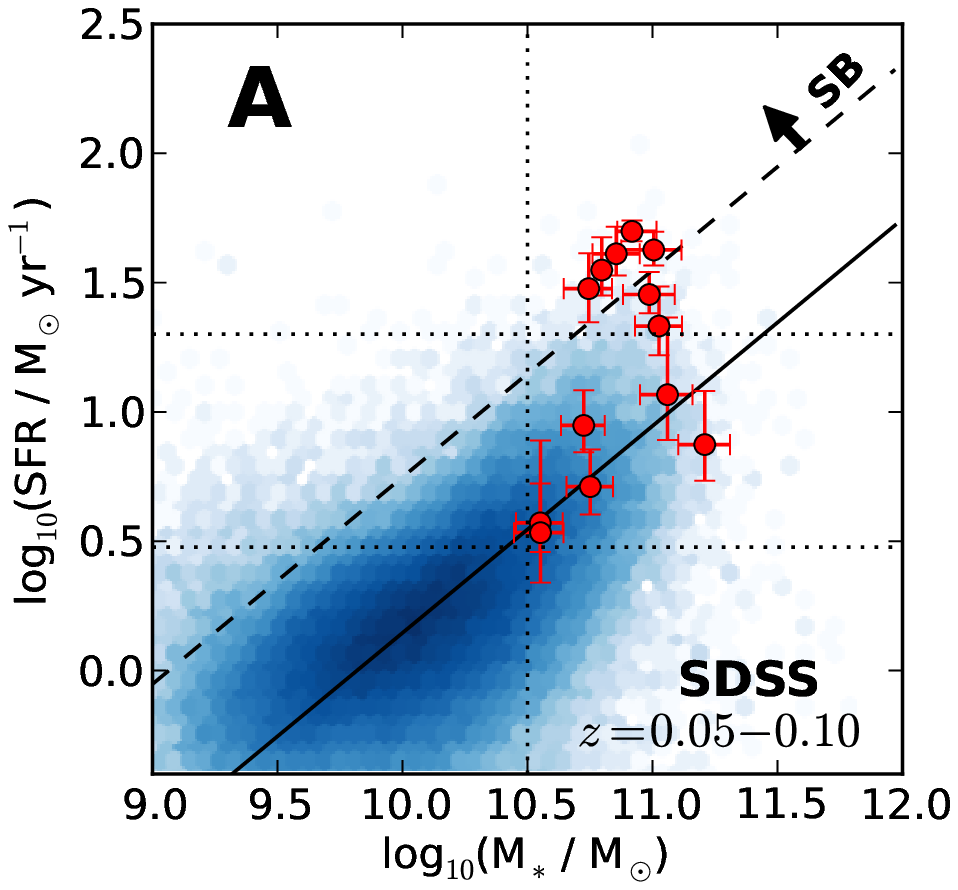}
\includegraphics[width=0.24\linewidth]{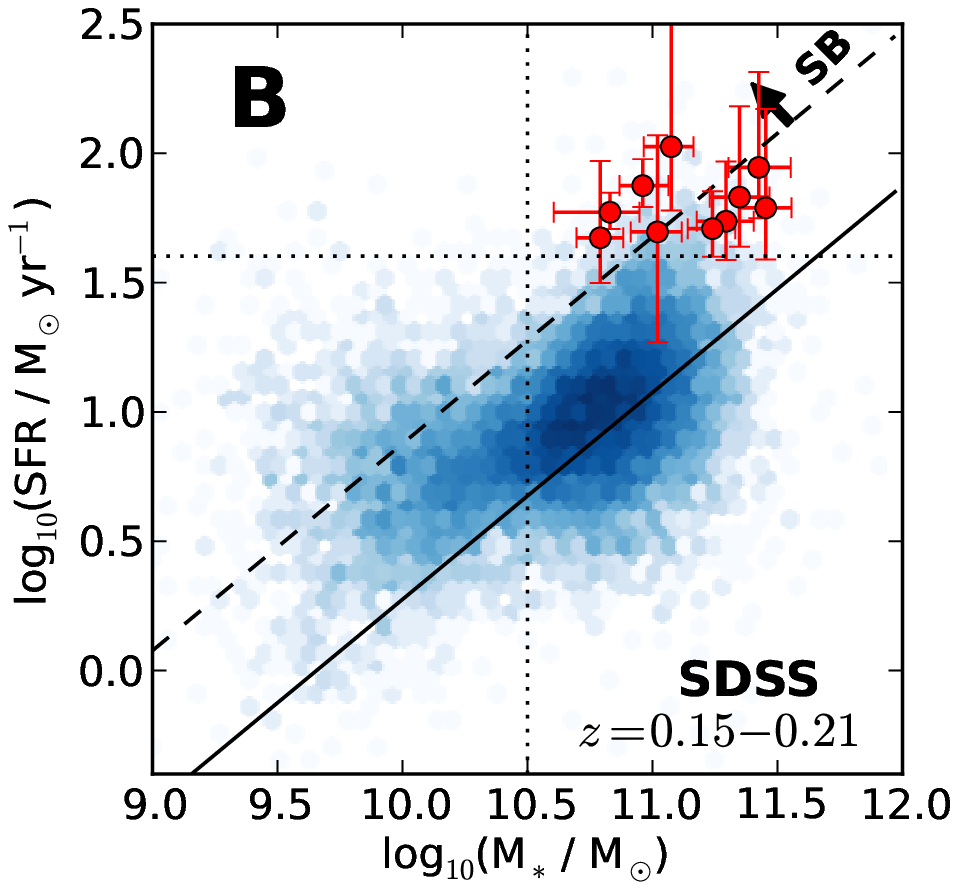}
\includegraphics[width=0.24\linewidth]{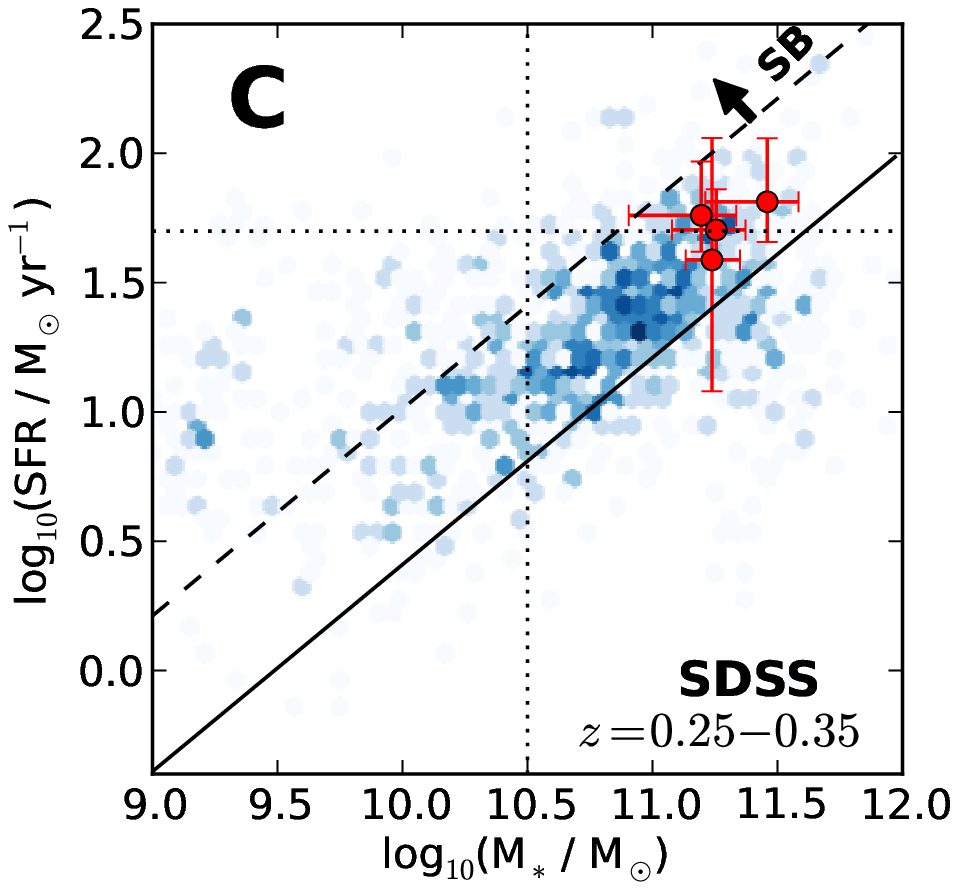}
\includegraphics[width=0.24\linewidth]{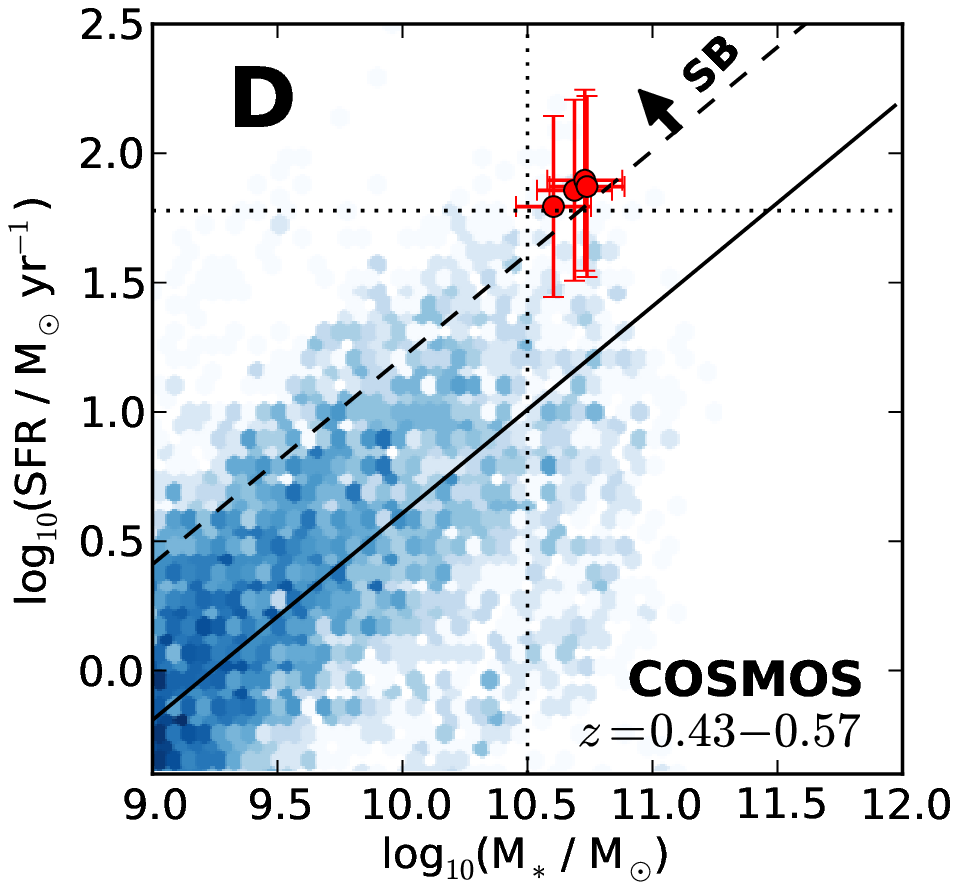}
\caption{Stellar mass versus SFR in each redshift bin. The redshift bin is indicated in the upper left corner of each panel. 
Red points with error bars show EGNoG galaxies. Dotted vertical and horizontal lines indicate the minimum $M_*$ and
SFR, respectively, required for sample selection in each redshift bin (see text for full description). 
Blue shading shows the logarithm of the density (in the $M_*$-SFR plane) of star-forming galaxies from 
the parent dataset in the specified redshift range (given in the lower right corner).
The redshift ranges plotted in bins B-D are slightly larger than the EGNoG redshift ranges, so that more 
data points may be included to better capture the behavior of the main sequence at each redshift.
The solid black line indicates the main sequence (Equation \ref{equ:myssfr})
at the average redshift of the bin. The starburst criterion (Equation \ref{equ:mySBssfr}) is indicated by the black dashed line.}
\label{fig:massvsfr}
\end{figure*}

For this work, we are only concerned with the behavior of sSFR($z$) out to $z\approx2.5$, but note
that there is evidence for a flattening of this relation at $z \gtrsim 2-2.5$ \citep[e.g.][]{Stark2009, Gonzalez2010, Reddy2012}.
For $z< 2.5$, we adopt a sSFR relation which roughly agrees with the three forms described above. 
Specifically, we define
\begin{equation}
\mathrm{sSFR}_\mathrm{MS}(\mathrm{Gyr}^{-1}) = 0.07 (1+z)^{3.2} \left(\frac{M_*}{10^{11}\ \mathrm{M}_\odot}\right)^{-0.2} 
\label{equ:myssfr}
\end{equation}
In Figure \ref{fig:sSFRmodels}, this is plotted in black for $M_*=10^{10}$ M$_\odot$ (dashed) and $10^{11}$ M$_\odot$ (solid).
We use this representative form of the sSFR to describe the main sequence of star-forming galaxies throughout this work.

To differentiate `starburst' (SB) galaxies from normal main sequence galaxies, we adopt the quantitative definition 
given by \cite{Rodighiero2011}, a study of PACS/Herschel observations of COSMOS and GOODS-South galaxies at $z=1.5-2.5$. 
\cite{Rodighiero2011} fit the main sequence population with a Gaussian distribution and define starbursts as
outliers of this distribution. They find deviations from the Gaussian main sequence starting at a sSFR four times that
of the main sequence. Therefore, we define starbursts as having 
\begin{equation}
\mathrm{sSFR}_\mathrm{SB} > 4 \times \mathrm{sSFR}_\mathrm{MS}
\label{equ:mySBssfr}
\end{equation}

In the selection of the EGNoG sample, we apply the following criteria in each redshift bin
in order to identify non-interacting, star-forming galaxies lying as close to the main sequence as possible.
At $z \lesssim 0.3$ (bins A-C), star-forming galaxies were selected (rejecting sources with AGN) using the cut 
from \cite{Kauffmann2003} in the BPT line-ratio diagram \citep{BPT1981}. We did not extend this criterion
to bin D at $z \approx 0.5$ since the \cite{Kauffmann2003} dataset only covers $z=0.02-0.3$.
Obviously interacting galaxies were excluded via visual inspection of the SDSS or COSMOS optical images.
However, some interacting galaxies may remain in the sample due to the difficulty of identification 
at the modest resolution of the SDSS images.

Practical considerations imposed the following further constraints.
From the SDSS dataset, we only used spectroscopically targeted galaxies, for which spectroscopic redshifts, 
stellar masses and SFRs were available from the MPA-JHU group.
We required a spectroscopic redshift so that the error in the redshift is
small enough to ensure that CO emission would be captured within the observed bandwidth.
We excluded galaxies with SFRs below a minimum value estimated from the instrument 
sensitivity in each redshift range, assuming a molecular gas depletion time 
($\tau_\mathrm{mgas} = \Mmgas / \mathrm{SFR}$) of 2 Gyr \citep{Leroy2008}, 
a Milky Way-like CO conversion factor \citep[e.g.][]{Dame2001} and
(for bins C and D) a CO\jthree to CO\jone line ratio ($r_{31}$) of 0.6 \citep{Aravena2010}.
Note that this is close to $r_{31} = 0.5$, found in the EGNoG bin C galaxies at $z\approx 0.3$
(see \citealp{Bauermeister2013a}). 
In bin A, we used two SFR cutoffs since this bin is composed of two subsamples: 
high-SFR galaxies (EGNoG A1-A6) observed as a pilot study for the EGNoG survey, 
and a lower-SFR cutoff sample (EGNoG A7-A13) that is more representative of the main sequence.
Finally, in order to restrict our selection to the main sequence
at the SFRs probed, we required a minimum stellar mass of $10^{10.5}$ M$_\odot$. 

In each redshift bin, the EGNoG galaxies were selected randomly from the subset of the parent sample
that satisfied the constraints discussed above.
Figure \ref{fig:massvsfr} shows stellar mass versus SFR in each of the four redshift bins. 
EGNoG galaxies are indicated by the red points with error bars. 
The vertical and horizontal dotted lines indicate the minimum $M_*$ and SFR, respectively, 
required for sample selection in each redshift bin. Galaxy C1 lies below the minimum SFR 
for bin C because it is a duplicate source in the SDSS catalog and was selected based on the higher
SFR. The average SFR of the duplicate entries lies below the minimum SFR for bin C. 
The blue shading
indicates the (logarithm of the) density of points in the $M_*$ - SFR plane for all star-forming galaxies 
(spectroscopic targets only for the SDSS data, bins A-C) in the redshift range indicated in the lower right of each panel.
The redshift ranges plotted for bins B-D are slightly larger than the EGNoG redshift ranges so that more points may be included 
to better capture the behavior of the main sequence at each redshift. The blue shading corresponds to approximately 100,000 galaxies at $z=0.05-0.1$, 
15,000 galaxies at $z=0.15-0.21$, 1300 galaxies at $z=0.25-0.35$, and 14,000 galaxies at $z=0.43-0.57$.
The main sequence of star-forming galaxies at each redshift
is indicated by the solid black line, with the starburst cutoff indicated by the dashed black line.
From bin A to bin C, the low-mass, low-SFR end of the main sequence becomes sparsely sampled as the 
number of spectroscopically targeted, star-forming galaxies available in the SDSS decreases. This effect is most
dramatic in the bin C panel, where the sample of spectroscopically-targeted star-forming galaxies in the SDSS 
is an order of magnitude smaller than in the bin B redshift range.  
Conversely, the small area covered by the COSMOS survey results in a sparsely sampled high-mass end of the
main sequence in the bin D panel. 

The result of our sample selection is that the EGNoG galaxies generally lie at the high-$M_*$, high-SFR end of the main sequence
of star-forming galaxies. Our starburst criterion suggests that roughly half of the galaxies in each of bins A and B,
and all of the bin D galaxies are starbursts. All four bin C galaxies lie below the starburst cut.
In each redshift bin, the EGNoG galaxies lie roughly within the width of the scatter around the main sequence. 
The starburst classification of some of the EGNoG galaxies will be considered further in Section \ref{sec:discussion}.

\section{CARMA Observations}
\label{sec:data}

Each of the 31 EGNoG galaxies was observed in at least one rotational transition of the CO
molecule: galaxies in bins A, B and C were observed in 
CO\jone ($\nu_\mathrm{rest} = 115.3$ GHz, $\nu_\mathrm{obs} \approx 107$, 98 and 88 GHz respectively);
galaxies in bins C and D were observed in 
CO\jthree ($\nu_\mathrm{rest} = 345.9$ GHz, $\nu_\mathrm{obs} \approx 266$ and 230 GHz respectively).
At each frequency, each galaxy was observed over several different days.
Each dataset includes observations of a nearby quasar for phase calibration (taken every 15-20 minutes), 
a bright quasar for passband calibration and either a planet (Uranus, Neptune or Mars) or MWC349 for flux calibration (in most cases). 

The reduction of all observations for this survey was carried out within the 
EGN\footnote{\url{http://carma.astro.umd.edu/wiki/index.php/EGN}} data reduction infrastructure
(based on the MIS pipeline; \citealp{PoundTeuben2012}) using the Multichannel Image Reconstruction, 
Image Analysis and Display (MIRIAD, \citealp{MIRIAD}) package for radio interferometer data reduction.
Our data analysis also used the {\tt miriad-python} software package \citep{miriadpython}.
The data were flagged, passband-calibrated and phase calibrated in the standard way. Final images
were created using {\tt invert} with {\tt options=mosaic} to properly handle and correct for the three different
primary beam patterns. All observations are single-pointing. We describe the CO\jone and CO\jthree observations
individually below. A full description of the data reduction and flux measurement is given in Appendix \ref{sec:datreducandflux}.
%The details of the individual observations (date, length, weather, calibrators, etc.) are given in Tables
%\ref{tab:3mmobs} and \ref{tab:1mmobs}.

\subsection{CO\jone}
The CO\jone transition lies in the 3~mm band of CARMA (single-polarization, linearly polarized feeds)
for galaxies in bins A, B and C. The line was observed with at least three overlapping 500 MHz bands, 
covering $\approx$ 4200, 4600 and 5000 \kmssp total, 
at 35, 39 and 42 \kmssp resolution for galaxies in bins A, B and C, respectively. 

Bin A galaxies were observed during three time periods: October to November 2010, April to May 2011 and February 2012.
All data were taken in CARMA's C configuration, with $26-370$ m baselines yielding a typical synthesized beam of 
$2.0\arcsec \times 1.5\arcsec$ at 107 GHz. These observations are sensitive to spatial scales up to $\approx 22\arcsec$,
which is sufficient for most galaxies in the sample but may resolve out some large scale structure in the largest galaxies. 
However, since molecular gas is centrally concentrated, we expect any under-estimation of the flux to be minimal.
(In D and E configuration in the 3mm band, we are sensitive to sufficiently large spatial scales so that we expect
to recover all of the flux.)
Each galaxy was observed for 2-3 hours (time on-source),
yielding final data cubes with a typical rms noise of $5-10$ mJy beam$^{-1}$ in 35 \kmssp channels. 

Bin B sources were observed from August to November 2011 and April 2012 in CARMA's D configuration, with
$11-150$ m baselines yielding a typical synthesized beam of $4.9\arcsec \times 3.9\arcsec$ at 98 GHz.
Supplementary observations were made of B1, B2, B3 and B7 in CARMA's C array in February 2012, 
which resulted in a synthesized beam of $3.0\arcsec \times 2.3\arcsec$ in the final maps combining data from both array configurations.
Each galaxy was observed in D configuration for approximately 7 hours (time on-source), resulting in 
a typical rms noise of 2.5 mJy beam$^{-1}$ in 39 \kmssp channels in the final cube.

Bin C observations were made from August to November 2011 in CARMA's D configuration, with a typical synthesized beam of 
$4.8\arcsec \times 3.9\arcsec$ at 88 GHz. Each galaxy was observed for 20 to 30 hours (time on-source),
yielding final cubes with a typical rms noise of $1.2$ mJy beam$^{-1}$ in 42 \kmssp channels. 

The flux scale in each dataset is set by the flux of the phase
calibrator, which is determined from the flux calibrator. For the fluxes used, see Appendix \ref{sec:datreducandflux}.

Of the 27 galaxies observed in the CO\jone line, we detected 24 easily, with 3 non-detections in bin A.
The $uv$-spectra of the detected galaxies are shown in Appendix \ref{sec:uvspectra}. 
%The 3 non-detections are discussed further in Appendix \ref{sec:nondetections}.

\subsection{CO\jthree}
The CO\jthree transition lies in the CARMA 1~mm band (dual-polarization, circularly polarized feeds)
for galaxies in bins C and D.
We again observed the line with at least three overlapping 500 MHz bands, covering $\approx$ 1500 and 1700 \kmssp total, 
at 14 and 15 \kmssp resolution, for galaxies in bins C and D respectively. 

Bin C observations were carried out in CARMA's E configuration during August 2011
and D configuration during April 2012. 
Source C4 was observed entirely in E configuration, while the other three sources
were observed mostly in D configuration. The E configuration has 
$8-66$ m baselines yielding a typical synthesized beam of 
$3.2\arcsec \times 2.5\arcsec$ at 266 GHz.
The D configuration has $11-150$ m baselines yielding a typical synthesized beam of 
$1.7\arcsec \times 1.5\arcsec$ at 266 GHz.
Each galaxy was observed for 2 to 6.5 hours (time on-source), yielding final cubes with a rms noise of 5-10 mJy beam$^{-1}$
in 42 \kmssp channels. 

Bin D observations were carried out in CARMA's E and D configuration during August 2011
and April to June 2012, respectively. In E configuration, 
a typical synthesized beam is $4.2\arcsec \times 3.6\arcsec$ at 230 GHz.
In D configuration, a typical synthesized beam is $2.5\arcsec \times 1.7\arcsec$ at 230 GHz.
Galaxies D1 and D2 were observed in both D and E configuration for approximately 10 hours (time on-source) total,
yielding final cubes with a rms noise of $\approx 7$~mJy beam$^{-1}$ in 48 \kmssp channels. 
Galaxies D3 and D4 were observed in D configuration for approximately 3 hours (time on-source), 
yielding final cubes with a rms noise of $\approx 10$~mJy beam$^{-1}$ in 48 \kmssp channels.

In bin C, sources C2, C3 and C4 were detected at the $\approx5\sigma$ level in the CO\jthree line.
However, C1, with its wide velocity profile (observed in the CO\jone line), was only marginally detected and we
provide an upper limit on the CO\jthree flux. The $uv$-spectra for the bin C galaxies are shown in Appendix \ref{sec:uvspectra}.
In bin D, we do not significantly detect any of the four galaxies, and list only upper limits. 
The bin D galaxies are discussed further in Section \ref{sec:binDdisc}

\section{Results}
\label{sec:results}

\subsection{CO Emission Maps and Molecular Gas Masses}
\label{sec:derivedprop}
Table \ref{tab:COprop} presents the quantities we derive from the CO observations.
%The derivation of the line flux ($S_{CO}$ in Jy \kms) is described in Section \ref{sec:fluxest}. 
The CO line luminosity is calculated from the line flux ($S_\mathrm{CO}$ in Jy \kms, calculated 
as described in Section \ref{sec:fluxest}) following 
\begin{equation}
\label{LCOdef}
L_\mathrm{CO}' = 3.25 \times 10^7 \ S_\mathrm{CO} \ \nu_\mathrm{obs}^{-2} \ r_\mathrm{com}^2 (1+z)^{-1}
\end{equation}
\citep[see the review by ][]{SolomonVandenBout2005},
where $\nu_\mathrm{obs}$ is in GHz and $r_\mathrm{com}$ is the comoving distance in Mpc. The units
of $L_\mathrm{CO}'$ are \Kkmspc. We report the measurement error 
for $S_\mathrm{CO}$. The error reported for $L_\mathrm{CO}'$ includes both the measurement error
of $S_\mathrm{CO}$ and a 30\% systematic error, added in quadrature (see Section \ref{sec:fluxest} for more details).

The center velocity, $v_\mathrm{center}$, is the flux-weighted average
velocity of the galaxy-integrated spectrum ($v=0$ at the redshift in Table \ref{tab:basicinfo}). The error reported
is the standard deviation of the $v_\mathrm{center}$ values found with the three flux measurement methods and
different channel averaging described in Section \ref{sec:fluxest}. 
The reported velocity width ($\Delta V$) is the full width of the emission, where `source' velocity channels are
selected by eye. We give the velocity width of a single channel as the error.

In the case of a non-detection (galaxies A4, A6, A12, D1, D2, D3, and D4; indicated by $^a$ in Table \ref{tab:COprop}), 
we estimate a $3\sigma$ upper limit on $S_\mathrm{CO}$ from the noise in the channel maps and a 
maximum velocity width motivated by detections at that redshift: 600 \kmssp in bin A and 400 \kmssp in bin D.
For the CO\jthree line in galaxy C1, while the CO\jthree channel maps
did not show evidence of a source upon visual inspection, an integrated spectrum made of a circular region $4.5\arcsec$
in radius at the center of the image suggests a $3\sigma$ detection. We calculate an upper limit on the line flux (and related
quantities) from this spectrum, over the velocities of the CO\jone line in C1. 

\begin{table*}[!t]
\centering
%                   name   trans           SCO                                         vcenter                                  DeltaV                                    LCO                                   Mmgas                    
%    fgas                                            tdep
\renewcommand{\arraystretch}{1.4}
\begin{tabular}{| c | c | x{20pt}@{$\ \pm\ $}x{16pt} | x{23pt}@{$\ \pm\ $}x{16pt} | x{22pt}@{$\ \pm\ $}x{20pt} | x{35pt}@{$\ \pm\ $}x{16pt} | c | x{22pt}@{$\ \pm\ $}x{20pt} | c | c |}
\hline
Name  & J$_\mathrm{up}$ &  \multicolumn{2}{c|}{$S_\mathrm{CO}$}  & \multicolumn{2}{c|}{$v_\mathrm{center}$} & \multicolumn{2}{c|}{$\Delta V$} & \multicolumn{2}{c|}{$L_\mathrm{CO}'$} & sSFR & \multicolumn{2}{c|}{$M_\mathrm{mgas}$} & $f_\mathrm{mgas}$ & $\tau_\mathrm{dep}$ \tn
 & & \multicolumn{2}{c|}{(Jy \kms)} & \multicolumn{2}{c|}{(\kms)} & \multicolumn{2}{c|}{(\kms)} & \multicolumn{2}{c|}{($10^9$ \Kkmspc)} & class &  \multicolumn{2}{c|}{($10^{9} M_\odot$) } &  &  (Gyr)  \tn 
 \hline 
A1 & 1 & 10.75 & 2.99 & 33.7 & 27.3 & 428.0 & 71.3 & 4.76 & 1.95  & SB & 5.18 & 2.12 & $0.09^{+0.04}_{-0.04}$ & $0.17^{+0.09}_{-0.08}$ \tn
A2 & 1 & 14.40 & 2.55 & 30.5 & 12.6 & 281.2 & 35.2 & 4.44 & 1.55  & norm & 19.32 & 6.73 & $0.17^{+0.06}_{-0.06}$ & $0.68^{+0.27}_{-0.27}$ \tn
A3 & 1 & 17.94 & 2.07 & 22.8 & 6.8 & 552.5 & 34.5 & 3.22 & 1.03  & norm & 14.01 & 4.50 & $0.12^{+0.04}_{-0.04}$ & $0.65^{+0.29}_{-0.29}$ \tn
A4$^a$ & 1 & 5.98 & 1.98 & 0.0 & 0.0 & 600.0 & 0.0 & 2.25 & 1.01  & SB & 2.45 & 1.09 & $0.03^{+0.01}_{-0.01}$ & $0.05^{+0.02}_{-0.02}$ \tn
A5 & 1 & 23.22 & 3.90 & -60.3 & 9.8 & 492.0 & 35.1 & 7.12 & 2.45  & SB & 7.75 & 2.66 & $0.07^{+0.03}_{-0.03}$ & $0.18^{+0.07}_{-0.07}$ \tn
A6$^a$ & 1 & 3.65 & 1.23 & 0.0 & 0.0 & 600.0 & 0.0 & 1.24 & 0.56  & SB & 1.35 & 0.61 & $0.02^{+0.01}_{-0.01}$ & $0.04^{+0.02}_{-0.02}$ \tn
A7 & 1 & 6.72 & 1.52 & 49.2 & 31.9 & 319.0 & 106.3 & 2.58 & 0.97  & SB & 2.81 & 1.05 & $0.04^{+0.02}_{-0.02}$ & $0.07^{+0.03}_{-0.03}$ \tn
A8 & 1 & 4.52 & 1.35 & -61.0 & 20.5 & 357.3 & 71.5 & 2.10 & 0.89  & norm & 9.14 & 3.87 & $0.05^{+0.03}_{-0.02}$ & $1.22^{+0.70}_{-0.70}$ \tn
A9 & 1 & 5.68 & 1.42 & -7.1 & 17.7 & 420.3 & 70.1 & 1.59 & 0.62  & norm & 6.92 & 2.70 & $0.16^{+0.06}_{-0.06}$ & $1.86^{+0.91}_{-0.91}$ \tn
A10 & 1 & 7.88 & 3.16 & 4.3 & 17.0 & 274.4 & 68.6 & 1.11 & 0.56  & norm & 4.83 & 2.42 & $0.08^{+0.04}_{-0.04}$ & $0.94^{+0.54}_{-0.54}$ \tn
A11 & 1 & 6.89 & 0.98 & -10.5 & 11.2 & 419.8 & 70.0 & 1.86 & 0.62  & norm & 8.09 & 2.69 & $0.13^{+0.05}_{-0.04}$ & $0.91^{+0.39}_{-0.39}$ \tn
A12$^a$ & 1 & 3.50 & 1.16 & 0.0 & 0.0 & 600.0 & 0.0 & 1.59 & 0.71  & norm & 6.92 & 3.09 & $0.16^{+0.07}_{-0.07}$ & $2.03^{+1.45}_{-1.45}$ \tn
A13 & 1 & 21.89 & 4.12 & -27.3 & 8.9 & 563.5 & 35.2 & 7.15 & 2.53  & norm & 31.12 & 11.02 & $0.21^{+0.08}_{-0.07}$ & $2.67^{+1.63}_{-1.63}$ \tn
\hline
B1 & 1 & 11.35 & 0.62 & -23.1 & 4.0 & 615.4 & 38.5 & 18.60 & 5.67  & norm & 80.95 & 24.68 & $0.23^{+0.08}_{-0.07}$ & $0.92^{+0.60}_{-0.60}$ \tn
B2 & 1 & 7.32 & 0.40 & 25.5 & 11.7 & 581.1 & 38.7 & 13.20 & 4.03  & norm & 57.45 & 17.52 & $0.23^{+0.08}_{-0.07}$ & $1.05^{+0.54}_{-0.54}$ \tn
B3 & 1 & 4.49 & 0.35 & -23.1 & 6.6 & 378.9 & 37.9 & 5.97 & 1.85  & SB & 6.50 & 2.01 & $0.10^{+0.03}_{-0.03}$ & $0.14^{+0.08}_{-0.08}$ \tn
B4 & 1 & 4.81 & 0.20 & -25.6 & 5.2 & 457.6 & 38.1 & 6.99 & 2.12  & SB & 7.61 & 2.30 & $0.07^{+0.03}_{-0.02}$ & $0.15^{+0.26}_{-0.10}$ \tn
B5 & 1 & 6.61 & 0.22 & 16.1 & 2.5 & 307.6 & 38.5 & 10.70 & 3.23  & norm & 46.57 & 14.06 & $0.21^{+0.07}_{-0.06}$ & $0.91^{+0.38}_{-0.38}$ \tn
B6 & 1 & 3.52 & 0.31 & -133.6 & 16.4 & 343.7 & 38.2 & 5.22 & 1.63  & SB & 5.68 & 1.78 & $0.08^{+0.05}_{-0.03}$ & $0.10^{+0.03}_{-0.03}$ \tn
B7 & 1 & 5.97 & 0.30 & -40.9 & 4.9 & 497.0 & 38.2 & 9.01 & 2.74  & norm & 39.21 & 11.93 & $0.12^{+0.04}_{-0.04}$ & $0.64^{+0.42}_{-0.42}$ \tn
B8 & 1 & 2.25 & 0.33 & -34.8 & 20.4 & 382.5 & 76.5 & 3.41 & 1.14  & SB & 3.71 & 1.24 & $0.03^{+0.01}_{-0.01}$ & $0.03^{+0.03}_{-0.03}$ \tn
B9 & 1 & 3.91 & 0.27 & -18.6 & 5.9 & 459.5 & 38.3 & 6.03 & 1.86  & norm & 26.24 & 8.08 & $0.11^{+0.04}_{-0.04}$ & $0.39^{+0.25}_{-0.25}$ \tn
B10 & 1 & 5.25 & 0.22 & 27.1 & 2.6 & 270.9 & 38.7 & 9.28 & 2.81  & SB & 10.10 & 3.06 & $0.10^{+0.03}_{-0.03}$ & $0.13^{+0.05}_{-0.05}$ \tn
\hline
C1 & 1 & 2.05 & 0.34 & -84.9 & 18.1 & 542.2 & 41.7 & 8.25 & 2.83  & norm & 35.90 & 12.31 & $0.17^{+0.06}_{-0.06}$ & $0.93^{+2.09}_{-0.69}$ \tn
C2 & 1 & 2.35 & 0.10 & -6.7 & 1.2 & 253.7 & 42.3 & 10.70 & 3.24  & norm & 46.57 & 14.11 & $0.23^{+0.18}_{-0.07}$ & $0.81^{+0.39}_{-0.39}$ \tn
C3 & 1 & 4.39 & 0.09 & -24.7 & 5.7 & 384.0 & 42.3 & 21.70 & 6.53  & norm & 94.44 & 28.40 & $0.25^{+0.15}_{-0.07}$ & $1.45^{+0.77}_{-0.77}$ \tn
C4 & 1 & 3.01 & 0.11 & 17.3 & 13.0 & 292.5 & 41.8 & 12.30 & 3.72  & norm & 53.53 & 16.18 & $0.23^{+0.10}_{-0.07}$ & $1.06^{+0.45}_{-0.45}$ \tn
\hline
C1$^a$ & 3 & 9.02 & 2.65 & -49.9 & 7.3 & 542.3 & 23.4 & 4.03 & 1.69  & norm & 35.08 & 14.73 & $0.17^{+0.07}_{-0.07}$ & $0.91^{+2.05}_{-0.71}$ \tn
C2 & 3 & 8.39 & 1.12 & 19.5 & 8.2 & 169.2 & 42.3 & 4.25 & 1.40  & norm & 36.99 & 12.15 & $0.19^{+0.16}_{-0.07}$ & $0.64^{+0.32}_{-0.32}$ \tn
C3 & 3 & 13.46 & 1.56 & -14.0 & 7.8 & 426.7 & 42.7 & 7.39 & 2.38  & norm & 64.32 & 20.69 & $0.18^{+0.12}_{-0.06}$ & $0.99^{+0.53}_{-0.53}$ \tn
C4 & 3 & 11.88 & 1.22 & 33.5 & 5.0 & 250.7 & 41.8 & 5.41 & 1.72  & norm & 47.09 & 14.93 & $0.21^{+0.10}_{-0.07}$ & $0.93^{+0.41}_{-0.41}$ \tn
\hline
D1$^a$ & 3 & 3.36 & 1.12 & 0.0 & 0.0 & 400.0 & 0.0 & 4.38 & 1.96  & SB & 9.53 & 4.27 & $0.19^{+0.09}_{-0.08}$ & $0.15^{+0.20}_{-0.11}$ \tn
D2$^a$ & 3 & 2.73 & 0.91 & 0.0 & 0.0 & 400.0 & 0.0 & 4.45 & 2.00  & SB & 9.68 & 4.34 & $0.17^{+0.08}_{-0.07}$ & $0.13^{+0.18}_{-0.10}$ \tn
D3$^a$ & 3 & 3.39 & 1.13 & 0.0 & 0.0 & 400.0 & 0.0 & 4.36 & 1.96  & SB & 9.49 & 4.25 & $0.15^{+0.08}_{-0.07}$ & $0.12^{+0.16}_{-0.09}$ \tn
D4$^a$ & 3 & 5.19 & 1.73 & 0.0 & 0.0 & 400.0 & 0.0 & 6.88 & 3.09  & SB & 14.97 & 6.71 & $0.21^{+0.10}_{-0.09}$ & $0.20^{+0.27}_{-0.14}$ \tn
\hline
\end{tabular}
\caption{Properties of the CO emission of the EGNoG galaxies (described in Section \ref{sec:derivedprop}). See Section 
\ref{sec:fluxest} for a description of the errors. \\
$^a$ CO quantities are upper limits.}
\label{tab:COprop}
\end{table*}

The molecular gas mass ($M_\mathrm{mgas}$) is calculated from $L_\mathrm{CO}'$ according to
\begin{equation}
\label{equ:Mmgasdef}
M_\mathrm{mgas} = 1.36~ \alpha_\mathrm{CO} ~L_\mathrm{CO(J-(J-1))}' ~/~ r_\mathrm{J1}
\end{equation}
where $\alpha_\mathrm{CO}$ is the CO\jone luminosity to \h2 mass conversion factor (akin to $X_\mathrm{CO}$) and
the factor of 1.36 accounts for Helium to yield the total molecular gas mass. $r_\mathrm{J1}$ is the
ratio of the CO line luminosities: $r_{J1} = L_\mathrm{CO}'(J\rightarrow J-1) / L_\mathrm{CO}'(1\rightarrow0)$. 
In the calculation of the
molecular gas mass from the CO\jthree luminosity (bin C and D observations), we use $r_{31} = 0.5$, 
observed in the EGNoG bin C galaxies \citep{Bauermeister2013a}.  
We use an $\alpha_\mathrm{CO}$ value selected according to the starburst classification
described by Equations \ref{equ:myssfr} and \ref{equ:mySBssfr}. 
For galaxies classified as starburst, we use the value observed in local ULIRGs:
$\alpha_\mathrm{CO}(\mathrm{ULIRG}) = 0.8$ M$_\odot$ (K km s$^{-1}$ pc$^2$)$^{-1}$ 
\citep{Scoville1997,DownSol1998}. For all other galaxies, we use a Milky Way-like conversion factor, 
$\alpha_\mathrm{CO}(\mathrm{Milky Way}) = 3.2$ M$_\odot$ (K km s$^{-1}$ pc$^2$)$^{-1}$
\citep[e.g.][]{Dame2001}. This classification (SB or norm), and the molecular gas mass calculated using
the corresponding $\alpha_\mathrm{CO}$ is reported for each galaxy in Table \ref{tab:COprop}.
There is of course considerable uncertainty in the appropriate $\alpha_\mathrm{CO}$ value, 
but we do not reflect this uncertainty in the error in the molecular gas mass. The error in $M_\mathrm{mgas}$ 
is calculated by propagating the error in $L_\mathrm{CO}'$. 
The classification of galaxies using the specific star formation rate is described in more detail
in Section \ref{sec:sbvnormal}. In Section \ref{sec:bimodalalpha}, we discuss the use of a bimodal
prescription for $\alpha_\mathrm{CO}$ in this work and the the possibility of a continuously varying conversion factor. 

We report the molecular gas fraction and the molecular gas depletion time in the last two columns of Table \ref{tab:COprop}. 
In keeping with previous work, we define the molecular gas fraction ($f_\mathrm{mgas}$) as the fraction of baryonic matter 
(ignoring atomic and ionized gas):
\begin{equation}
\label{equ:fmgasdef}
f_\mathrm{mgas} = \frac{M_\mathrm{mgas}}{M_* + M_\mathrm{mgas}}
\end{equation}
The molecular gas depletion time ($\tau_{mgas}$) represents the time it would take to consume all the available
molecular gas at the current star formation rate, defined:
\begin{equation}
\label{equ:tdepdef}
\tau_\mathrm{dep} = \frac{M_\mathrm{mgas}}{SFR}
\end{equation}
The errors in both of these quantities include both the error in the CO luminosity and the error in the 
stellar mass or SFR.

\begin{figure*}
\includegraphics[width=7in]{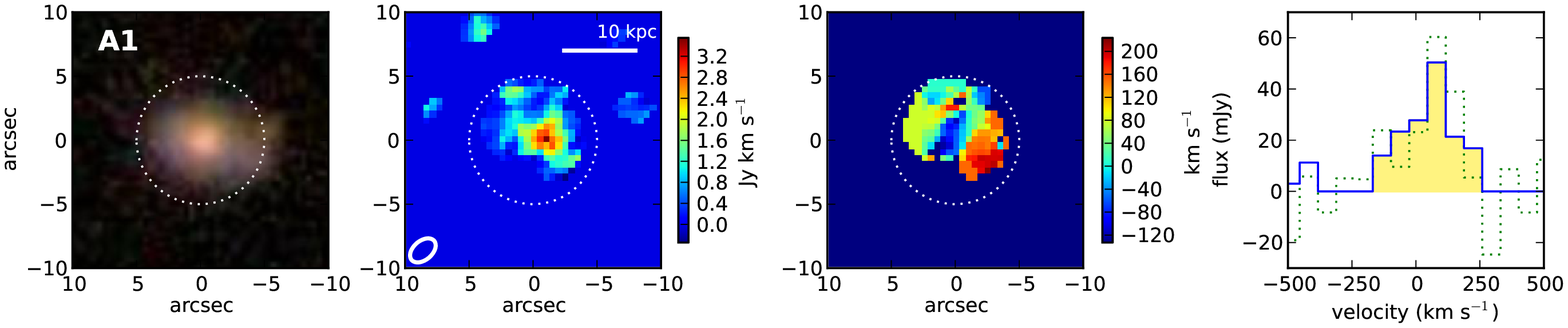}
\includegraphics[width=7in]{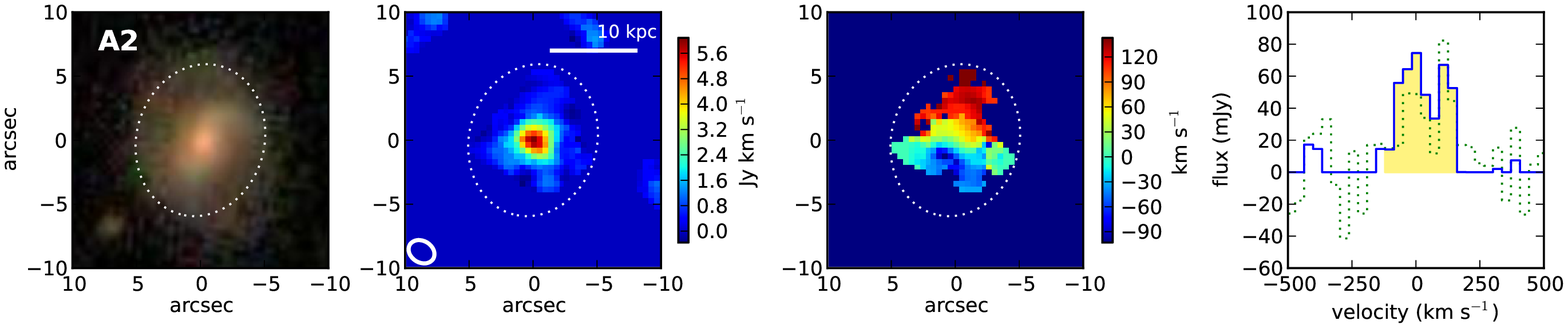}
\includegraphics[width=7in]{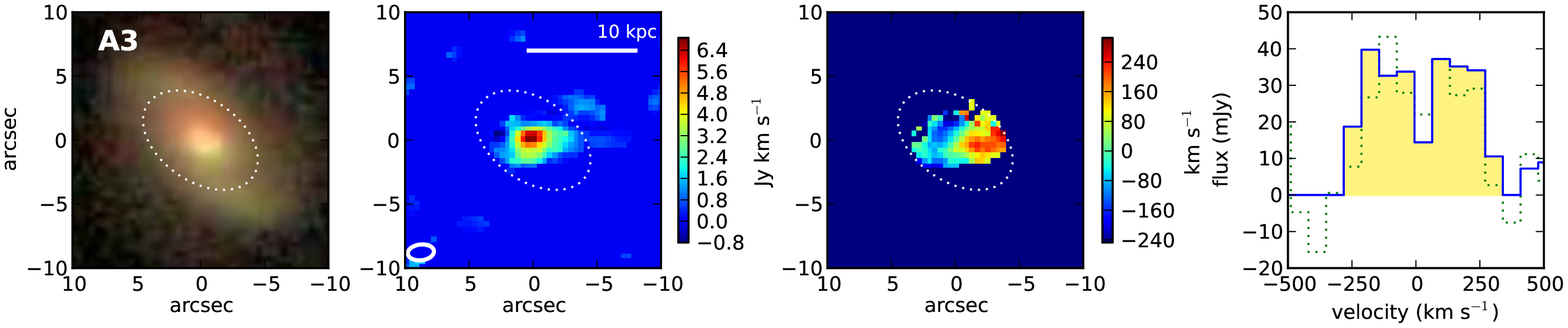}
\includegraphics[width=7in]{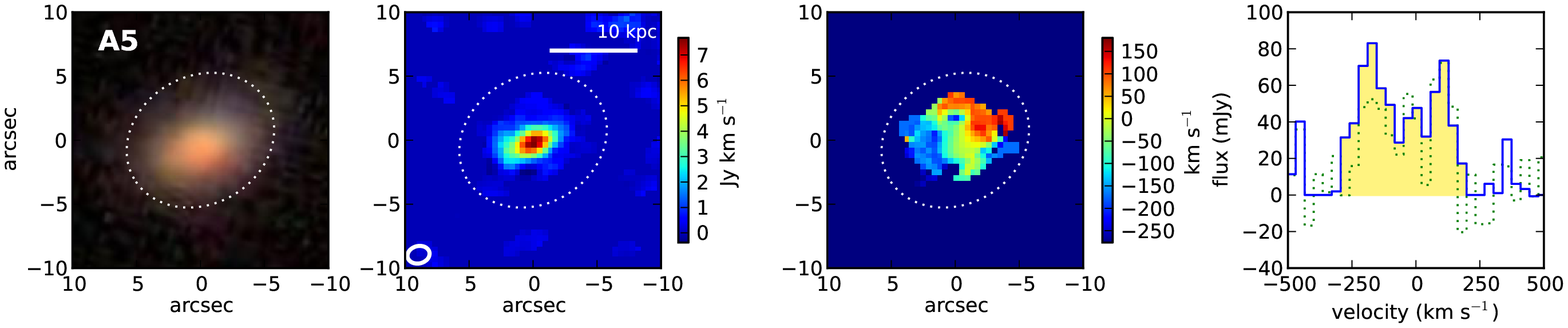}
\includegraphics[width=7in]{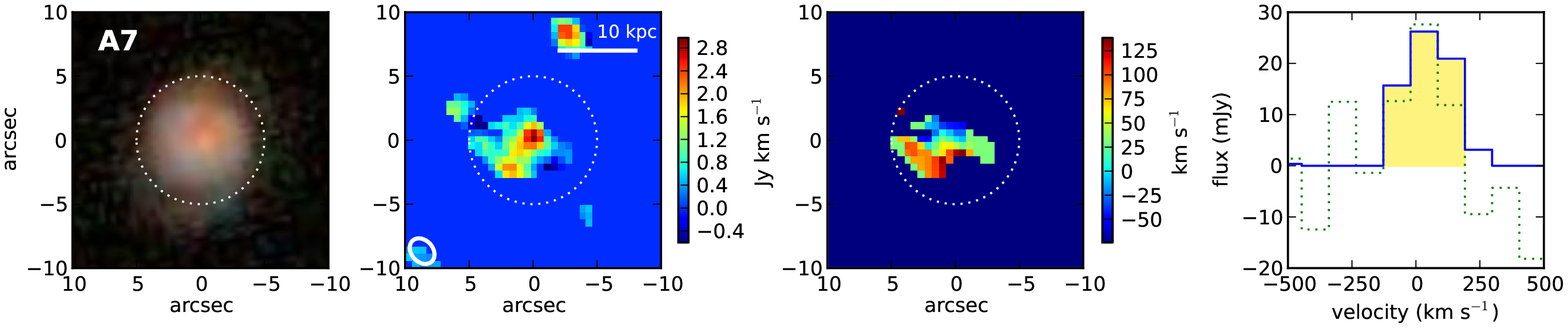}
\caption{Integrated CO emission maps for detected bin A galaxies, part 1. 
The left panel shows the optical image with the galaxy name in the upper left.
The moment 0 and moment 1 maps are displayed in the left middle and right 
middle panels respectively. The dotted white ellipse indicates
the source region. In the moment 0 map, the synthesized beam is indicated by the solid 
white ellipse in the lower left corner and a 10 kpc scale bar is given in the top right.
The moment 1 map is masked outside of the source region ellipse for clarity.
The far right panel shows the integrated spectrum of the galaxy: the solid blue line is calculated 
with masking, the dotted green line is without.}
\label{fig:binAmaps1}
\end{figure*}
\begin{figure*}
\includegraphics[width=7in]{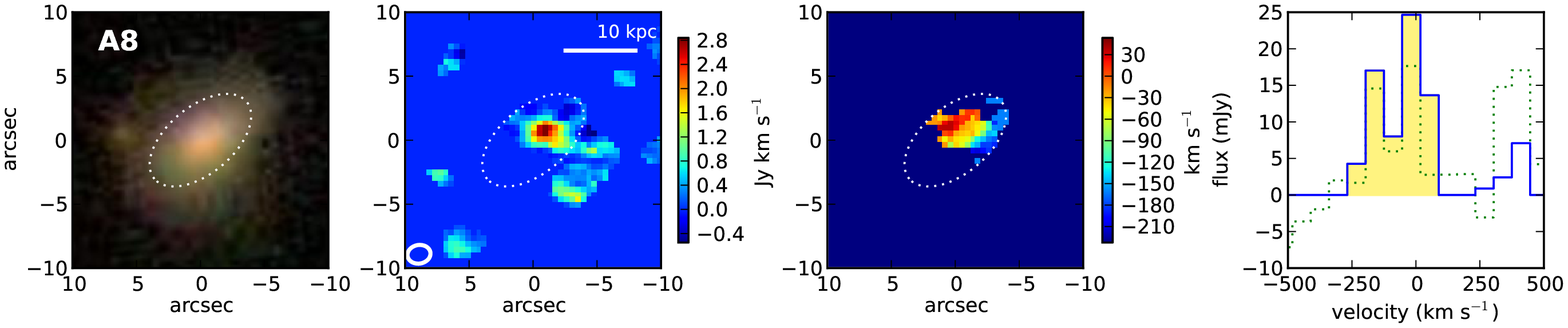}
\includegraphics[width=7in]{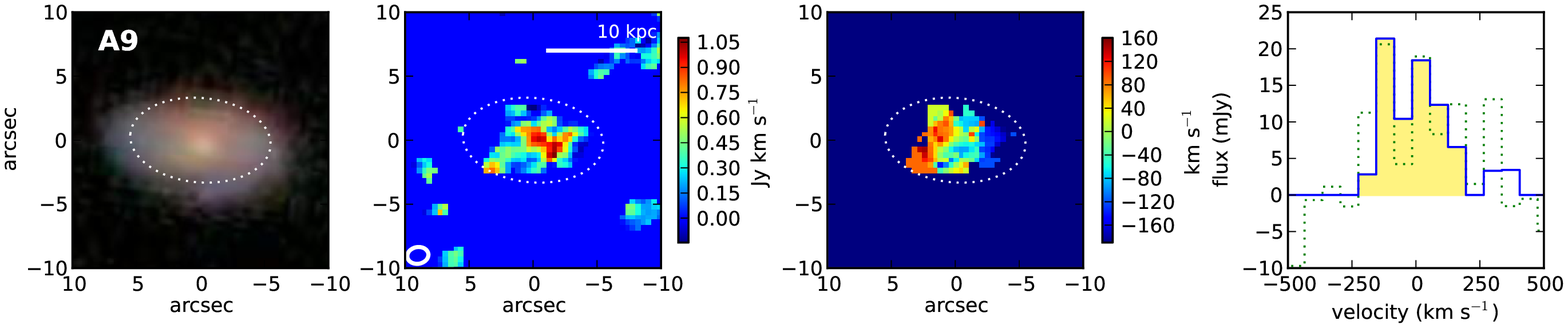}
\includegraphics[width=7in]{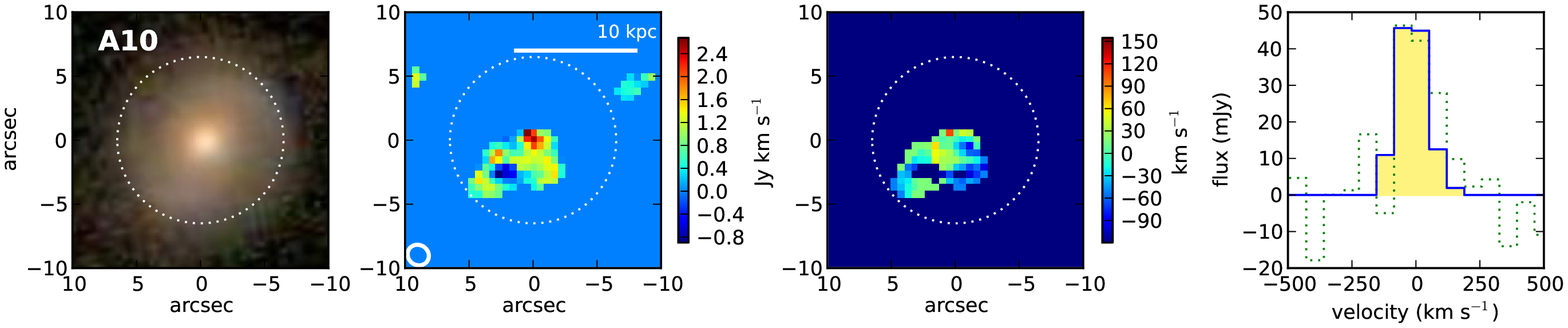}
\includegraphics[width=7in]{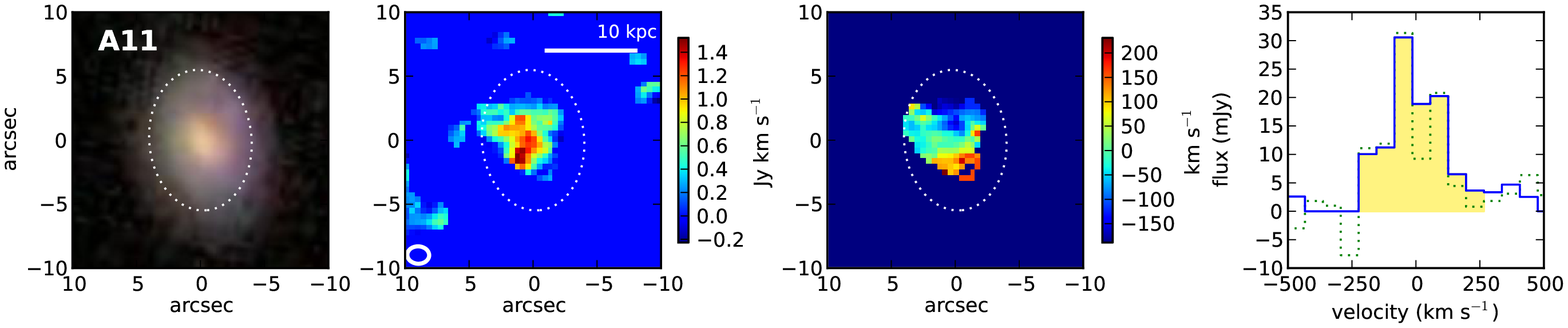}
\includegraphics[width=7in]{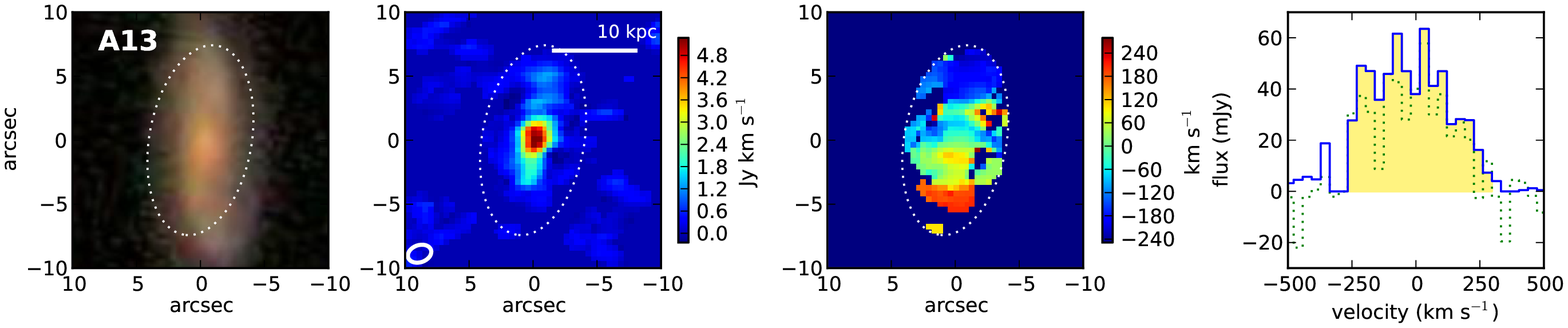}
\caption{Integrated CO emission maps for detected bin A galaxies, part 2. Same as Figure \ref{fig:binAmaps1}.}
\label{fig:binAmaps2}
\end{figure*}
\begin{figure*}
\includegraphics[width=7in]{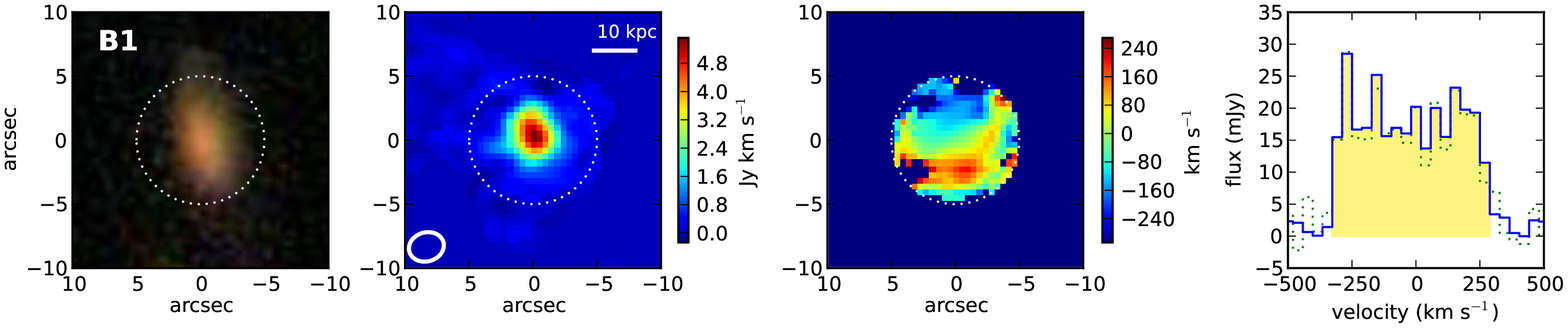}
\includegraphics[width=7in]{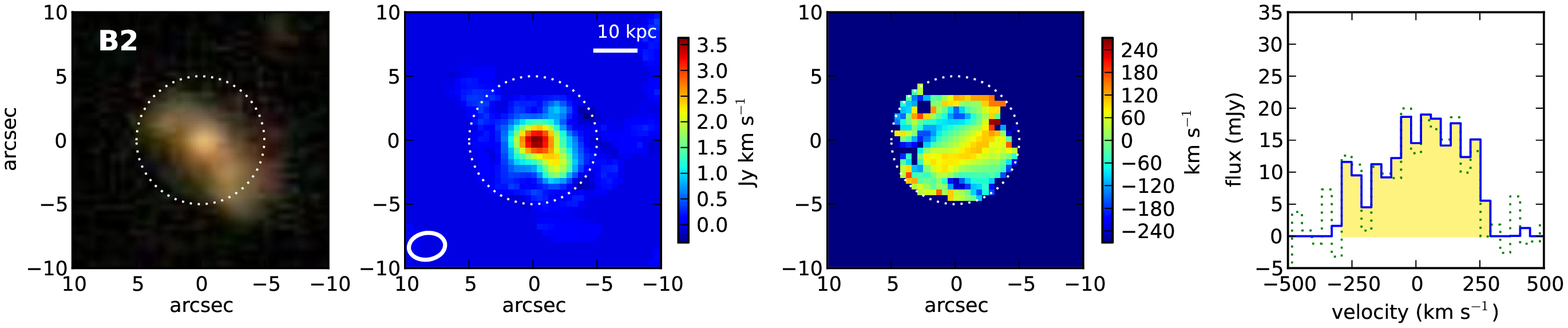}
\includegraphics[width=7in]{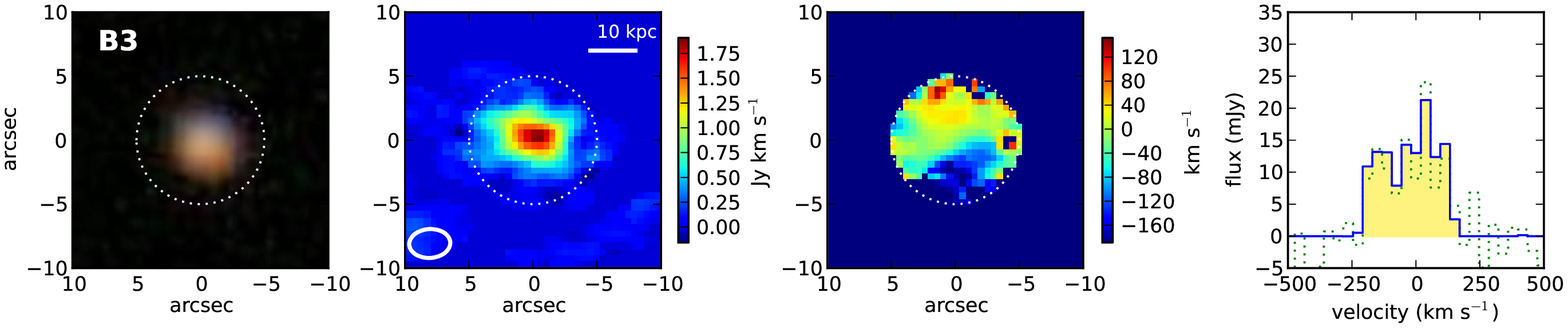}
\includegraphics[width=7in]{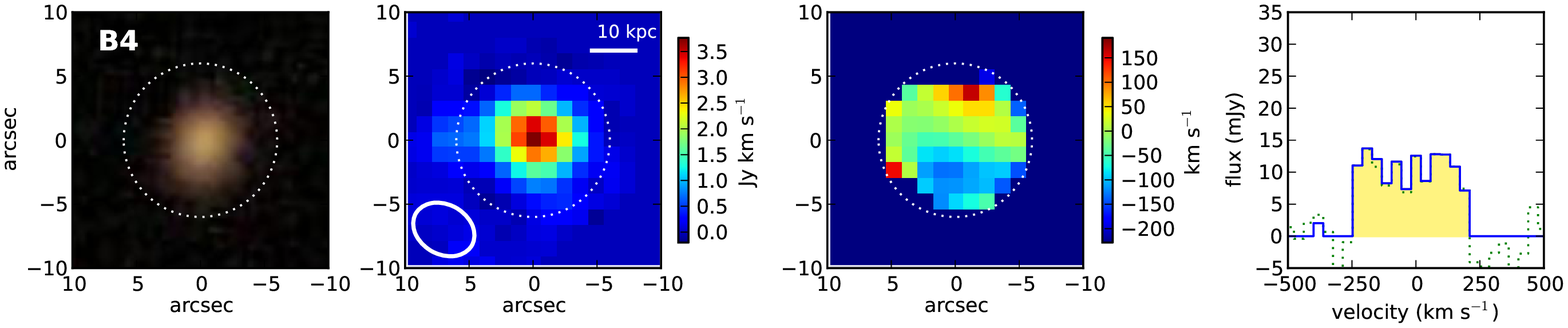}
\includegraphics[width=7in]{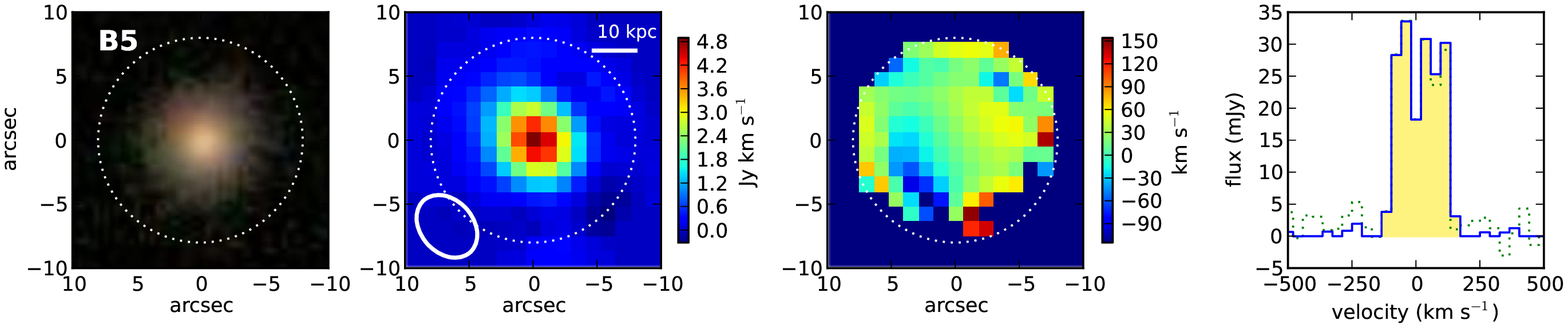}
\caption{Integrated CO emission maps for bin B galaxies, part 1. Same as Figure \ref{fig:binAmaps1}.}
\label{fig:binBmaps1}
\end{figure*}
\begin{figure*}
\includegraphics[width=7in]{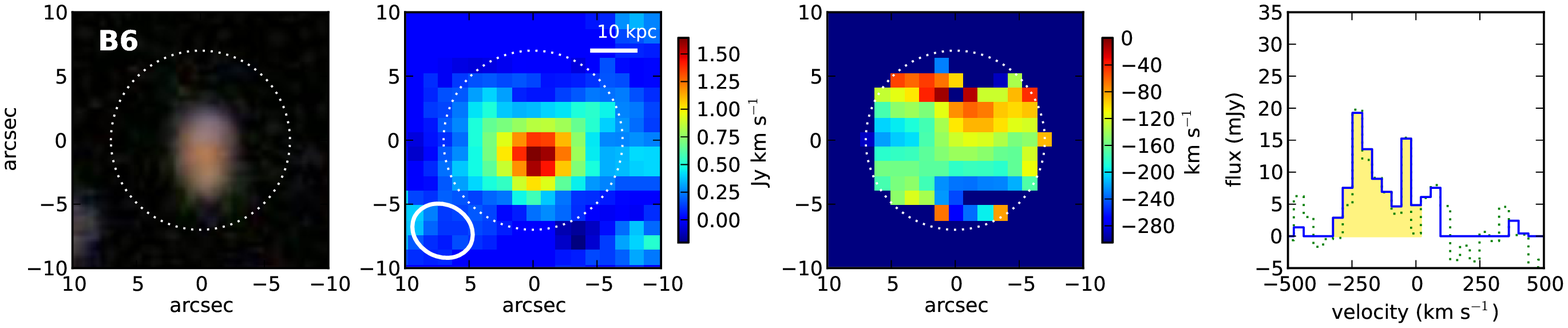}
\includegraphics[width=7in]{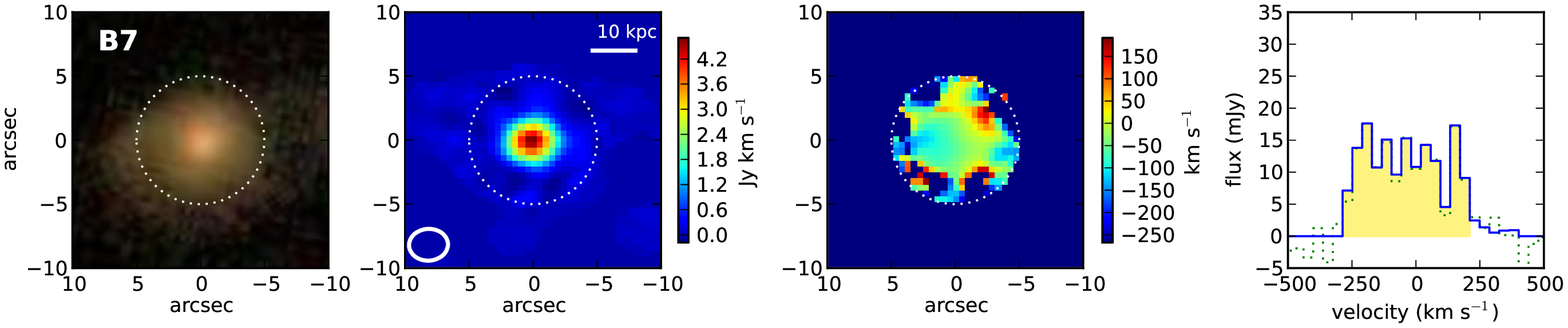}
\includegraphics[width=7in]{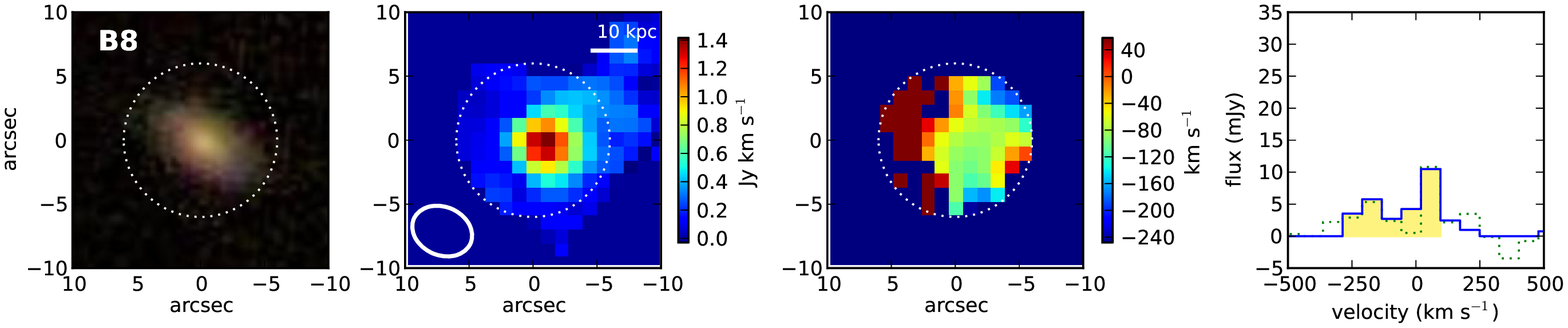}
\includegraphics[width=7in]{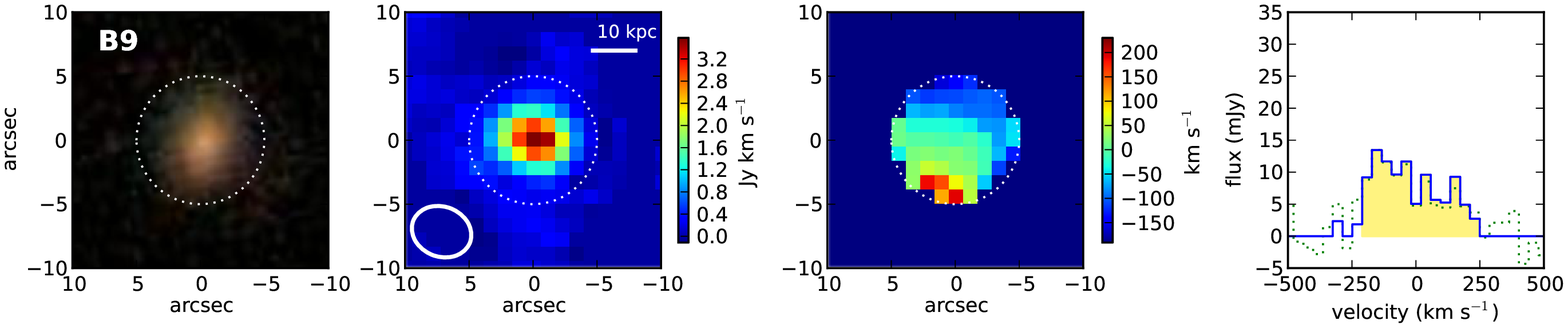}
\includegraphics[width=7in]{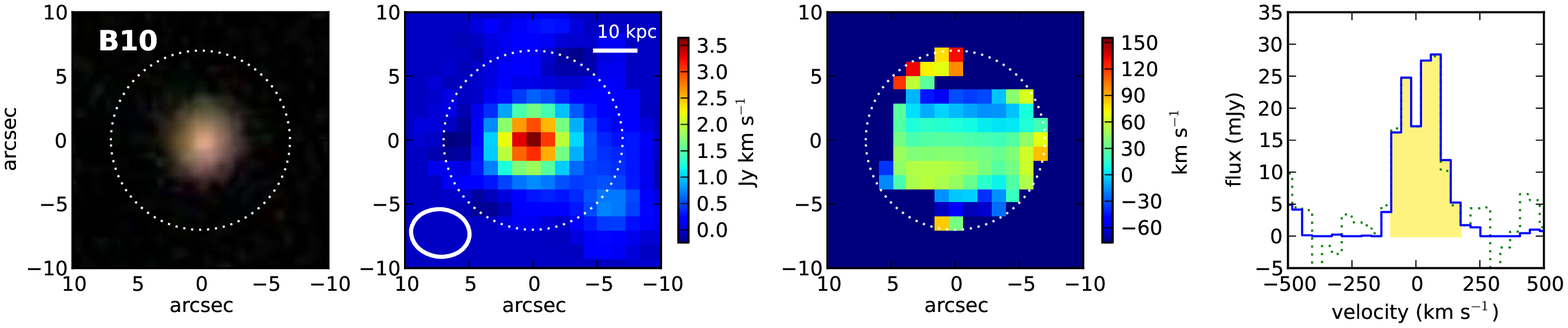}
\caption{Integrated CO emission maps for bin B galaxies, part 2. Same as Figure \ref{fig:binAmaps1}.}
\label{fig:binBmaps2}
\end{figure*}

\begin{figure*}[t]
\includegraphics[width=7in]{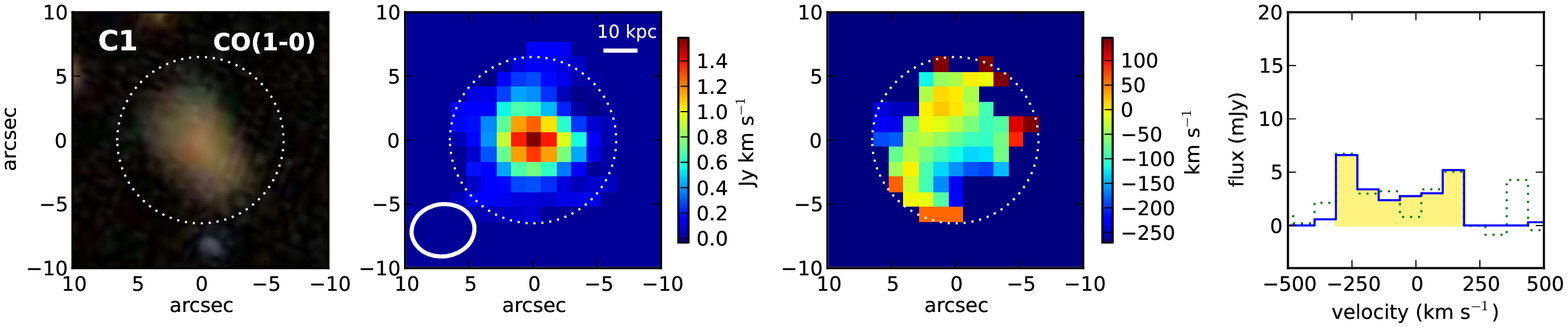}
\includegraphics[width=7in]{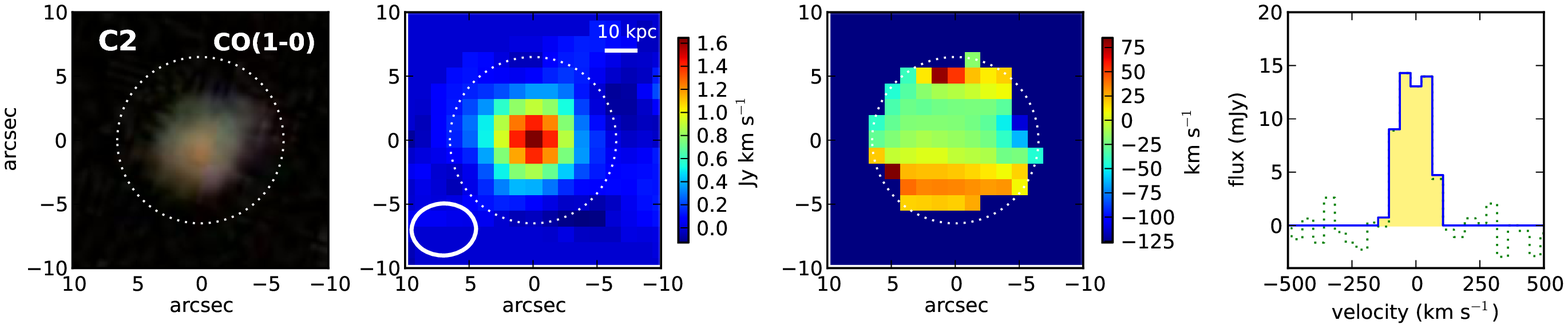}
\includegraphics[width=7in]{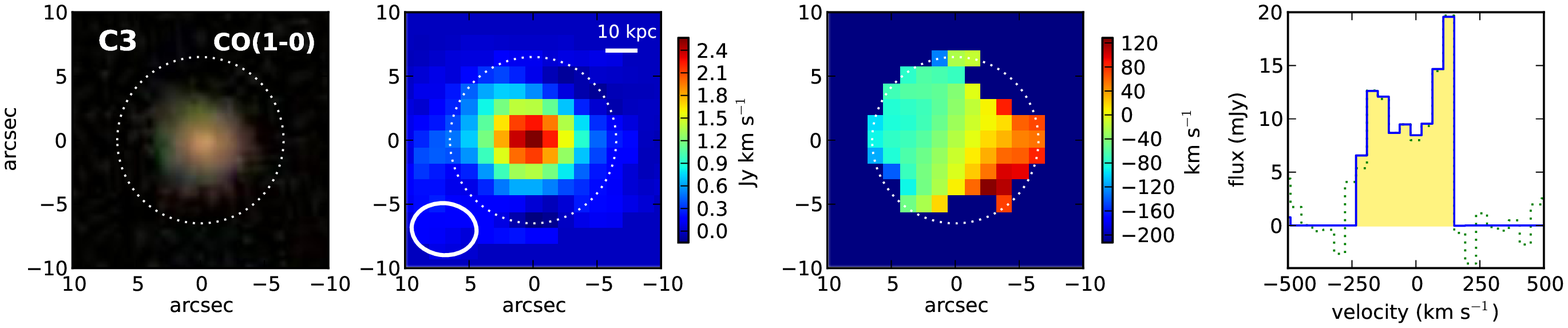}
\includegraphics[width=7in]{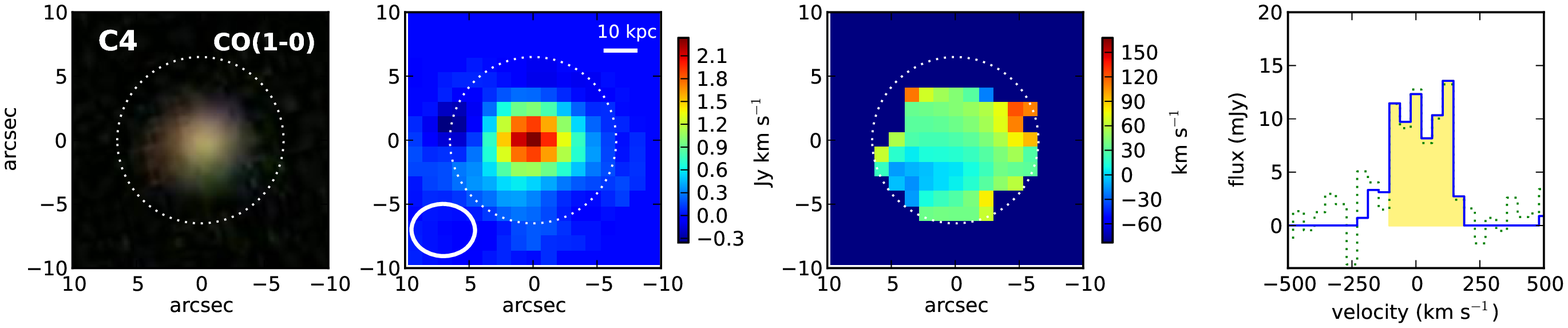}
\caption{Integrated CO emission maps for bin C galaxies in the CO\jone line. Same as Figure \ref{fig:binAmaps1}.}
\label{fig:binCmaps3mm}
\end{figure*}

\begin{figure*}[t]
\includegraphics[width=7in]{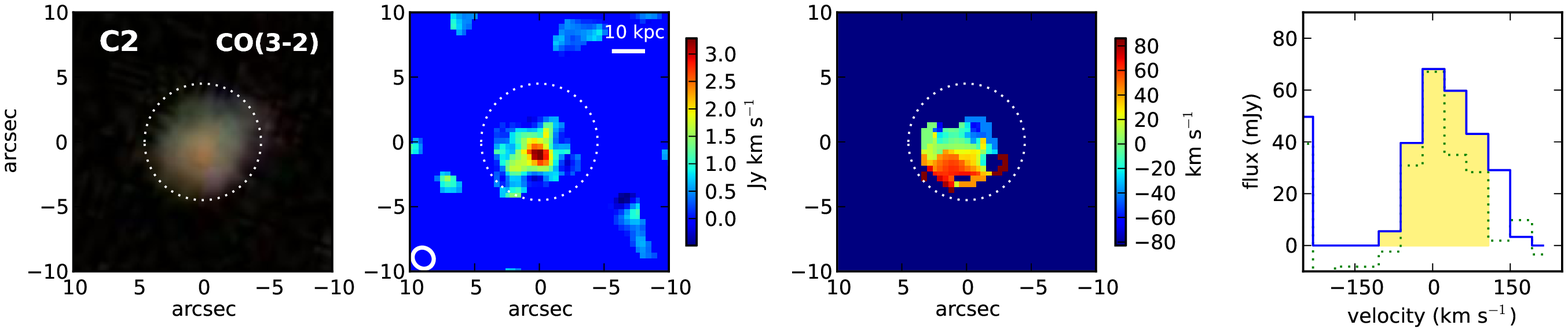}
\includegraphics[width=7in]{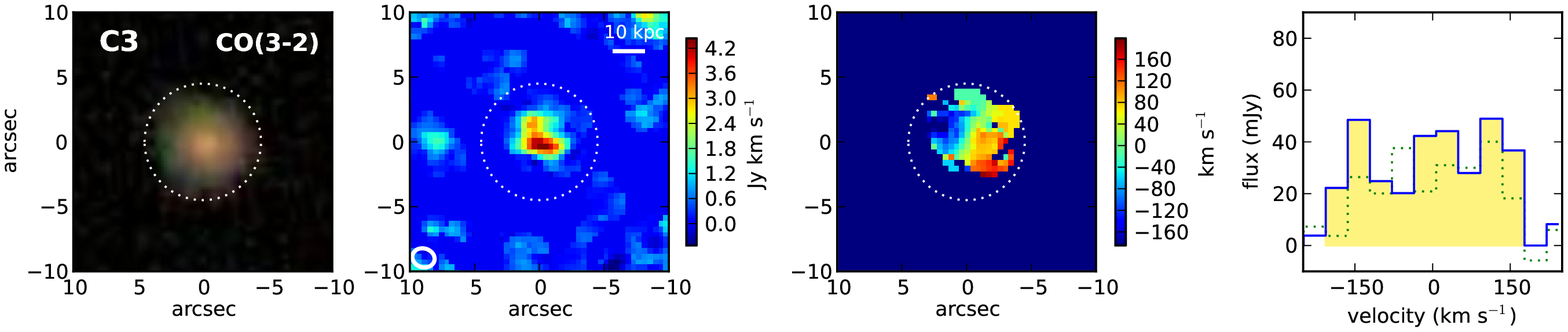}
\includegraphics[width=7in]{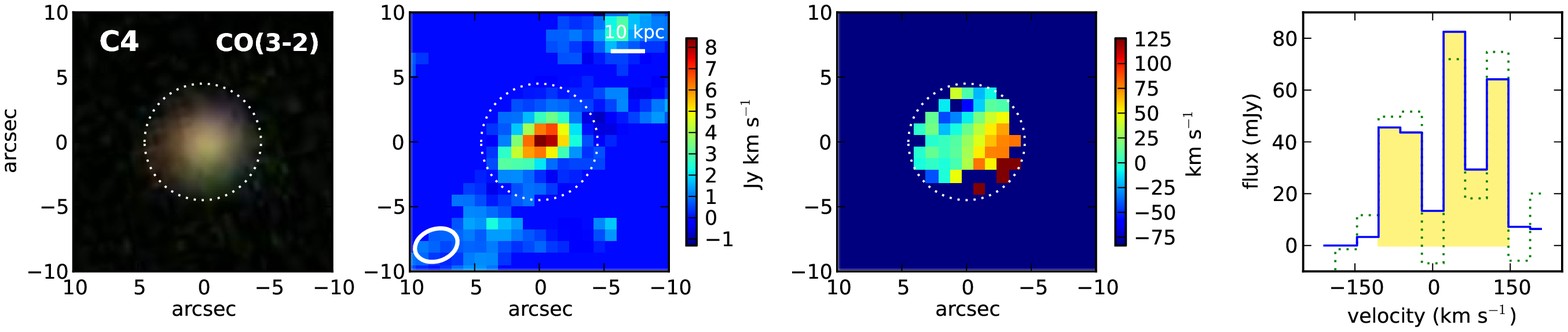}
\caption{Integrated CO emission maps for (detected) bin C galaxies in the CO\jthree line. Same as Figure \ref{fig:binAmaps1}.}
\label{fig:binCmaps1mm}
\end{figure*}

The integrated CO emission maps of all detected galaxies are presented in Figures \ref{fig:binAmaps1} - \ref{fig:binCmaps1mm}.
For each galaxy, the optical image (from SDSS) is in the left-most panel, the moment 0 (total intensity) map
is in the left middle panel, the moment 1 (intensity-weighted mean velocity) map is in the right middle panel,
and the integrated spectrum is in the far right panel. The dotted white ellipse shows the region in which
pixels were summed (if not masked; see Section \ref{sec:fluxest} for details). 
For clarity, we have additionally masked the moment 1 map outside of the source region (dotted white ellipse)
in Figures \ref{fig:binAmaps1} - \ref{fig:binCmaps1mm}.
The white bar in the moment 0 map
shows the angular size of 10 kpc at the redshift of the galaxy. In the far right panel, the solid blue line shows
the masked spectrum and the dotted green line shows the unmasked spectrum. Channels containing
source emission are shaded yellow.

The moment maps are produced using the $2\sigma$ clipped smooth mask method (see Section \ref{sec:fluxest}). 
The moment 0 map is a simple sum of the masked cube in the `source'
velocity channels. The moment 1 map is produced by summing the masked cube multiplied by 
the velocity in each channel, then normalizing by the moment 0 value. Thus, the color in the moment 1 map
indicates the intensity-weighted mean velocity in each pixel. The moment 1 map is only
calculated in pixels which have a positive moment 0 value.
Moment 1 pixels with values outside the velocity range covered by the source channels are masked.

In most of the bin A galaxies, we spatially resolve the CO disk, illustrated by a velocity gradient across the 
emission region in the moment 1 map (indicative of a rotating disk). The morphology of the bin A CO
emission is discussed further in Section \ref{sec:binamorph}. The bin B galaxies which were observed
only in D configuration are spatially unresolved. However, galaxies B1, B2, B3 and B7 were observed in C configuration
as well, increasing the resolution by roughly a factor of 2. Of these four, galaxies B1 and B3 show evidence for ordered rotation
but are only marginally resolved. Bin C galaxies are spatially 
unresolved in the CO\jone line and marginally resolved in the CO\jthree line.
The CO\jthree moment 1 maps of the galaxies C2, C3 and C4 are suggestive of a rotating disk of gas.
Bin D galaxies are not detected, but would be unresolved based on the sizes of the stellar disks.

\subsection{Morphology of the Low-Redshift Sample}
\label{sec:binamorph}

For the EGNoG galaxies in redshift bin A, the spatial resolution of the CO observations
allows us to comment on the morphology of the molecular gas disk at $z=0.05-0.1$.
The integrated CO intensity maps in Figures \ref{fig:binAmaps1} and \ref{fig:binAmaps2} are suggestive
of irregular morphology at first glance. However, due to the modest signal to noise ratio
of these observations, it is difficult to infer the true shape of the underlying CO emission. 

\begin{figure*}[t]
\includegraphics[width=\linewidth]{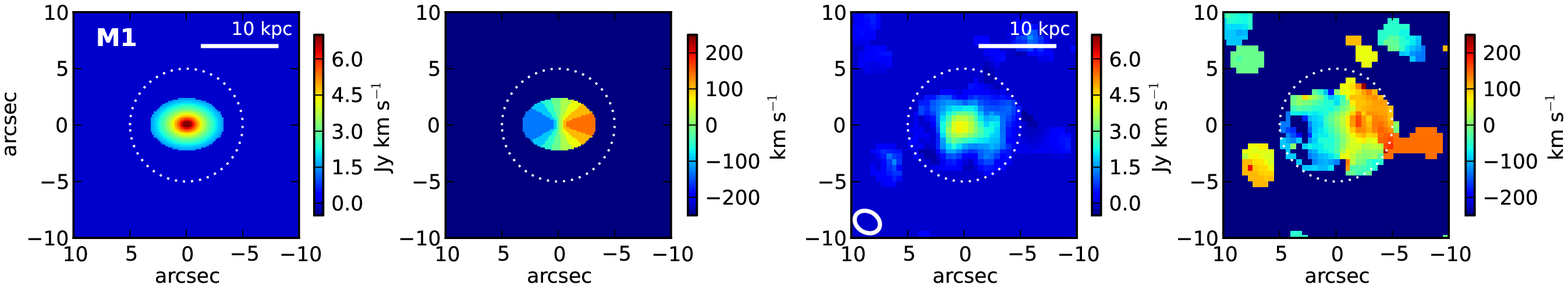}
\includegraphics[width=\linewidth]{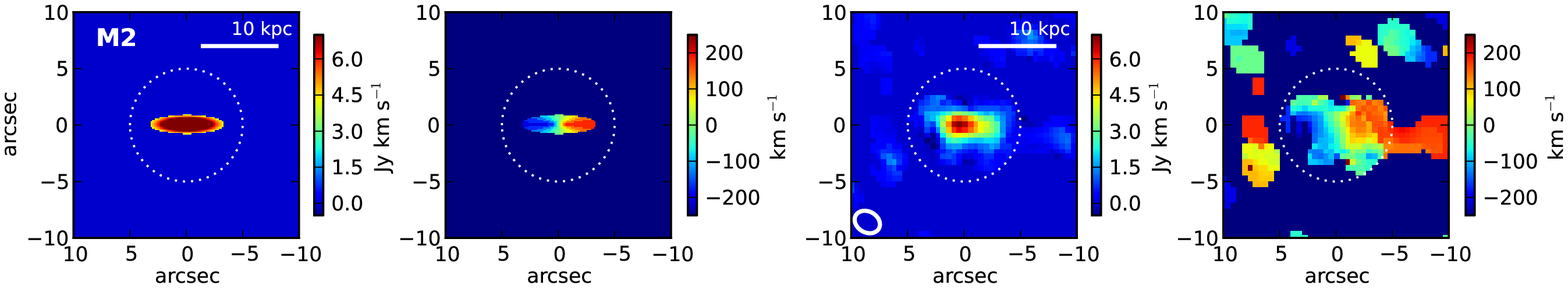}
\includegraphics[width=\linewidth]{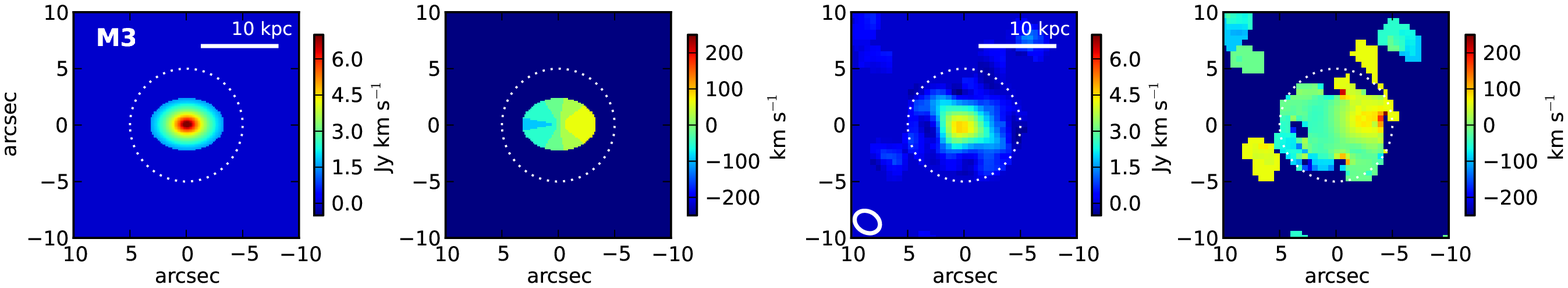}
\includegraphics[width=\linewidth]{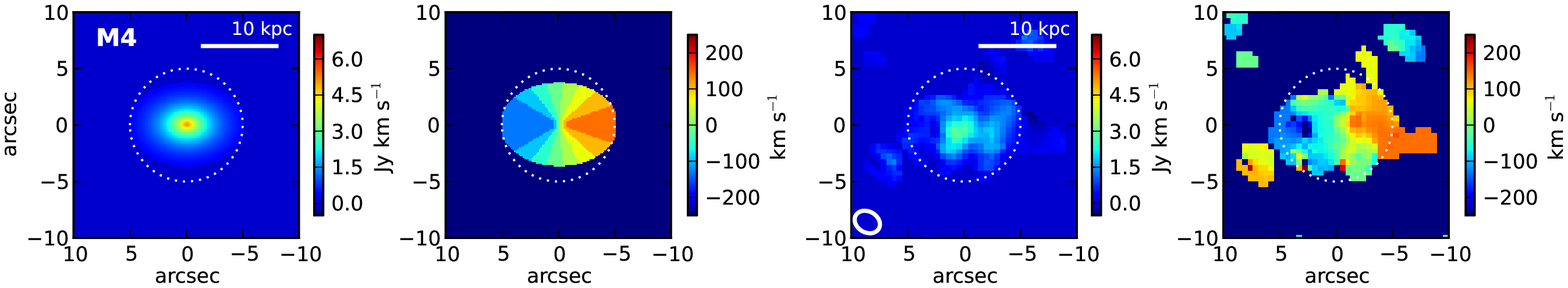}
\caption{Model galaxy maps. For each model (M1 - M4, parameters given in Table \ref{tab:modelpar}), 
the far left and left middle panels show the model galaxy moment 0 and moment 1 maps respectively.
In the right middle and far right panels are the moment 0 and moment 1 maps from normal processing
of the model galaxy inserted into noise-only data. }
\label{fig:morphmodels}
\end{figure*}

To determine the effect of a modest signal to noise ratio on our CO emission maps, 
we produced a simple model disk galaxy at $z=0.08$,
inserted it into noise-only $uv$ data channels from a bin A galaxy dataset and processed the data as we did the
real data. We used a Milky-Way-like rotation curve, linear in the central 1 kpc, flat outside.
The distribution of the CO emission is exponential, with a scale radius of 3 kpc.
We varied the total flux, rotation velocity of the flat part of the rotation curve ($v_\mathrm{flat}$), 
inclination angle ($i$) and physical radius of the
molecular gas disk ($R_\mathrm{CO}$). Models M1 to M4, shown in Figure \ref{fig:morphmodels}, 
are representative of the results of varying all four parameters within reasonable ranges.
The parameters used in the four models are listed in Table \ref{tab:modelpar}. 

The `observed' model maps (right two panels in Figure \ref{fig:morphmodels}) 
suggest that at the signal to noise level typical of our bin A data, 
irregularities in the CO emission maps do not necessarily indicate an underlying irregular structure
in the molecular gas disk. In most cases, our observations are consistent with the CO in the bin A galaxies
being ordered in a rotating disk. We conclude that deeper observations would be required to investigate 
the detailed morphology of the molecular gas disks in these systems. 

Two exceptions are galaxies A3 and A8, which exhibit disturbed morphologies that are not likely
the result of low signal to noise ratios. Both galaxies show rotating molecular gas disks misaligned
with the major axis of the optical disk, as well as CO emission outside the optical disk. These two galaxies
are discussed in detail in Appendix \ref{sec:A3andA8}.

\subsection{Non-Detections of the Bin D Galaxies}
\label{sec:binDdisc}

\begin{figure}[t]
\includegraphics[width=3.5in]{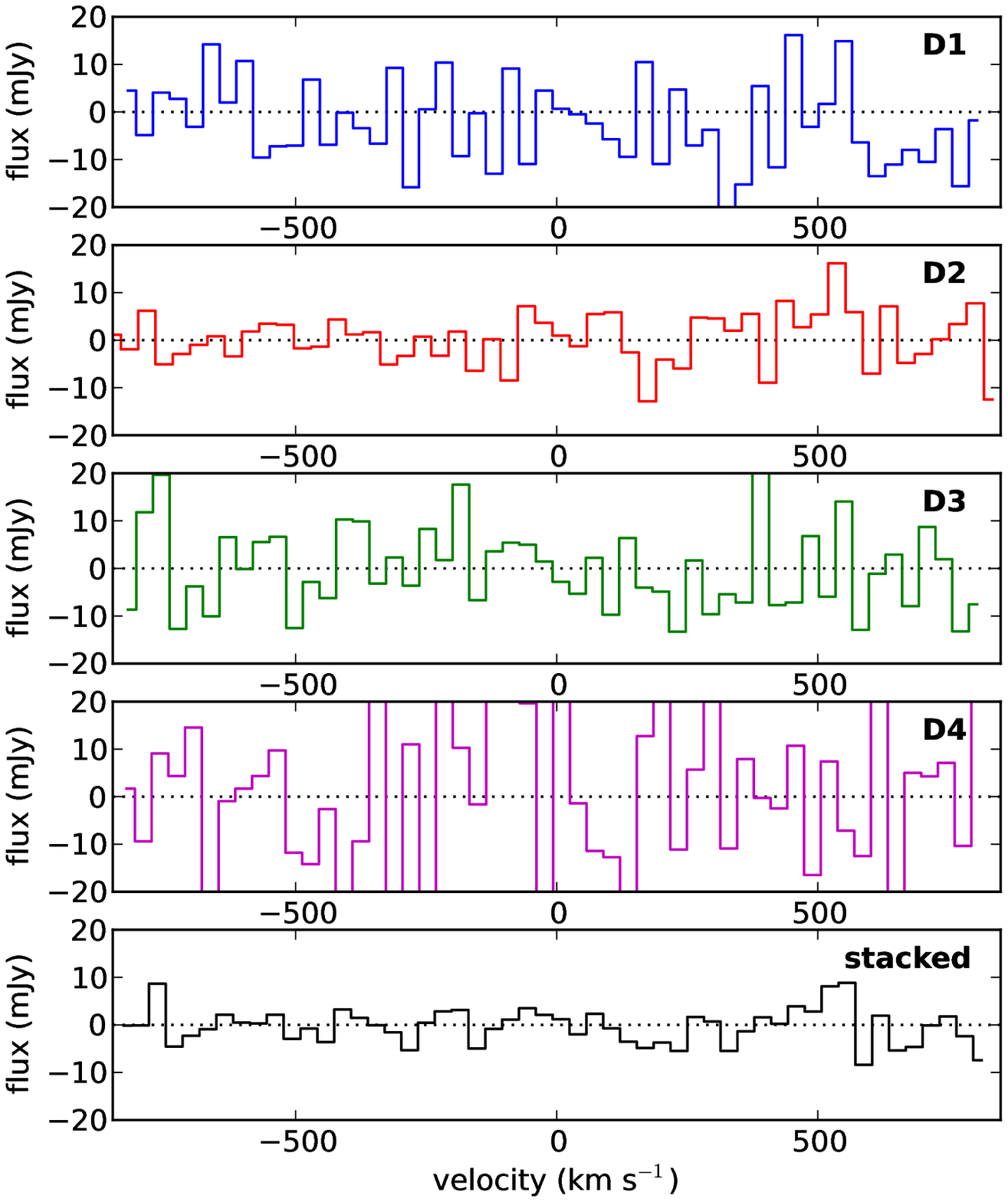}
\caption{Spectra of the four bin D sources within a circular region $2\arcsec$ in radius (top four panels), plus the 
stacked (average) spectrum, weighted by the inverse of the variance, in the bottom panel. }
\label{fig:binDstack}
\end{figure}

We do not detect CO\jthree in any of the four bin D galaxies at $z\approx0.5$.
The $3\sigma$ upper limits on $S_\mathrm{CO}$ are between 2.7 and 5.2 Jy \kms, 
assuming a total velocity width of 400 \kms. The spectra of the 
four bin D galaxies are shown in the top four panels of Figure \ref{fig:binDstack}
at a velocity resolution of $\approx 32$ \kms.
The bottom panel shows the stacked spectrum, where each individual spectrum
is weighted by the inverse of the variance. Note that the spectra are stacked centered at the spectroscopic
redshifts given in Table \ref{tab:COprop}, which have a typical uncertainty of 110 \kms.
The stacked spectrum does not indicate a detection; 
we calculate a 3$\sigma$ upper limit of 1.5 Jy \kmssp for a 400 \kmssp wide signal.

\begin{table}[t]
\centering
\begin{tabular}{| c | c | c | c | c |}
\hline
 &      $i$  & $v_\mathrm{flat}$ & $R_\mathrm{CO}$ & $S_\mathrm{CO}$ \\
Model & (deg) & (\kms) & (kpc) & (Jy \kms) \\
\hline
M1 &  45 & 200 & 5  & 14 \\
M2 &  75 & 200 & 5  & 14 \\
M3 &  45 & 100 & 5  & 14 \\
M4 &  45 & 200 & 8  & 14 \\
\hline
\end{tabular}
\caption{Model parameters: inclination angle ($i$), rotation velocity of the flat part of the velocity curve ($v_\mathrm{flat}$), 
radius of the CO disk ($R_\mathrm{CO}$) and total CO flux ($S_\mathrm{CO}$). }
\label{tab:modelpar}
\end{table}

To place these limits in context, we calculate the expected CO flux of a galaxy
with SFR $\approx 70$ M$_\odot$ yr$^{-1}$, typical for the bin D galaxies. 
First we consider these galaxies to be normal, using a Milky way-like $\alpha_\mathrm{CO}$
and $r_{31} = 0.5$. For $\tau_\mathrm{mgas} = 0.77 \pm 0.57$ Gyr (average from our
literature sample; see Section \ref{sec:tdep}), we expect a CO flux  
$S_\mathrm{CO} = 1.2-8.4$ Jy \kms. 
Performing the same calculation using a starburst $\alpha_\mathrm{CO}$ and the starburst
average $\tau_\mathrm{mgas} = 0.07\pm0.04$ Gyr (Section \ref{sec:tdep}), we expect
$S_\mathrm{CO} = 0.76-2.7$ Jy \kms. We have used $r_{31}=0.5$ in this calculation but
note that a larger value may be appropriate in a starburst system. We can further constrain the expected CO flux
using the typical stellar mass ($\log(M_*) = 10.7$) and the expected molecular gas fraction (see Section \ref{sec:fgasevol}).
For $f_\mathrm{mgas} = 0.03-0.4$, we expect $S_\mathrm{CO} = 0.13-2.6$ Jy \kmssp
for normal galaxies or $S_\mathrm{CO} = 0.5-11.0$ Jy \kmssp for starburst galaxies. 
Note also that these estimates have an uncertainty of a factor of 1.5-2.5 from the 
uncertainty in the SFR and $M_*$ of the bin D galaxies, in addition to further uncertainty
in the $r_{31}$ value.

Considering the range of expected CO fluxes, four non-detections at the depth of these
observations is not unreasonable. Given the large, overlapping ranges of expected 
CO fluxes for starburst and normal galaxies detailed above, we do not draw any conclusions
about the gas consumption properties of the bin D galaxies based on these upper limits. 

%Averaging 2 channels together, we calculate the spectrum for each of the bin D galaxies 
%within a radius of 2\arcsec. We stack the spectra, weighting by the variance of each individual 
%spectrum. The results are shown in Figure \ref{fig:binDstack}. In the stacked spectrum, we find 3.7 mJy
%noise, which corresponds to a 3$\sigma$ upper limit of 1.48 Jy km/s for a 400 km/s wide signal. 
%At the redshift of the bin D sample, this corresponds to $L_\mathrm{CO}' \approx 2 \times 10^9$ \Kkmspc.
%Using $r_{31} = 0.5$ (Bauermeister et al. 2012a), this yields an upper limit on the average molecular
%gas mass in the four bin D galaxies of $1.7 \times 10^{10}$ M$_\odot$ using a Milky Way conversion factor
%or $0.4 \times 10^{10}$ M$_\odot$ using a starburst conversion factor. Taking the typical stellar mass
%of the bin D galaxies as $5\times10^{10}$ M$_\odot$, this corresponds to 
%$f_\mathrm{mgas} = 0.25$ considering them as normal galaxies or 0.07 for starbursts.

\section{Discussion}
\label{sec:discussion}

We now look at the EGNoG galaxies in the context of starburst and normal galaxies 
at low and high redshift from the literature. We present the literature sample in the following
section (Section \ref{sec:litdata}), describe our identification and treatment of starburst and normal
galaxies in the literature dataset (Section \ref{sec:sbvnormal}), then investigate the molecular gas depletion times
in normal and starburst galaxies (Section \ref{sec:tdep}) and the evolution of
the molecular gas fraction with redshift (Section \ref{sec:fgasevol}).

\subsection{Literature Data}
\label{sec:litdata}

The local galaxies of our literature data compilation include normal spiral galaxies, LIRGs and ULIRGs.
We include 22 galaxies from \cite{Leroy2008}, 10 of which are defined by the authors to be dwarf
galaxies ($M_*< 10^{10}$ M$_\odot$). \cite{Leroy2008} present CO data from
the HERACLES and BIMA SONG surveys, stellar masses estimated from the SINGS survey {\it K}-band 
luminosities, and star formation rates calculated from a combination of GALEX FUV and Spitzer 24 $\mu$m. 
We augment this dataset with 47 galaxies (26 of which have $M_* < 10^{10}$ M$_\odot$) from the overlap of the
H$\alpha$ survey by \cite{Kennicutt2008} and the CO sample (taken from the literature) of \cite{OR2009}.
We estimate stellar masses, star formation rates and CO luminosities for these galaxies following \cite{Bothwell2009}.
For each galaxy, the stellar mass is calculated from the {\it B}-band luminosity using a mass-to-light ratio  
estimated using the {\it B-V} color \citep{BelldeJong2001}. The SFR is calculated
from the H$\alpha$ flux via the relation given by \cite{Kennicutt1998ARAA}, using an extinction correction
calculated from the {\it B}-band luminosity following \cite{Lee2009b}. For more details, see \cite{Bothwell2009}.

We include a sample of 56 local LIRGs and ULIRGs, 
with CO data compiled from \cite{GS2004}, \cite{Solomon1997} and \cite{Sanders1991}.
We use stellar masses and SFRs from the following: the GOALS survey \citep{Howell2010}, with stellar masses
calculated from {\it K}-band luminosties and SFRs estimated from the combination of IRAS infrared and 
GALEX FUV luminosities; \cite{daCunha2010}, with stellar masses and SFRs from SED
fitting of mid-IR Spitzer/IRS spectra and multi-band photometry available from the literature; 
and \cite{Hou2011} (table of values via private communication),
with stellar masses from SED fitting of SDSS spectra. We estimate SFRs for the \cite{Hou2011} galaxies 
(for which only stellar masses are given) from the total infrared luminosity ($L_\mathrm{IR}$; 8-1000 $\mu$m) using 
SFR $= 1.7\times 10^{-10}~ L_\mathrm{FIR}$ \citep{Kennicutt1998ARAA}, where $L_\mathrm{FIR}$ is
the far-infrared luminosity in $L_\odot$) and $L_\mathrm{IR} / L_\mathrm{FIR} = 1.3$ \citep{GraciaCarpio2008}. 

At intermediate redshifts ($z=0.05-1$), the EGNoG sample is augmented by the CO\jone observations 
of 7 LIRGs (2 are upper limits) at $z\approx0.4$ by \cite{Geach2011}, as well as CO\jone, CO\jtwo, and 
CO\jthree observations of 56 ULIRGs at $0.2 < z < 1.0$ by \cite{Combes2011} and \cite{Combes2013}.
We follow Combes et al. and use $r_{21} = r_{32} = 1$,
motivated by the fact that these are extremely star-forming systems and thus the gas is likely highly excited.
We exclude galaxies that are only observed in the CO\jfour line due to the uncertainty of $r_{41}$. 
For the Geach et al. galaxies, stellar masses are from
the {\it K}-band luminosities using a mass-to-light ratio determined from SED fitting using {\it BVRIJK} imaging.
SFRs are from the total infrared luminosities, estimated from the PAH 7.7 $\mu$m
line (see \citealt{Geach2009b}). 
\cite{Combes2011} estimate stellar masses for the $0.2 < z < 0.6$ sample from K-corrected optical and near-IR luminosities. 
For the $0.6 < z < 1.0$ sample, \cite{Combes2013} estimate stellar masses from SED-fitting of 
optical and near-IR luminosities from public catalogs. 
We estimate SFRs from the \cite{Combes2011, Combes2013} far-IR luminosities using the conversion of
\cite{Kennicutt1998}, as described above for a few of the LIRGs and ULIRGs.

At high redshifts ($z > 1$), we include 67 SFGs presented by \cite{Tacconi2012} 
in the IRAM Plateau de Bure High-z Blue Sequence CO\jthree Survey (PHIBSS) 
%(from \citealt{Tacconi2010}), 6 SFGs from \cite{Daddi2010a},
and the 16 of 26 SMGs from \cite{Bothwell2012} with stellar masses, at $z < 3$ (tables for both datasets via private communication). 
The SFGs were observed in the CO\jthree \citep{Tacconi2010} and CO\jtwo \citep{Daddi2010a, Magnelli2012} lines
and their luminosities were converted to CO\jone luminosities using $r_{31} = 0.5$ \citep{Bauermeister2013a}
and $r_{21} = 0.8$ \citep{Dannerbauer2009}.
The SMGs from \cite{Bothwell2012} were observed in various rotational transitions of the CO molecule ($J_\mathrm{upper} =$ 2-7)
and their luminosities were converted to CO\jone luminosity using an excitation model calculated by Bothwell et al. 
from their entire sample, normalized to a common far-IR luminosity.
The stellar masses for the \cite{Tacconi2012} SFGs are estimated from SED modeling
and the SFRs are derived from a combination of the UV and IR luminosities or extinction-corrected H$\alpha$ luminosity.
%synthesis fitting to rest-frame UV to optical/near-IR SEDs and H$\alpha$ luminosities. 
%\cite{Daddi2010a} use the average of 3 estimators for the SFR (dust-corrected UV luminosities, 
%mid-IR continuum luminosities from 24 $\mu$m Spitzer imaging, and VLA 1.4 GHz radio continuum)
%and derive stellar masses from stellar template fitting to multicolor photometry in the 
%rest-frame UV, optical, and near-IR bands.
\cite{Bothwell2012} use stellar masses from \cite{Hainline2011} (not available for all 26 galaxies), 
which come from stellar population synthesis fitting to observed-frame optical to mid-IR SEDs.
We estimate SFRs for the SMG sample galaxies from the \cite{Bothwell2012} far-IR luminosities
(which are calculated from the 1.4 GHz continuum utilizing the far-IR-radio correlation) using the conversion of
\cite{Kennicutt1998}, as described above for a few of the LIRGs and ULIRGs.

\begin{figure*}[t]
\centering
\includegraphics[height=3.7in]{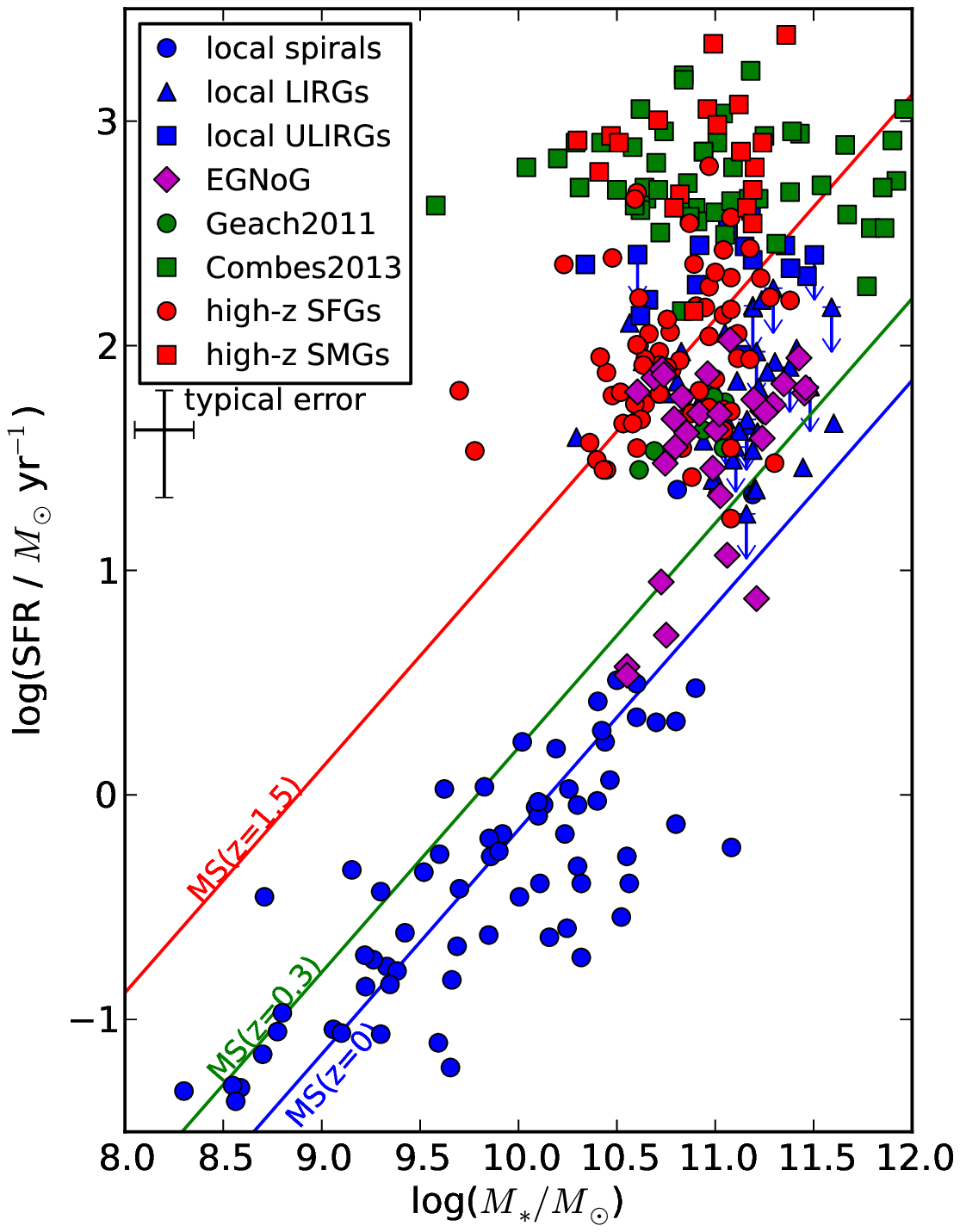}
\includegraphics[height=3.7in]{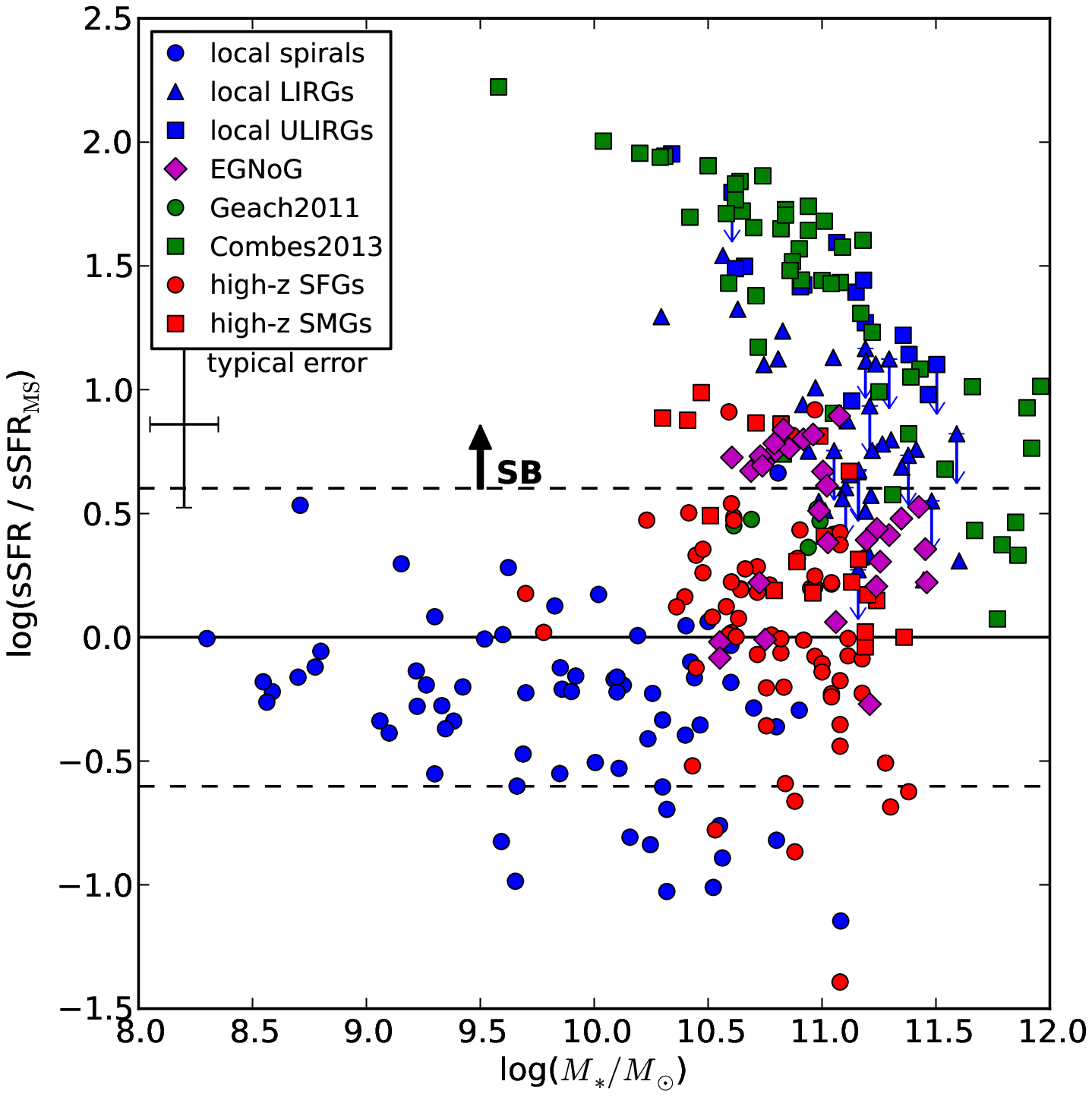}
\caption{Star formation rate (left panel) and normalized specific star formation rate (right panel) versus
stellar mass for our literature sample and the EGNoG galaxies. Colors indicate approximate redshifts and plotting
symbols show the galaxy type (see text for full description). In the left panel, the main sequence at three representative
redshifts is indicated by the blue, green and red lines. In the right panel, the sSFR is normalized by the main sequence
sSFR (for the relevant $z$ and $M_*$). Starburst (SB) galaxies lie above the higher dashed line. }
\label{fig:sSFRlit}
\end{figure*}

While we have attempted to make this compilation of literature data as uniform as possible, 
the derived properties are not all calculated in a consistent fashion.
In converting CO luminosities to molecular gas masses, we have standardized the selection of $\alpha_\mathrm{CO}$ value
using the starburst classification of Equations \ref{equ:myssfr} and \ref{equ:mySBssfr} (described in the next section). 
The $J_\mathrm{upper} > 1$ CO luminosities are converted to CO\jone luminosities using 
the ratios prescribed by the individual authors in the case of ULIRGs and SMGs. We use standard
ratios for all other galaxies: $r_{31} = 0.5$ and $r_{21} = 0.8$. 
The $M_*$ and SFR values, however, are calculated by a variety of methods, as described above.
This may have systematic effects on the analysis in the following sections. In the interest of
comparing with as large a sample as possible, we include all the data described here in
our analysis, but note that systematic effects may be present.

\subsection{Identifying Starburst Galaxies}
\label{sec:sbvnormal}

In Figure \ref{fig:sSFRlit}, we present the stellar mass and star formation properties of
the literature sample and the EGNoG survey galaxies. For the literature sample,
the colors indicate approximate redshifts ($z\approx0$ in blue, 
$z\sim0.05-0.5$ in green, and $z > 1$ in red) and the symbols indicate galaxy type: normal galaxies
(local spirals, \citealp{Geach2011} galaxies at $z=0.4$ and high-redshift SFGs) are marked with circles, 
local LIRGs with triangles, and starbursts (local ULIRGs, Combes et al. intermediate-redshift ULIRGs, 
and high-redshift SMGs) are marked with squares.
The EGNoG galaxies are plotted with a unique symbol and color (magenta diamond) for emphasis.
This color and symbol scheme will be used for the rest of this work.

\begin{figure*}[t]
\centering
\includegraphics[height=3.7in]{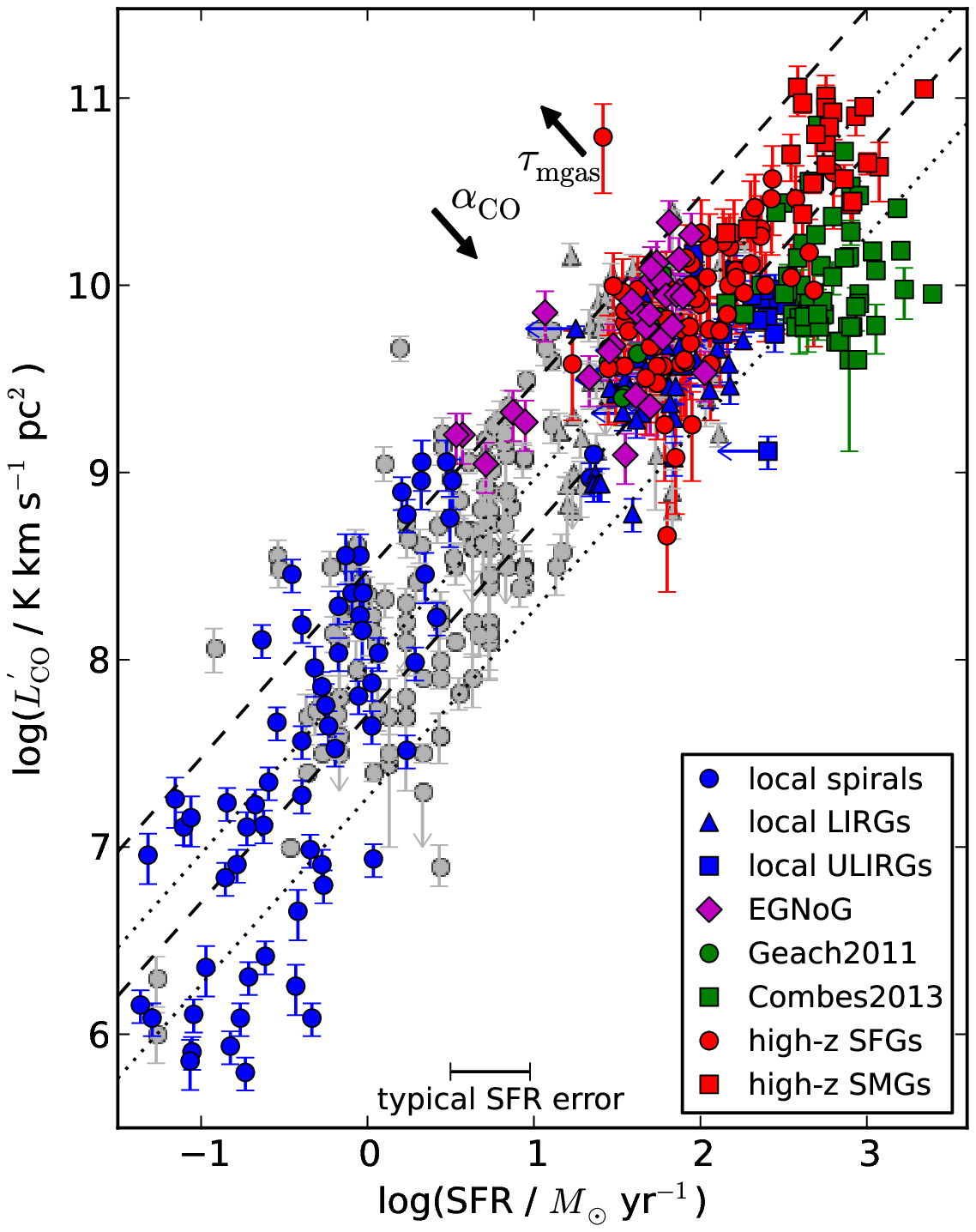}
\includegraphics[height=3.7in]{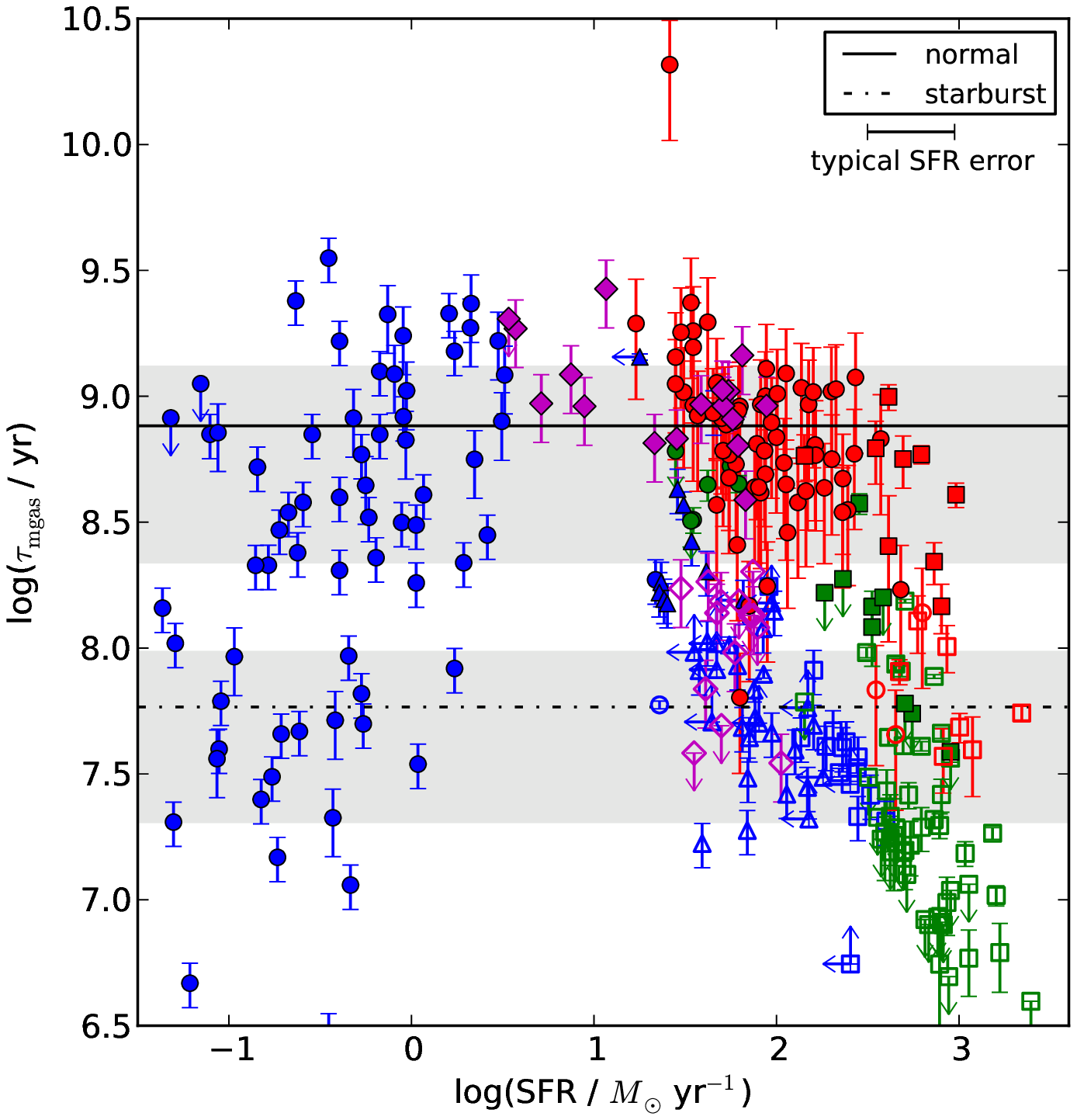}
\caption{CO luminosity ($L_\mathrm{CO}'$; left panel) and molecular gas depletion time ($\tau_\mathrm{mgas}$; right panel)
versus SFR. In the left panel, galaxies from \cite{Yao2003}, \cite{Mao2010}, and \cite{Papa2011} 
are plotted (gray points) in addition to the literature
sample. Dashed (dotted) lines show the expected region in $L_\mathrm{CO}'$-SFR space occupied by normal (starburst) galaxies.
In the right panel, the horizontal solid (dot-dashed) line indicates the average $\tau_\mathrm{mgas}$ value for normal (starburst)
galaxies, the gray region indicating $\pm1\sigma$. Starburst galaxies are plotted as un-filled symbols in the right panel. }
\label{fig:tdep}
\end{figure*}

The left panel of Figure \ref{fig:sSFRlit} shows the SFR versus $M_*$, with the main sequence at 
three representative redshifts indicated by the blue ($z=0$), green ($z=0.3$), and red ($z=1.5$) lines. 
Normal galaxies lie near the main sequence curve at the corresponding redshift. 
In order to more precisely discern starburst galaxies from normal galaxies, 
in the right panel of Figure \ref{fig:sSFRlit} we plot the sSFR of each galaxy normalized by the main sequence 
sSFR (sSFR$_\mathrm{MS}$; Equation \ref{equ:myssfr}) at the redshift and stellar mass of the galaxy. 
The horizontal black line indicates the main sequence relation, with the dashed lines showing a spread of 0.6 dex.
Starburst galaxies lie above the higher dashed line, with sSFRs
larger than 4 times the main sequence sSFR (as defined in Equation \ref{equ:mySBssfr}).
In most cases, our starburst classification agrees well with expectations (e.g. ULIRGs are 
mostly classified as starbursts), with the exception of the SMG population, which we find to be mostly
consistent with the main sequence defined in Equations \ref{equ:myssfr} and \ref{equ:mySBssfr}. 
However, we note that since most of the SMGs included here are at $z \gtrsim 2$, 
this result depends sensitively on the high-redshift behavior of the sSFR$_\mathrm{MS}(z)$ prescription,
which is more uncertain and may flatten at these redshifts (see discussion in Section \ref{sec:sampselect}).

Using this starburst criterion, we standardize the conversion of CO luminosity to molecular gas mass in 
our literature sample, as we did for the EGNoG galaxies in Section \ref{sec:derivedprop}. 
For galaxies that are classified as starburst based on their specific star formation rates, 
we use the value observed in starburst systems: $\alpha_\mathrm{CO}(\mathrm{ULIRG}) = 0.8$ M$_\odot$ (K km s$^{-1}$ pc$^2$)$^{-1}$
(local ULIRGs: \citealt{Scoville1997}, \citealt{DownSol1998}; high-redshift SMGs: \citealt{Tacconi2008}).
For all other galaxies, we use a Milky Way-like conversion factor, 
$\alpha_\mathrm{CO}(\mathrm{Milky Way}) = 3.2$ M$_\odot$ (K km s$^{-1}$ pc$^2$)$^{-1}$
\citep[e.g.][]{Dame2001}.
These conversion factors do not explicitly include He; see Equation \ref{equ:Mmgasdef}.
This bimodal prescription for the conversion factor is discussed further in Section \ref{sec:bimodalalpha}.

\subsection{Gas Depletion Time}
\label{sec:tdep}

Using our compilation of data, we investigate the molecular gas depletion time ($\tau_\mathrm{mgas}$)
for starburst and normal galaxies. Excluding dwarf galaxies, the average molecular
gas depletion time for the normal (non-starburst) galaxies is $0.76 \pm 0.54$ Gyr. 
The average for the starburst galaxies is $0.06 \pm 0.04$ Gyr. (In calculating these averages, 
we have excluded galaxy Q2343-MD59 from \citealp{Tacconi2012}, as it is an obvious outlier.)
These averages generally agree
with previous work (see references for the literature sample in Section \ref{sec:litdata}). 
However, we note that a depletion time calculated from molecular gas and SFR surface densities 
may differ from ours (calculated from the total molecular gas mass and SFR)
due to different scale lengths of the CO and SFR distributions and trends in gas depletion
time with galaxy size (larger galaxies will dominate an area-weighted average gas depletion time).
This is the case for the sample of \cite{Leroy2008}, in which the authors
calculate an average molecular gas depletion time of 1.9 Gyr using surface densities, 
while the average depletion time using total quantities would be 1.3 Gyr (which lies within
the error of our average value for normal galaxies).

\begin{figure*}[t]
\centering
\includegraphics[width=0.49\linewidth]{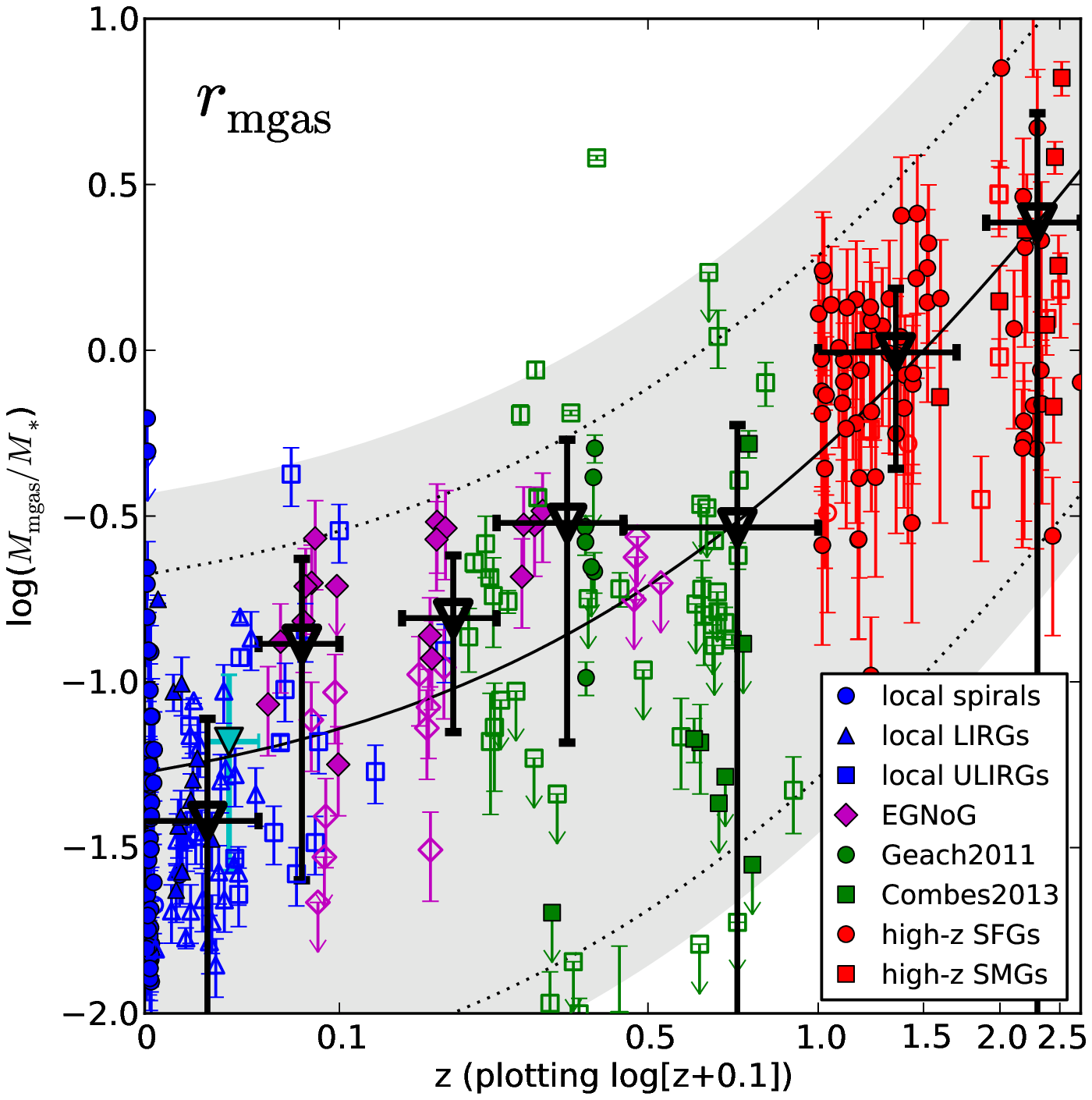}
\includegraphics[width=0.49\linewidth]{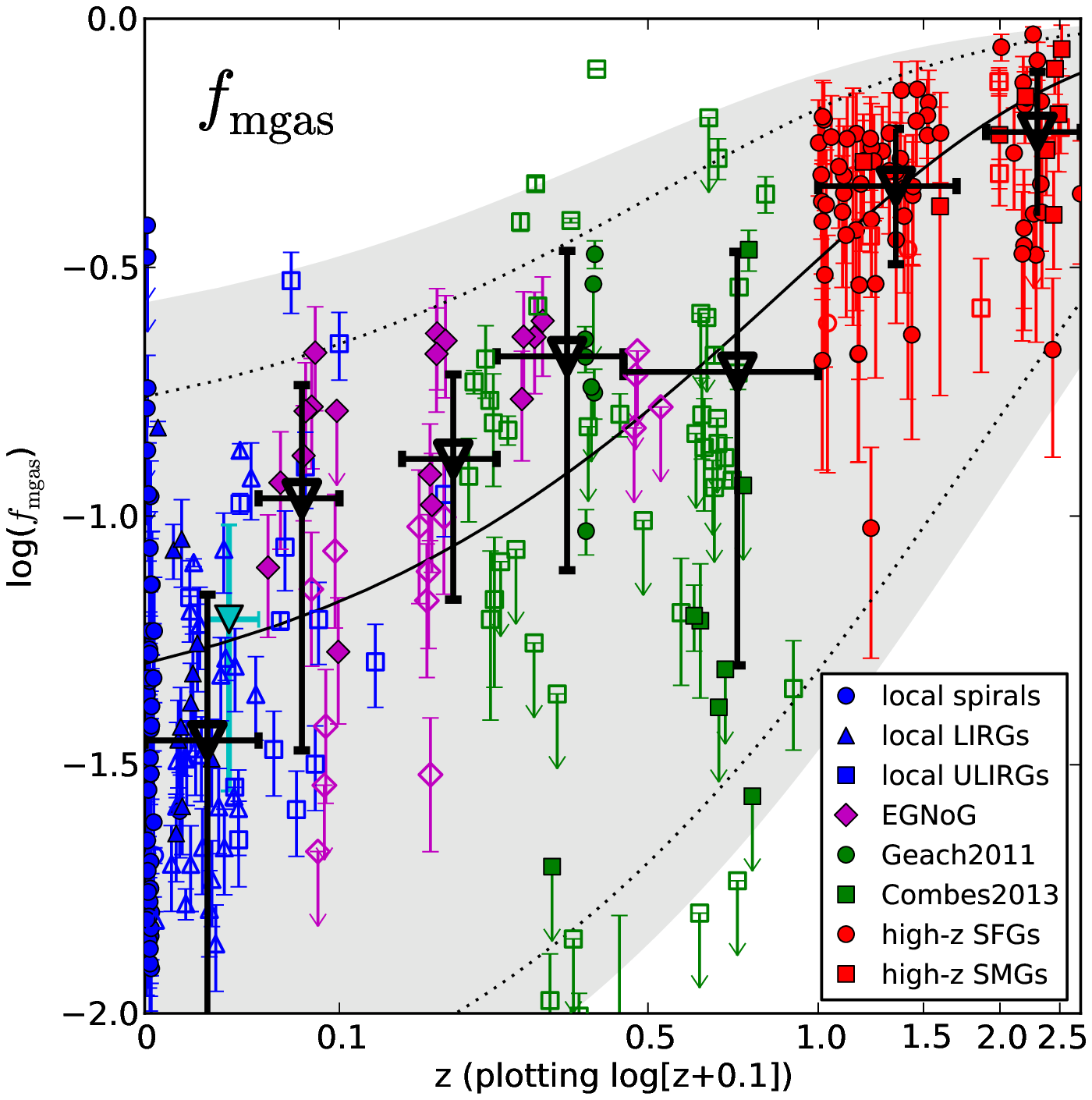}
\caption{Molecular gas ratio ($r_\mathrm{mgas}$) and molecular gas fraction ($f_\mathrm{mgas}$) versus redshift. We
plot $log(z+0.1)$ on the x-axis. The average value from the COLDGASS survey is plotted as a cyan upside-down triangle. Starburst
galaxies are plotted as un-filled symbols. Average values (with $1\sigma$ errors) for each of 6 redshift bins are plotted
as un-filled black upside-down triangles. The shaded gray area shows the expected behavior for normal galaxies,
with the solid black curve indicating the average. The dotted black lines bound the expected behavior for starburst galaxies.}
\label{fig:rgasevol}
\end{figure*}

In the left panel of Figure \ref{fig:tdep}, we plot CO luminosity versus SFR to illustrate that 
starburst and normal galaxies fill the same space in this plane, making the differentiation
of starburst from normal galaxies impossible using only CO luminosity and SFR. 
In addition to the literature sample described in Section \ref{sec:litdata}, we include here
local spirals, LIRGs and ULIRGs from \cite{Yao2003}, \cite{Mao2010} and \cite{Papa2011}
as gray circles, triangles and squares respectively. These datasets do not include stellar masses 
and thus are only included in this plot for illustrative purposes
(we estimate SFRs from the far-IR luminosities using the \citealt{Kennicutt1998} conversion).
The dashed lines indicate the area of the plane occupied by normal galaxies using 
$\alpha_\mathrm{CO}(\mathrm{Milky Way})$ and the average $\tau_\mathrm{mgas}$ for
normal galaxies given above. The dotted lines indicate the area occupied by starburst galaxies
(using $\alpha_\mathrm{CO}(\mathrm{ULIRG})$ and the starburst $\tau_\mathrm{mgas}$),
which overlaps the normal galaxy area significantly. This overlap occurs because the 
differences in $\alpha_\mathrm{CO}$ and $\tau_\mathrm{mgas}$ between starburst and normal
galaxies cancel each other out in this plane: as the effective $\alpha_\mathrm{CO}$ increases, 
a galaxy moves toward the lower right and as the effective $\tau_\mathrm{mgas}$ increases, 
a galaxy moves toward the upper left.

In the right panel of Figure \ref{fig:tdep}, the molecular gas depletion time is plotted against
SFR, with the average value for normal (starburst) galaxies indicated by the solid (dot-dashed)
line. 1$\sigma$ error bars on the averages are indicated by the gray shaded regions. 
The CO to H$_2$ conversion factor used for each data point is indicated by the fill of the symbol:
filled symbols are normal ($\alpha_\mathrm{CO}($Milky Way$)$) and 
un-filled symbols are starburst ($\alpha_\mathrm{CO}($ULIRG$)$) according to the sSFR 
criterion in Equation \ref{equ:mySBssfr} (plotted in the right panel of \ref{fig:sSFRlit}. Only galaxies that can be classified this way
(galaxies for which the $M_*$ and SFR have been estimated) are included in this plot.
Note that the $\tau_\mathrm{mgas}$ error bars plotted only represent the error associated with 
the CO measurement and not the SFR error
since we do not have errors on all the SFR measurements in the literature sample. 
We estimate that the typical SFR error is $\pm 50\%$.

The population of non-starbursting, local galaxies in the lower left corner of the right panel are dwarfs ($M_* < 10^{10}$), for which
a larger conversion factor may be more appropriate due to decreasing metallicity. 
\cite{Leroy2011} find that the conversion factor begins to increase significantly as the metallicity falls below 
$12 + \log(\mathrm{O}/\mathrm{H}) = 8.2-8.4$, which is expected
for $M_* \lesssim 10^9$ M$_\odot$ \citep{Tremonti2004}. This change in conversion factor would shift 
the dwarf galaxy points to higher $\tau_\mathrm{mgas}$. 
Therefore, we do not include dwarf galaxies in the calculation of the average $\tau_\mathrm{mgas}$ values.

\subsection{Evolution of the Molecular Gas Fraction}
\label{sec:fgasevol}

In order to calculate the expected evolution of the molecular gas fraction, we first
consider the ratio of the molecular gas mass to the stellar mass
($r_\mathrm{mgas}$), which can be written as the product of 
the molecular gas depletion time and the specific star formation rate: 
\begin{equation}
r_\mathrm{mgas} = \frac{M_\mathrm{mgas}}{M_*} = \frac{M_\mathrm{mgas}}{\mathrm{SFR}} \frac{\mathrm{SFR}}{M_*} = \tau_\mathrm{mgas} \times \mathrm{sSFR}
\label{equ:rmgas}
\end{equation}
In Section \ref{sec:sampselect}, we motivated and defined the sSFR of the main sequence as a function 
of redshift and stellar mass. As illustrated in the right panel of Figure \ref{fig:sSFRlit}, main sequence galaxies lie roughly within
a factor of 4 ($\pm 0.6$ dex) of the main sequence relation and starburst galaxies lie at sSFRs 4 to 30 times
larger than the main sequence value. 
In the previous section, we found an average $\tau_\mathrm{mgas}$ of $0.76 \pm 0.54$ Gyr for normal
galaxies and $0.06 \pm 0.04$ Gyr for starburst galaxies. Using these typical $\tau_\mathrm{mgas}$ values
and sSFR ranges (relative to the sSFR$_\mathrm{MS}$; see Equation \ref{equ:myssfr}), we get:
\begin{multline}
r_\mathrm{mgas} (\mathrm{normal}) = \\
(0.05 \pm 0.04) \eta_\mathrm{norm} (1+z)^{3.2} M_{*,11}^{-0.2}%\left( \frac{M_*}{10^{11}\ \mathrm{M}_\odot} \right)^{-0.2} 
\label{equ:rgasnorm}
\end{multline}
\begin{multline}
r_\mathrm{mgas} (\mathrm{starburst}) = \\
(0.04 \pm 0.03) \left(\frac{\eta_\mathrm{SB}}{10}\right) (1+z)^{3.2} M_{*,11}^{-0.2}%\left( \frac{M_*}{10^{11}\ \mathrm{M}_\odot} \right)^{-0.2}
\label{equ:rgasSB}
\end{multline}
where $M_{*,11} = M_* / (10^{11}\ \mathrm{M}_\odot)$ and $\eta_\mathrm{norm}$ and $\eta_\mathrm{SB}$
are the ranges of sSFR/sSFR$_\mathrm{MS}$ values appropriate for normal and starburst galaxies,
respectively: $\eta_\mathrm{norm} = 1/4$ to $4$ and $\eta_\mathrm{SB} = 4$ to $30$. 
In each of Equations \ref{equ:rgasnorm} and \ref{equ:rgasSB}, 
the scatter in $\tau_\mathrm{mgas}$ is reflected in the error of the first term ($\pm 0.04$ and $\pm 0.03$).
The corresponding molecular gas fraction may be calculated from $r_\mathrm{mgas}$: $f_\mathrm{mgas} = (1+r^{-1}_\mathrm{mgas})^{-1}$. 
The result is that the expected behavior of $r_\mathrm{mgas}$ and $f_\mathrm{mgas}$ is remarkably
similar for normal and starburst galaxies.
%\begin{multline}
%r_\mathrm{mgas} (\mathrm{normal}) = \\
%(0.05 \pm 0.04) \left[ \frac{1}{4} - 4 \right] (1+z)^{3.2} M_{*,11}^{-0.2}%\left( \frac{M_*}{10^{11}\ \mathrm{M}_\odot} \right)^{-0.2} 
%\label{equ:rgasnorm}
%\end{multline}
%\begin{multline}
%r_\mathrm{mgas} (\mathrm{starburst}) = \\
%(0.04 \pm 0.03) \left(\frac{[4-30]}{10}\right) (1+z)^{3.2} M_{*,11}^{-0.2}%\left( \frac{M_*}{10^{11}\ \mathrm{M}_\odot} \right)^{-0.2}
%\label{equ:rgasSB}
%\end{multline}
%where $M_{*,11} = M_* / (10^{11}\ \mathrm{M}_\odot)$. In each of Equations \ref{equ:rgasnorm} and \ref{equ:rgasSB}, 
%the scatter in $\tau_\mathrm{mgas}$ is reflected in the error of the first term ($\pm 0.04$ and $\pm 0.03$)
%and the appropriate range of sSFR values (relative to sSFR$_\mathrm{MS}$) is given in square brackets.
%The corresponding molecular gas fraction may be calculated from $r_\mathrm{mgas}$: $f_\mathrm{mgas} = (1+r^{-1}_\mathrm{mgas})^{-1}$. 
%The result is that the expected behavior of $r_\mathrm{mgas}$ and $f_\mathrm{mgas}$ is remarkably
%similar for normal and starburst galaxies.

Figure \ref{fig:rgasevol} shows the data and expectation for $r_\mathrm{mgas}$ (left panel) and 
$f_\mathrm{mgas}$ (right panel) versus redshift, plotting $log(z+0.1)$ on the x-axis in order to spread
out the low to intermediate redshifts for clarity. As in the right panel of Figure \ref{fig:tdep}, 
the starburst galaxies (identified using Equation \ref{equ:mySBssfr}) are plotted as un-filled symbols. 
Note that the error bars plotted only represent the error associated with 
the CO measurement (and not the stellar mass error), 
since we do not have errors on all the stellar mass measurements in the literature sample. 
We estimate that a typical stellar mass error is 0.1-0.2 dex.
We also plot the average value
at $z=0.025-0.05$ reported by the COLDGASS survey \citep{Saintonge2011a} as a cyan upside-down triangle.
We plot the expected $r_\mathrm{mgas}$ and $f_\mathrm{mgas}$ 
versus redshift for normal galaxies (Equation \ref{equ:rgasnorm}) as the solid black curve, with the gray shaded area indicating the range of possible values
(we take a nominal $M_*$ of $10^{11}$ M$_\odot$). The range for 
starburst galaxies (Equation \ref{equ:rgasSB}), indicated by the dotted black curves, overlaps
the expected range for normal galaxies.

The data points are binned into 7 redshift ranges (number of detected galaxies in parenthesis):
$z=0-0.05$ (110), $0.05-0.1$ (19), $0.15-0.25$ (18), $0.25-0.45$ (18), $0.45-1.0$ (15), $1.0-1.7$ (53), and $1.9-2.7$ (28).
The average values (for all detections) are plotted as un-filled black upside-down triangles, with vertical lines showing
the standard deviations 
(for the $z=0.25-0.45$ bin, the obvious outlier from \citealp{Combes2011} at $r_\mathrm{mgas} = 3.8$ is excluded).
All redshift bin averages lie within the expected range for both normal (gray shaded region) and
starburst (dotted lines) galaxies, and the average expectation curve
($r_\mathrm{mgas} = \mathrm{sSFR}_\mathrm{MS} \times (0.76$ Gyr)) lies
within the dispersion (vertical black lines). 

We note that the three intermediate redshift bins
($z=0.05-0.01$, $0.15-0.25$, and $0.25-0.45$) lie systematically higher than the average expectation curve. 
However, the number of galaxies in each of these bins is small and is
dominated by SFGs (EGNoG and Geach et al. galaxies), which we have shown to lie close to the starburst cutoff (see Figure \ref{fig:sSFRlit}). 
Therefore, these points are sensitive to the exact starburst criterion used (which determines $\alpha_\mathrm{CO}$).
For example, if we consider galaxies with sSFR $> 2$ sSFR$_\mathrm{MS}$
to be starbursts, the three intermediate redshift bin average values decrease by $\approx0.2$ dex, 
falling more in line with the average expectation curve. 

In summary, while the EGNoG galaxy redshift bins lie systematically higher than the average expectation curve
for the starburst criterion (and resulting $\alpha_\mathrm{CO}$ choices) adopted in this work, a reasonable
variation of this criterion results in a decrease of $\approx 0.2$ dex, bringing the averages in line with expectation. 
We also note, but do not investigate, that systematic effects may be present due to the use of different methods 
for calculating $M_*$ and SFR, and from instrumental sensitivity limiting CO detections to gas-rich 
galaxies at intermediate and high redshift. 
Overall, we conclude that the data agree well with the behavior of $r_\mathrm{mgas}$ and $f_\mathrm{mgas}$
expected from a simple prescription of sSFR and $\tau_\mathrm{mgas}$ with $z$ and $M_*$ in star-forming galaxies.
While the EGNoG bin D galaxies (non-detections) have not been included in the redshift bin averages discussed above, 
we note that the four upper limits are included in Figure \ref{fig:rgasevol} and agree with expectations as well.

\subsection{A Bimodal Conversion Factor Prescription}
\label{sec:bimodalalpha}

In this work, we have used a bimodal prescription for the conversion factor $\alpha_\mathrm{CO}$
in normal and starburst galaxies. We have included some local low-mass galaxies in our literature sample,
but we do not include these systems in the calculation of average values since
a different $\alpha_\mathrm{CO}$ is likely appropriate due to decreasing metallicity \citep{Leroy2011, Narayanan2012}.
While our bimodal prescription (excluding low-mass galaxies) likely does not describe the true conversion factor in all of these galaxies
(e.g. simulations by \citealt{Narayanan2012} suggest a continuous variation of $\alpha_\mathrm{CO}$ from normal to starburst galaxies),
it appears to capture the typical behavior. A recent study by \cite{Magnelli2012} using observed dust masses to calculate
the conversion factor in $z>1$ star-forming galaxies found smaller conversion factors in starburst galaxies (classified by sSFR, as in this work),
but it was unable to distinguish between a step function and a continuous variation of $\alpha_\mathrm{CO}$ as a function of sSFR/sSFR$_\mathrm{MS}$. 
In this work, we have used only total measurements of the galaxy properties, and we note that
with resolved CO measurements, it would be possible to apply a value of $\alpha_\mathrm{CO}$
more tailored to local conditions (like the prescriptions of \citealp{Narayanan2012, Bolatto2013}),
which could tighten the observed relations.

Independent of this issue of bimodal or continuous conversion factor is the question of whether
the local values for normal and starburst conversion factors may be extended to high redshift. 
Both simulations by \cite{Narayanan2012} and observations by \cite{Daddi2010a} suggest that the
conversion factor is lower than the Milky Way value in normal star-forming galaxies at high redshift.
We cannot answer this question with the current dataset, but note that 
using a reduced $\alpha_\mathrm{CO}$ in high-redshift SFGs or a continuous variation of $\alpha_\mathrm{CO}$ 
from normal to starburst galaxies would not change the general conclusions.

\section{Conclusions}
\label{sec:conclusions}

In this paper, we present the EGNoG survey: CO observations of 31 galaxies from $z=0.05$ to $z=0.5$.
We detect 24 of the 31 observed galaxies, providing integrated CO maps, spectra, and luminosities for
detections and $3\sigma$ upper limits for non-detections. 

We place the EGNoG galaxies in context with a sample
of 185 normal and starburst galaxies at low and high redshift collected from the literature. 
We standardize the comparison of the EGNoG and literature galaxies using a simple prescription for molecular gas mass calculation.
Each galaxy is classified as normal or starburst based on its sSFR relative
to the sSFR of the main sequence of star-forming galaxies at that redshift
and stellar mass (see Equations \ref{equ:myssfr} and \ref{equ:mySBssfr}).
We then use a bimodal prescription for $\alpha_\mathrm{CO}$ based on this classification 
(a Milky Way-like value for normal galaxies, a ULIRG-like value for starbursts), and calculate
the molecular gas mass, molecular gas depletion time, and molecular gas fraction of each galaxy.

Using this prescription, we find an average molecular gas depletion time of $0.76 \pm 0.54$ Gyr for normal
galaxies and $0.06 \pm 0.04$ Gyr for starburst galaxies. We calculate an average molecular gas
fraction of 10-20\% at the intermediate redshifts probed by the EGNoG survey ($z=0.05-0.5$). 
By expressing the molecular gas fraction in terms of the sSFR and molecular gas depletion time
($f_\mathrm{mgas} = (1+ [\tau_\mathrm{mgas}\times\mathrm{sSFR}]^{-1})^{-1}$), and using
typical ranges of sSFR and $\tau_\mathrm{mgas}$ for starburst and normal galaxies, we
calculate the expected evolution of the molecular gas fraction with redshift. 
We find that the expected behavior for normal and starburst galaxies is remarkably similar
and agrees well with the EGNoG and literature data from $z=0$ out to $z\sim2.5$.

-------------------------------------------------------------------
\acknowledgements
The authors thank the anonymous referee for helpful comments, 
Kevin Bundy for useful discussions and assistance with the COSMOS
dataset and Linda Tacconi, Fran\c{c}oise Combes, Matt Bothwell and L.G. Hou for 
providing their datasets for comparison with the EGNoG results.
A. Bauermeister thanks Statia Cook, Dick Plambeck and Peter Williams 
for useful discussions on the reduction and analysis of the EGNoG data.
A. Bolatto wishes to acknowledge partial support from
grants CAREER NSF AST-0955836, NSF AST-1139998, as well as a Cottrell
Scholar award from the Research Corporation for Science Advancement.
We thank the OVRO/CARMA staff and the CARMA observers for their assistance in obtaining the data. 
Support for CARMA construction was derived from the Gordon and Betty Moore Foundation, 
the Kenneth T. and Eileen L. Norris Foundation, the James S. McDonnell Foundation, the 
Associates of the California Institute of Technology, the University of Chicago, the states of 
California, Illinois, and Maryland, and the National Science Foundation. Ongoing CARMA 
development and operations are supported by the National Science Foundation under a 
cooperative agreement, and by the CARMA partner universities.

\bibliographystyle{yahapj}
%\bibliography{/Users/amber/Documents/research/myrefs}
\bibliography{myrefs}

\newpage
\begin{appendix}

\section{Stellar Mass and SFR Probability Distribution Functions}
\label{sec:msfrPDFs}

\begin{figure*}[b]
\centering
\includegraphics[width=\linewidth]{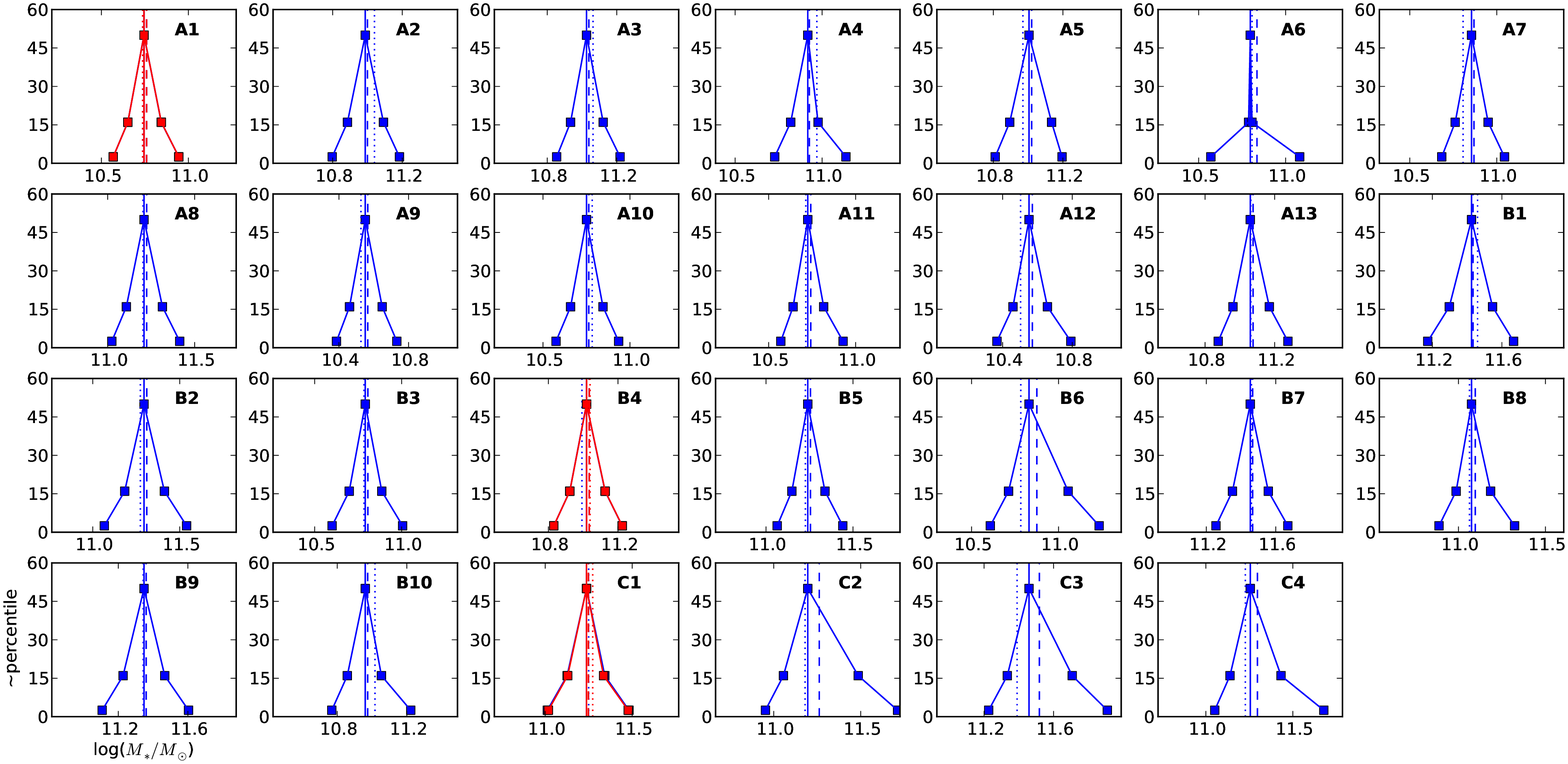}
\includegraphics[width=\linewidth]{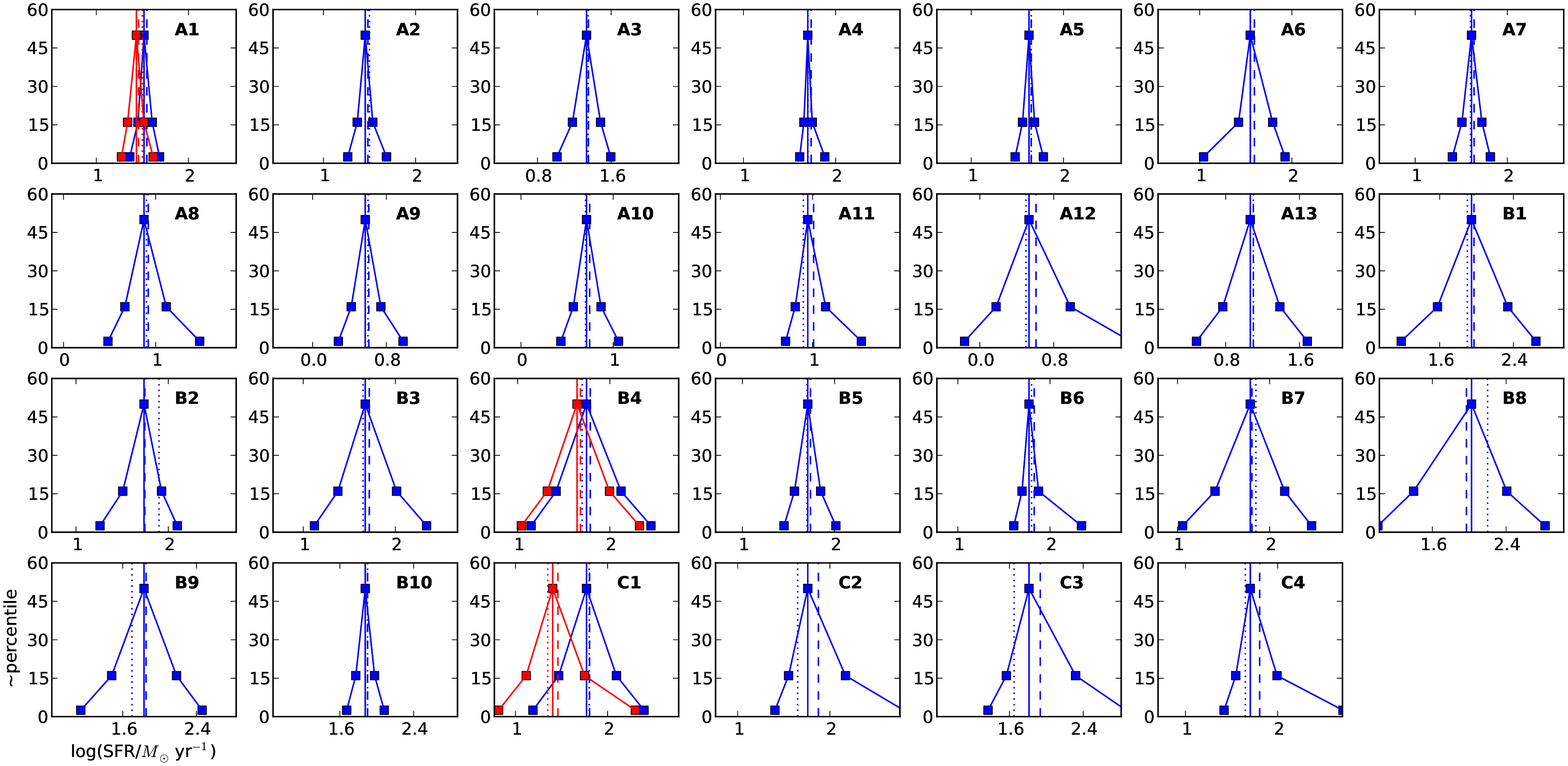}
\caption{Stellar mass (top) and SFR (bottom) probability distribution functions of EGNoG 
galaxies selected from SDSS (bins A, B and C). 
For galaxies A1, B4 and C1, a duplicate SDSS source exists, over-plotted in red.}
\label{fig:masssfrPDFs}
\end{figure*}

For each of the EGNoG galaxies in redshift bins A-C, we show the stellar mass and SFR probability distribution functions
in Figure \ref{fig:masssfrPDFs} in the top and bottom panels respectively. These distributions come 
from the MPA-JHU group (see text) for SDSS DR7 galaxies.
Data points give the 2.5, 16, 50, 84 and 97.5 percentiles of the distribution. The median, 
mean and mode are indicated by the vertical solid, dashed and dotted lines respectively. 
For galaxies A1, B4 and C1, a duplicate SDSS source exists (as a result of SDSS automated source-finding). 
The duplicate source is plotted in red in Figure \ref{fig:masssfrPDFs}.
For each of the quantities plotted in Figure \ref{fig:masssfrPDFs} (stellar mass in the top panels, 
SFR in the bottom panels), the x-axis range is constant.

\section{Data Reduction and Flux Measurement}
\label{sec:datreducandflux}

\begin{table*}[t]
\centering
\renewcommand{\arraystretch}{1.1}
\begin{tabular}{| l | c | r | r | r | l |}
\hline
Calibrator & Obs. & Freq. &  & Flux & Primary Calibrator \\
Name & Band & (GHz) & Date Range & (Jy) & (Flux in Jy) \\
\Cline{0.8pt}{1-6}
0006--063 & 3mm & 105 & Oct 2010 & 1.6 & MWC349 (1.28 Jy) \\
\hline
0010+109 & 3mm & 109 & Nov 2010 & 0.2 & Uranus \\
\hline
0108+015 & 3mm & 97 & Sep 2011 & 2.0 & Uranus; Neptune \\
0108+015 & 3mm & 97 & Feb 2012 & 1.7 & Uranus \\
\hline
0854+201$^p$ & 3mm & 107 & Nov 2010 & 5.2 & 3C84 (11 Jy) \\
0854+201$^p$ & 3mm & 98 & Aug-Oct 2011 & 4.0 & Mars \\
0854+201$^p$ & 3mm & 98 & Nov 2011 & 5.2 & Mars \\
0854+201$^p$ & 3mm & 98 & Feb 2012 & 5.2 & Mars \\
0854+201 & 1mm & 258 & Aug 2011 & 2.0 & Mars \\
0854+201 & 1mm & 254 & Apr 2012 & 4.0 & Mars \\
\hline
0920+446 & 3mm & 106 & Nov 2010 & 1.9 & 3C84 (11 Jy) \\
0920+446 & 3mm & 106 & Feb 2012 & 1.0 & 3C84 (15 Jy) \\
\hline
0958+655 & 3mm & 105 & Nov 2010 & 1.1 & 3C273 (12 Jy) \\
0958+655 & 3mm & 105 & Feb 2012 & 1.7 & MWC349 (1.28 Jy) \\
\hline
1058+015$^p$ & 3mm & 107 & May 2011 & 4.6 & 3C84 (10 Jy) \\
1058+015$^p$ & 3mm & 99 & Nov 2011 & 4.1 & Mars \\
1058+015$^p$ & 3mm & 99 & Feb 2012 & 3.6 & Mars \\
1058+015$^p$ & 3mm & 99 & Apr 2012 & 2.8 & Mars \\
1058+015 & 1mm & 266 & Aug 2011 & 2.6 & Mars \\
1058+015 & 1mm & 226 & Apr 2012 & 2.0 & Mars \\
\hline
1159+292 & 3mm & 105 & Apr-May 2011 & 1.4 & 3C84 (10 Jy); 3C273 (10 Jy) \\
1159+292 & 3mm & 105 & Feb 2012 & 0.7 & Mars \\
\hline
1224+213 & 1mm & 252 & Apr 2012 & 0.6 & Mars \\
\hline
1310+323 & 3mm & 90 & Sep-Nov 2011 & 1.7 & Mars \\
1310+323 & 3mm & 98 & Apr 2012 & 1.4 & Mars \\
1310+323 & 1mm & 266 & Aug 2011 & 0.6 & MWC349 (2.0 Jy) \\
\hline
1357+193 & 3mm & 105 & Apr-May 2011 & 0.7 & MWC349 (1.28 Jy) \\
1357+193 & 3mm & 105 & Sep-Nov 2011 & 0.8 & Mars; MWC349 (1.28 Jy) \\
\hline
2134--018 & 3mm & 106 & Apr 2011 & 1.2 & Uranus \\
\hline
2232+117 & 3mm & 106 & Apr 2011 & 1.3 & MWC349 (1.28 Jy) \\
2232+117 & 3mm & 97 & Aug-Sep 2011 & 1.5 & Uranus; MWC349 (1.23 Jy) \\
2232+117 & 3mm & 97 & Feb 2012 & 1.7 & Uranus; Neptune; MWC349 (1.23 Jy) \\
\hline
3C273 & 3mm & 98 & Apr 2012 & 6.8 & Mars \\
\hline
3C454.3 & 3mm & 107 & Oct 2010 & 39.5 & Uranus \\
\hline
\end{tabular}
\caption{Summary of calibrator fluxes used to set the flux scale of EGNoG data. For each flux, the observing band
(1 or 3 mm), frequency, and date range are listed. The last column lists the primary calibrator used to set the flux.
For non-planet calibrators, the flux used (at the frequency given in the third column) is given in parentheses. \\
$^p$ Indicates the calibrator is linearly polarized, requiring additional reduction steps (only applicable for 3 mm data) 
detailed in Appendix~\ref{sec:polcal}. } 
\label{tab:calfluxes}
\end{table*}

\subsection{Data Reduction}
Each dataset was reduced and calibrated as follows.
The data were flagged for antenna - antenna shadowing and
any other issues present during the observation (e.g. high system temperature, high gain, tracking problem, etc.). 
The instrument bandpass was calibrated with
{\tt mfcal} on a bright passband calibrator. The time-dependent antenna gains (from atmospheric variation)
were derived by performing a {\tt selfcal} on the phase calibrator with an averaging interval of 18 minutes 
(the timescale of switching between the source and phase calibrator). For galaxies A5, A11, B3, B4, B7, C1 and C2, 
the phase calibrator (0854+201 or 1058+015) was up to 10\% polarized, which required additional steps in 
the reduction of the 3mm data (observed with linearly polarized feeds). 
This is described in detail in Appendix \ref{sec:polcal}.

For each dataset, the flux of the phase calibrator was set during the antenna
gain calibration in order to properly set the flux scale of the data. The flux of each phase
calibrator was assumed to be constant over timescales of weeks, and was 
therefore determined from the best datasets of the survey using
{\tt bootflux} on bandpass-calibrated, phase-only gain-calibrated data using a planet (Uranus, Neptune or Mars)
or MWC349 as a primary flux calibrator. In some cases, none of these flux calibrators was available
so a bright quasar (3C84 or 3C273) was used instead, with the flux estimated using historical flux monitoring data
at CARMA\footnote{See CARMA Memo \#59 \citep{CARMAmemo59} for a description of flux monitoring at CARMA}. 
Table \ref{tab:calfluxes} provides the 
flux used for each phase calibrator and the
primary calibrator used. For non-planet primary calibrators, the flux used is given in parenthesis.
The flux of MWC349 was calculated assuming 1.2 Jy at 92 GHz with a spectral index of 0.5, the typical value from historical flux
monitoring at CARMA\footnotemark[\value{footnote}]$^,$\footnote{A history of MWC349 fluxes measured at CARMA is maintained
at \url{http://cedarflat.mmarray.org/fluxcal/primary_sp_index.htm}}.  
The brightness temperature ($T_b$) of the planets was set by the CARMA system. The brightness temperature
of Mars comes from the Caltech thermal model of Mars (courtesy of Mark Gurwell), which includes seasonal variations in temperature.
This model can be accessed in MIRIAD using {\tt marstb}. For Uranus and Neptune, the CARMA system uses the following power laws:
\begin{eqnarray}
T_b(\mathrm{Uranus}) = 134.7 \left(\frac{\nu}{\mathrm{100~GHz}}\right)^{-0.337} ~\mathrm{K}\\
T_b(\mathrm{Neptune}) = 129.8 \left(\frac{\nu}{\mathrm{100~GHz}}\right)^{-0.350}~\mathrm{K}
\end{eqnarray}

Data cubes were produced combining all fully calibrated datasets of each source. 
We used {\tt invert}, weighting the visibilities by the system temperature as well as using
a Briggs' visibility weighting robustness parameter \citep{Briggs1995} of 0.5. 
Since CARMA is an inhomogeneous array (these data use both the 10 and 6 m dishes), 
we also use {\tt options=mosaic} in the {\tt invert} step to properly handle the three
different primary beams patterns (10m-10m, 10m-6m and 6m-6m). All observations consisted of a single pointing.
The resulting cubes are primary-beam-corrected. If channel averaging was required (based on the strength
of the CO line), this was done in the {\tt invert} step.

We deconvolved each image with {\tt mossdi} (the mosaic version of {\tt clean}), 
cleaning down to the rms noise within a single channel, within a cleaning box selected by eye to include
only source emission. We cleaned only channels containing visible source emission.
Cleaning down to a specified noise level is preferred to using a set number of clean iterations
due to the nature of the spectral line emission: some channels will contain more flux and
therefore require more clean iterations. In tests using a model source of known flux inserted
into real data (emission-free channels), we found a $1\sigma$ cutoff to best extract
the true source emission without overestimating the flux over a range of detection significance levels, with
a 10-30\% uncertainty in the recovered flux (depending on the significance of the signal).
The final cleaned cubes were produced by {\tt restor}, which convolves the clean component cube with 
a clean beam (calculated by fitting a Gaussian to the combined mosaic beam given 
by {\tt mospsf}) and adds the residuals from the cleaning process.

\subsection{Flux Estimation}
\label{sec:fluxest}
Total source fluxes in the CO lines are calculated by summing `source' pixels
in each `source' velocity plane of the final, calibrated, cleaned cubes. The source velocity planes are 
selected by eye. The source pixels are those within the `source' region
which are not masked by our smooth mask, adapted from the masked moment method of \cite{Dame2011}. 
The smooth mask is created by applying a $2\sigma$ clip to a smoothed version of the image: 
Hanning smoothing is done along the velocity axis, and each velocity plane
is convolved spatially with a Gaussian beam twice the size of the original synthesized beam. This smoothed
mask is used to exclude noise pixels but still capture low-level emission that a simple $\sigma$-clip (without smoothing) would miss.
In our own testing (using a model source of known flux inserted
into real data), we found a $2\sigma$ clip to
best reproduce the true flux over a range of detection significance levels. 

\begin{figure*}
\centering
\includegraphics[width=6.5in]{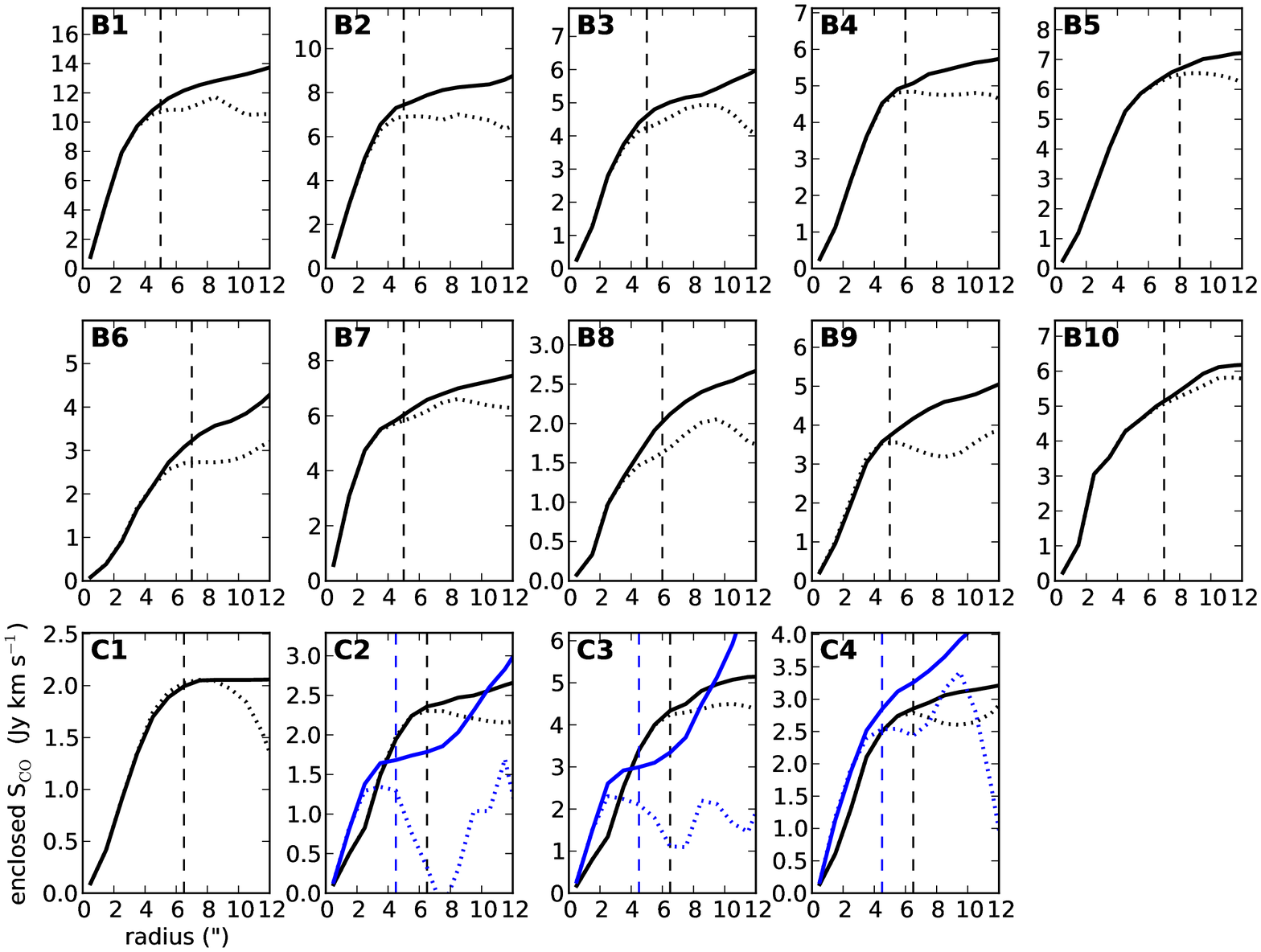}
\caption{Enclosed flux versus radius for bin B and C sources. The masked (unmasked) data are
shown by the solid (dotted) line. The vertical dashed line indicates the source region radius adopted
for each galaxy. For galaxies C2, C3 and C4, the CO\jone line data are shown in black and the CO\jthree line data in blue.}
\label{fig:binBCenclflux}
\end{figure*}

An appropriate source region size was selected to recover all of the flux without including the negative bowl. 
Since we do not have single-dish data to complement the interferometric data presented here, 
our datasets are missing the shortest $uv$ spacings. As a result, the emission in the clean maps
sits in a negative bowl, which will affect the measured flux (calculated by summing pixels within a given radius).
The bin A galaxies are observed with a sufficiently small beam so that the galaxies are resolved. In this case, 
we use an elliptical source region selected by eye to enclose the spatial extent of the CO emission of each galaxy. 
In the case of the bins B and C galaxies, the molecular gas is unresolved, so we use 
a circular source region with radius set to 
the radius at which the radial profile of the enclosed flux first peaks, thereby excluding the negative bowl. 
The radial profiles of the enclosed flux for bin B and C sources are shown in Figure \ref{fig:binBCenclflux}.
The negative bowl is most evident in the unmasked profiles (dotted lines in Figure \ref{fig:binBCenclflux}), in which the enclosed flux 
peaks and then decreases with radius. We have selected the following source region radii by eye (approximately
the radius at which the enclosed flux first peaks):
$4.5 \arcsec$ for bin C sources in the CO\jthree line (1 mm data); $5 \arcsec$ for sources B1, B2, B3, B7 and B9; 
$6 \arcsec$ for B4 and B8; $6.5 \arcsec$ for bin C sources in the CO\jone line (3 mm data); 
$7 \arcsec$ for B6 and B10; $8 \arcsec$ for B5.
The source region radius of each galaxy in bins B and C is indicated by the vertical dashed line in each panel of Figure \ref{fig:binBCenclflux}.
For all galaxies, the source region is indicated by the dotted white ellipse in the moment 0 and moment 1 maps
of Figures \ref{fig:binAmaps1} - \ref{fig:binCmaps1mm}.

The error in each flux measurement is estimated from the standard deviation of the measured fluxes
using different velocity channel averaging and starting channels, using three different methods 
of calculating the flux in each case.
The three methods are: the $2\sigma$ masking technique described above, 
the same masking technique with a $3\sigma$ clip, and the simple addition of all pixels (no mask) within the
source region. 

We performed extensive testing of our analysis technique to choose the parameters of the reduction
to eliminate systematic offsets and minimize the uncertainty 
due to noise (as described above). 
We find a 10-30\% error in the flux measurement in each channel coming from noise in the data and 
the adopted reduction and analysis steps. 
From this, we adopt an average uncertainty of 20\% in the flux estimated in each channel, which results in an
uncertainty in the total flux of 20\%$N_\mathrm{ch}^{-0.5}$, where $N_\mathrm{ch}$ is the number of velocity
channels in which the flux is summed. For the total flux values reported in Table \ref{tab:COprop}, we estimated
the errors from the variations in the fluxes calculated using different channel averaging, flux measurement methods, etc. 
(as described above), consistent with the 20\%$(N_\mathrm{ch})^{-0.5}$ expected. 

Further, these data suffer from systematic effects due to the absolute flux calibration and primary beam correction.
We set the flux scale in our dataset using a primary flux calibrator, the flux of which
is only known to $\approx20\%$. In the primary beam correction of the dataset, pointing and focus errors at the 
time of the observations as well as errors in the primary beam model can significantly reduce image fidelity, leading
to errors in the measured fluxes of $\approx20\%$ (see Square Kilometer Array (SKA) Memo \#103, \citealt{SKAmemo103}). 
Combining these errors in 
quadrature, we estimate that our flux measurements suffer from additional uncertainties of up to $\approx30\%$. 
We consider all these factors in the presentation of our data in Table \ref{tab:COprop}: 
for the line flux ($S_\mathrm{CO}$), we report the measured error; for $L_\mathrm{CO}'$, we include an additional $30\%$ error
(added in quadrature to the measured error in $S_\mathrm{CO}$).

\section{Polarized Calibrators}
\label{sec:polcal}

In this section, we describe how we calibrated datasets which used a strongly polarized
source for gain calibration. 
This applies to the 3mm data only, which were taken using single-polarization, linearly polarized
receivers. The 1mm data used dual-polarization, circularly polarized feeds and are thus not affected
by the strong linear polarization present in some calibrators. 

For Stokes I intensity $I$, linear polarization fraction $p_{qu}$, and
polarization angle $PA$, the observed YY intensity as a function of parallactic angle $\chi$
is described by
\begin{equation}
YY = I \{ 1 - p_{qu} \cos[2(PA - \chi)]\}
\label{equ:polflux}
\end{equation}
Thus, the observed YY amplitude varies sinusoidally with parallactic angle in a dataset.
Any amplitude self-calibration on the gain calibrator would assume constant amplitude
with time and yield higher or lower gains according to this amplitude variation. Since this effect can
be as large as 10\% in some cases, we corrected for it in the following way. After a phase-only {\tt selfcal}, 
we used a modified version of {\tt uvcal} to remove the known polarization effect from the calibrator
data and then performed an amplitude {\tt selfcal}. The resulting gains should represent the atmospheric 
variations during the track, with the influence of the polarization of the calibrator removed. 

\begin{figure*}[t]
\begin{minipage}[h]{0.5\linewidth}
\centering
\includegraphics[width=3.5in]{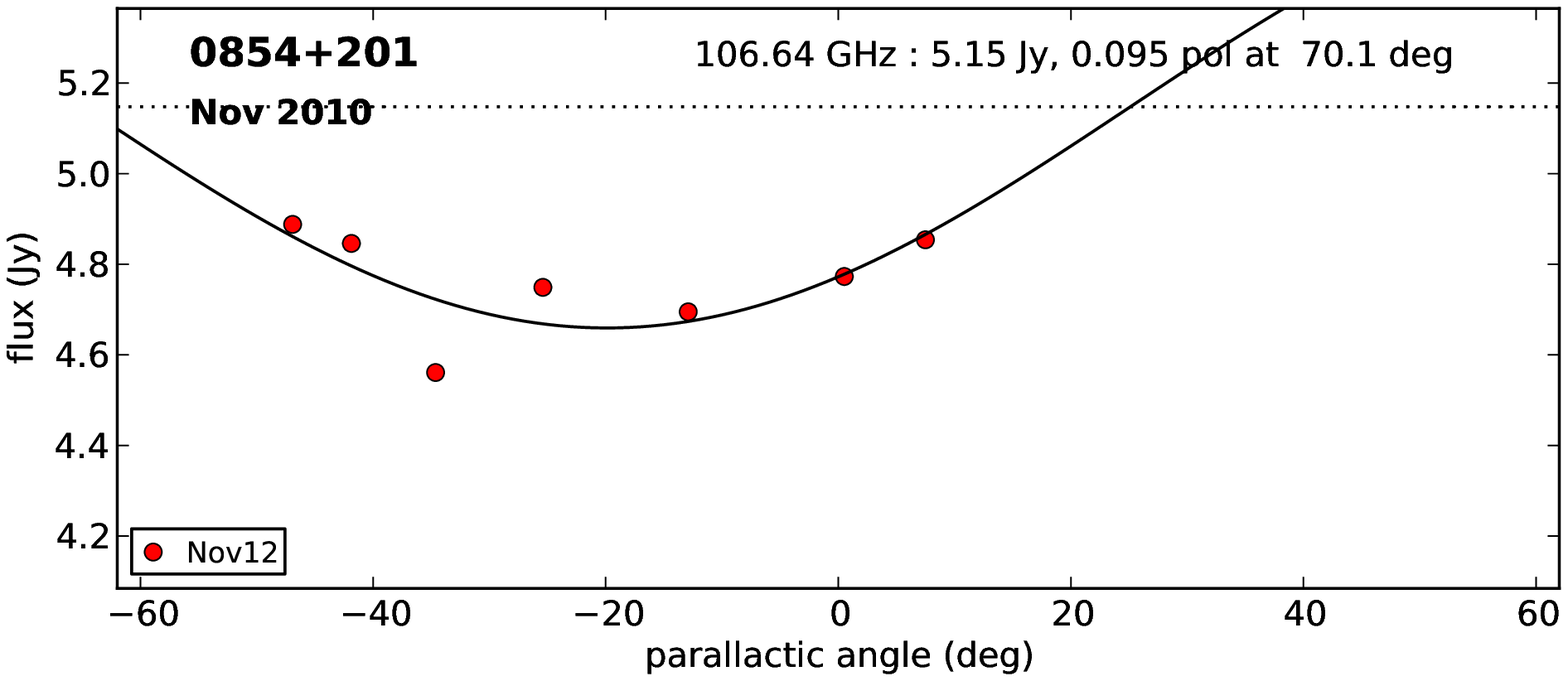}
\includegraphics[width=3.5in]{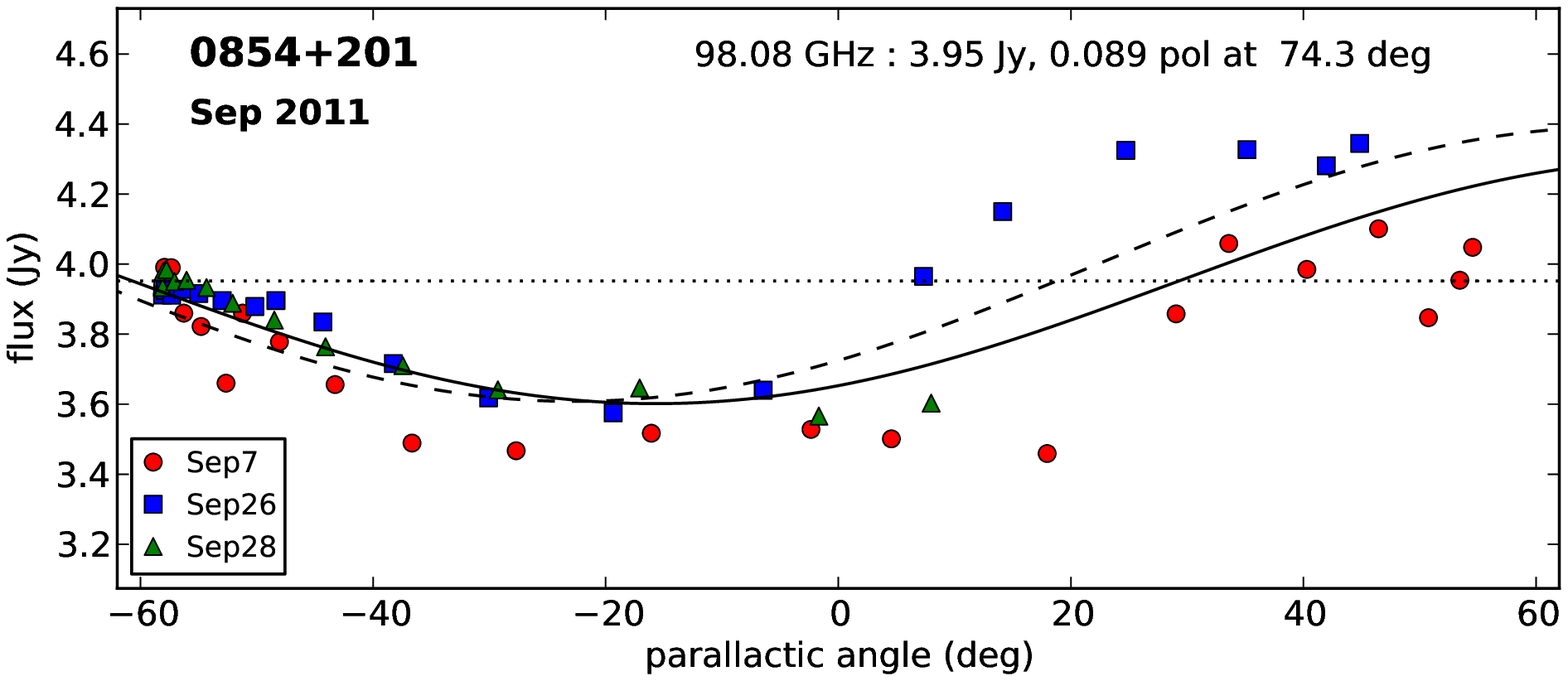}
\includegraphics[width=3.5in]{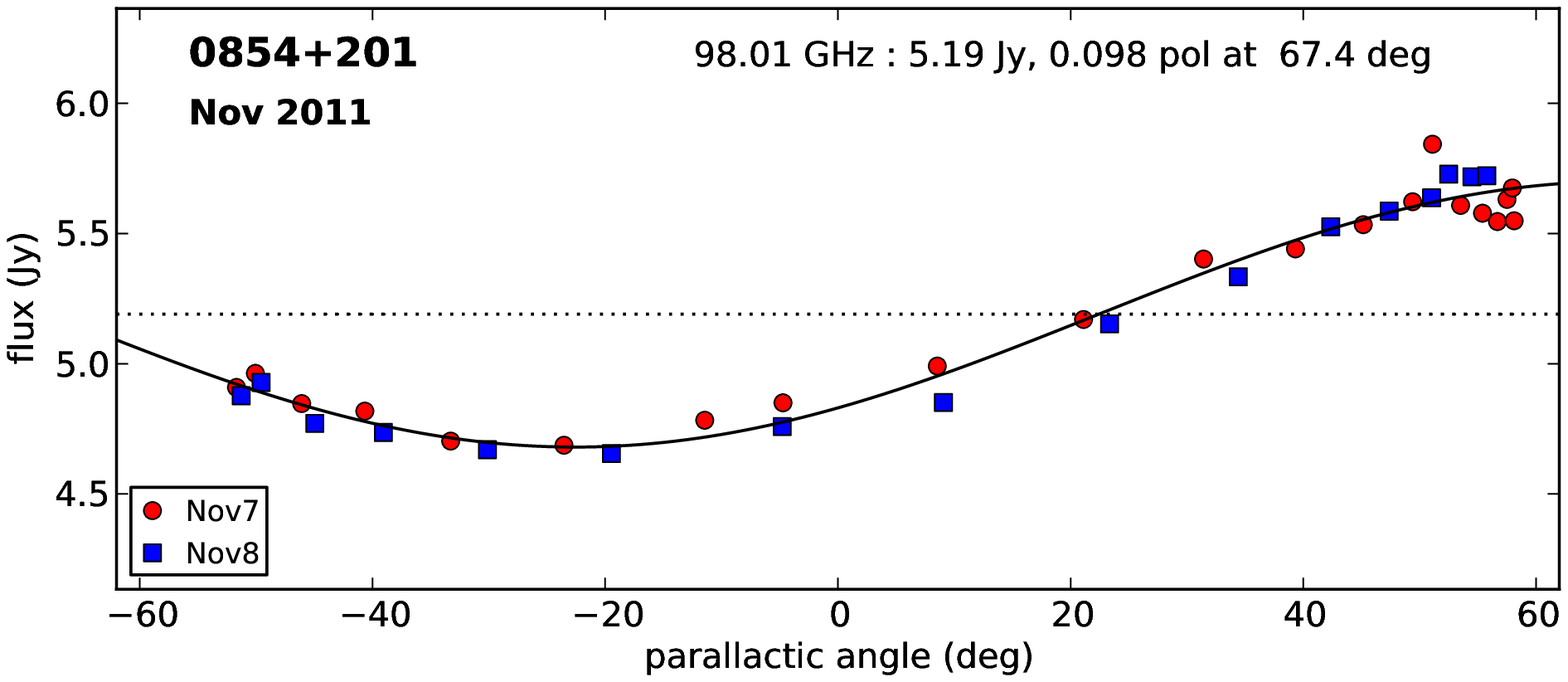}
\includegraphics[width=3.5in]{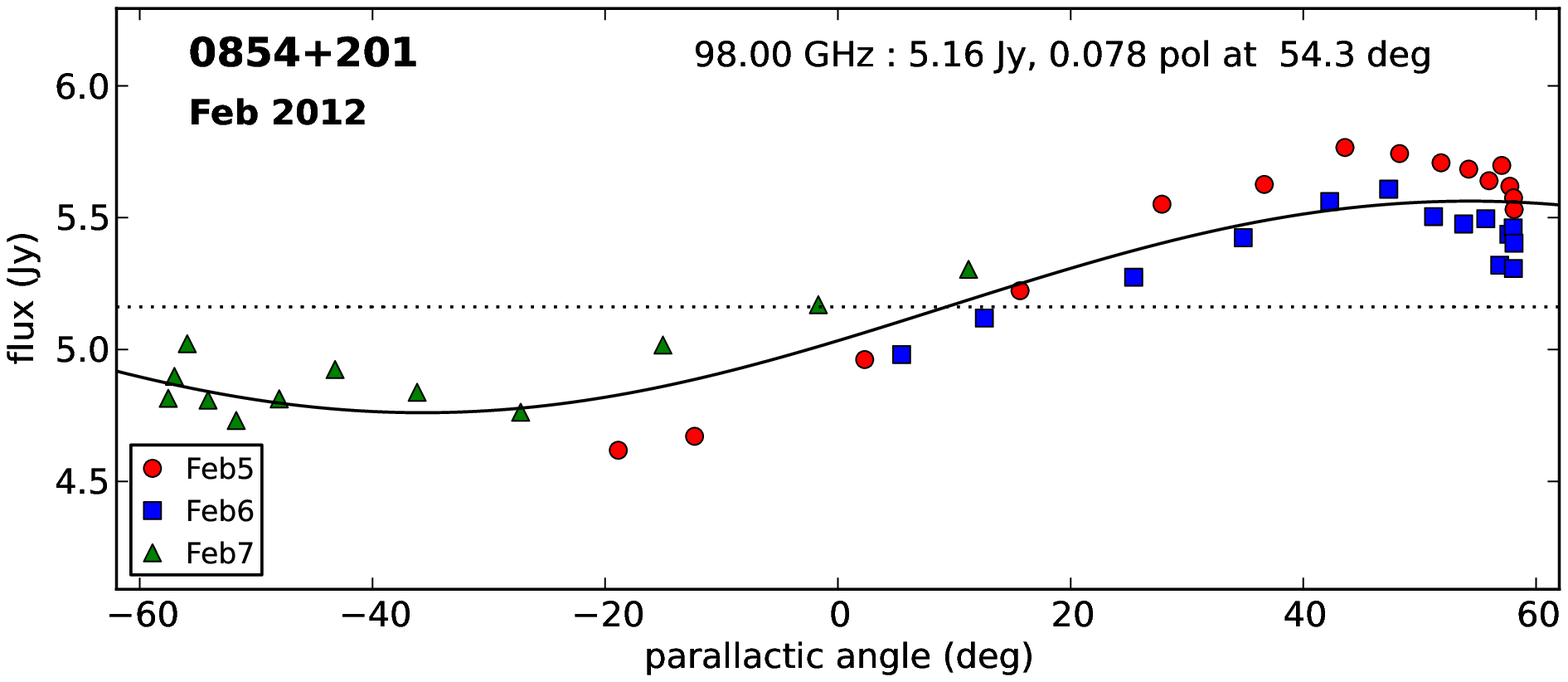}
\end{minipage}
\begin{minipage}[h]{0.5\linewidth}
\centering
\includegraphics[width=3.5in]{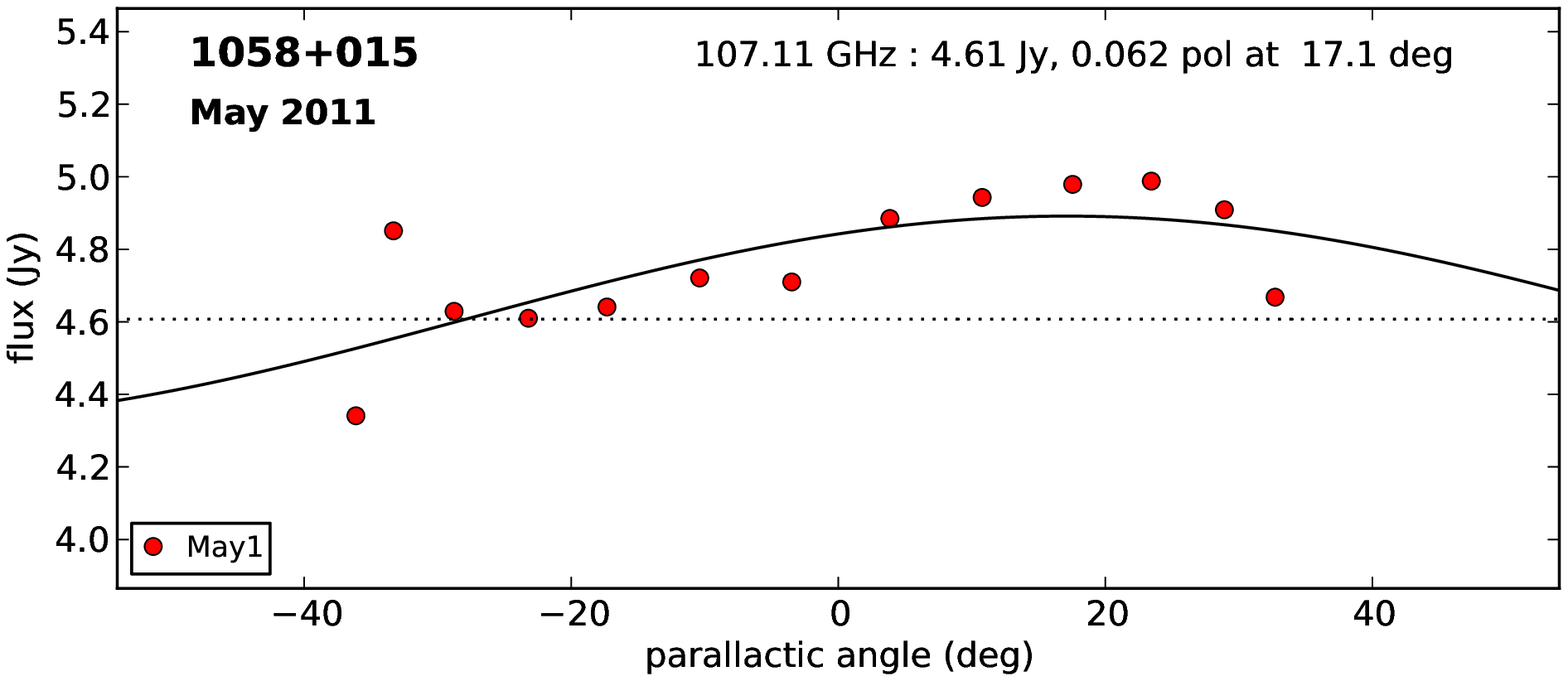}
\includegraphics[width=3.5in]{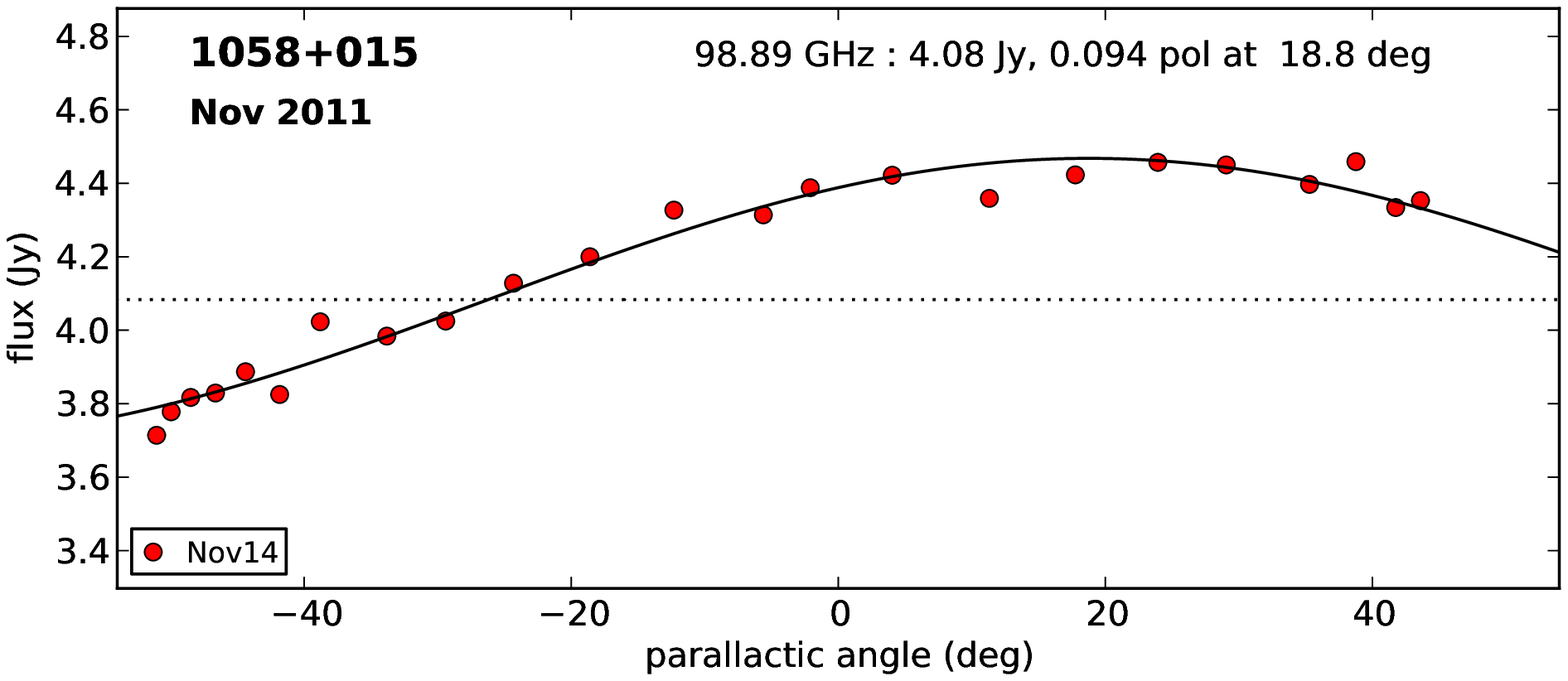}
\includegraphics[width=3.5in]{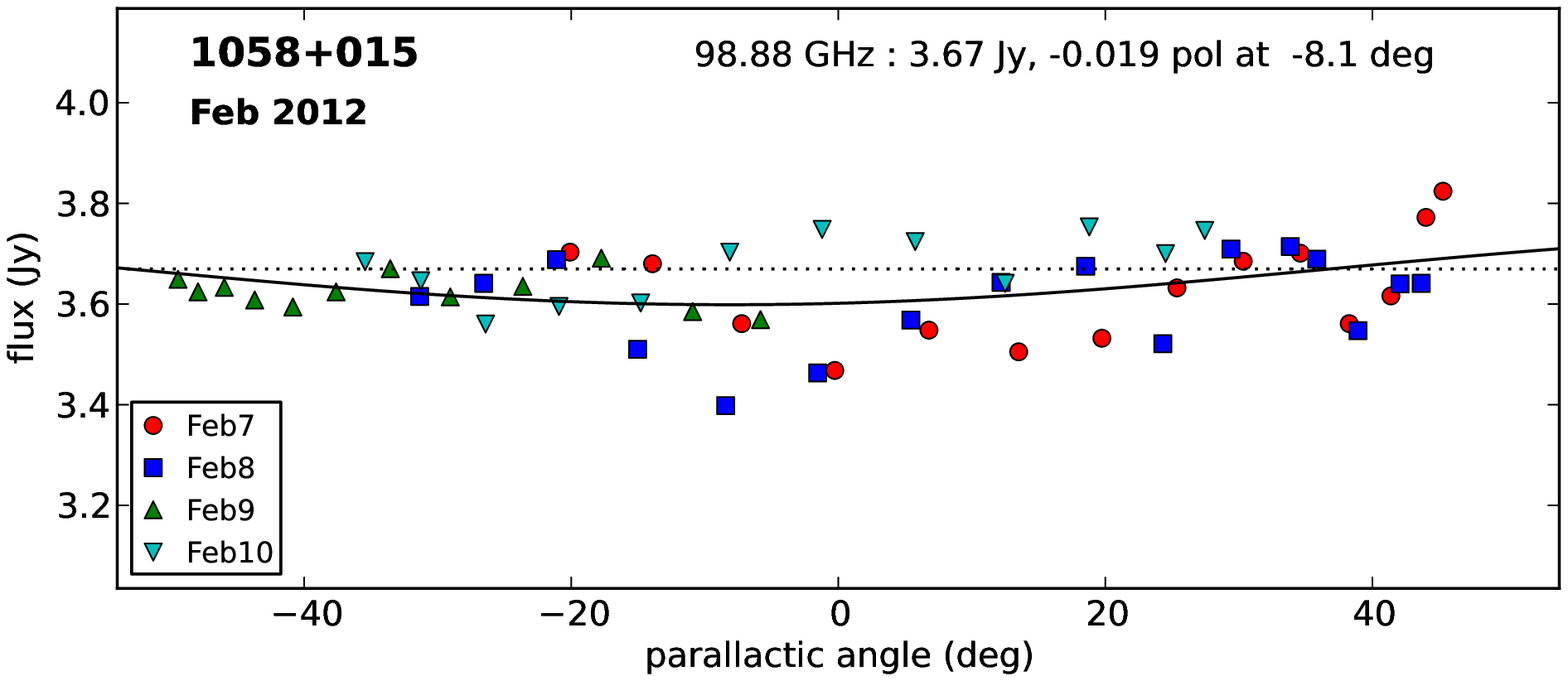}
\includegraphics[width=3.5in]{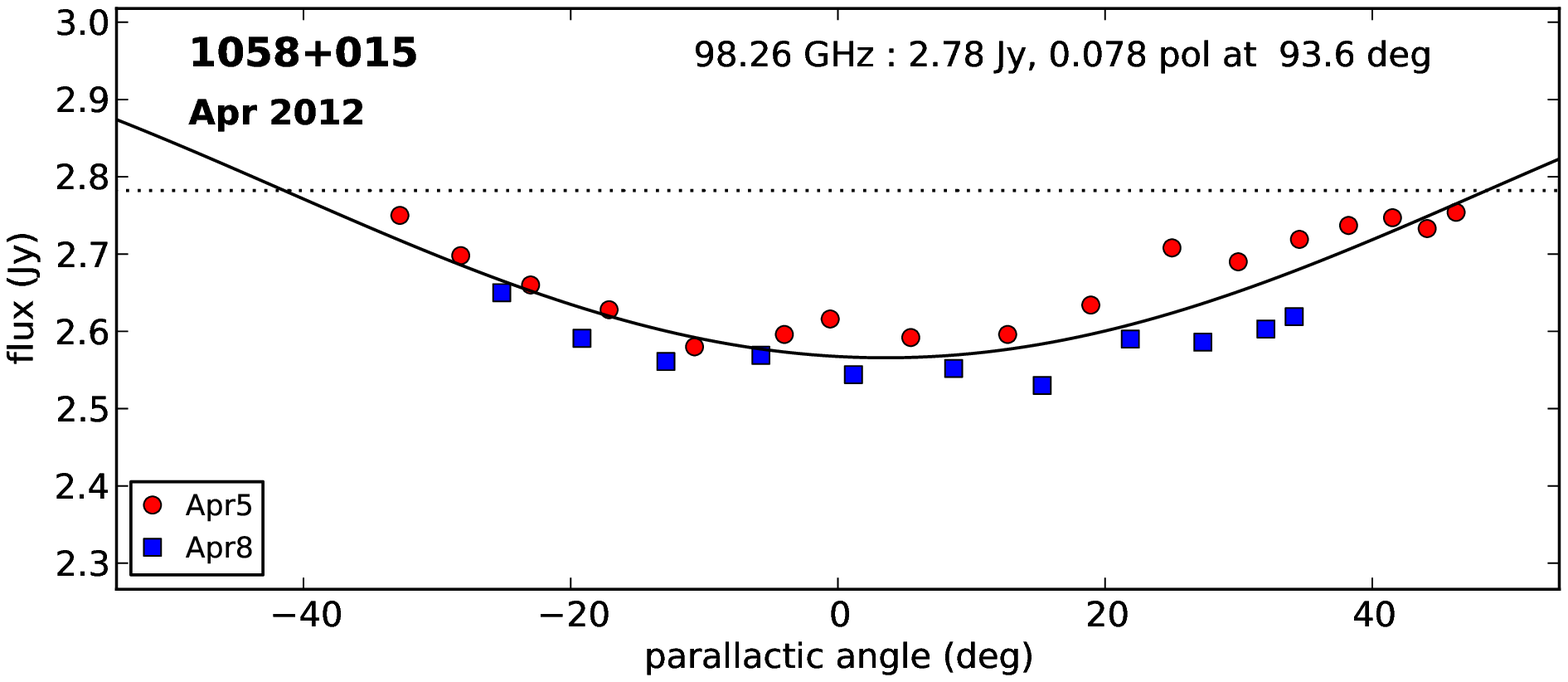}
\end{minipage}
\caption{Polarization of calibrators 0854+201 (left) and 1058+015 (right). In each panel, flux versus parallactic angle is plotted for 
the best datasets at each epoch (indicated in the upper left). The solid black line shows the best fit, with the dotted black line indicating the best-fit Stokes I flux. 
Calibrator 0854+201 (left) was observed at four epochs: at 107 GHz in November 2010 and at 98 GHz in September 2011, November 2011 and February 2012. 
Calibrator 1058+015 (right) was observed at four epochs as well: at 107 GHz in May 2011 and at 99 GHz November 2011, February 2012 and April 2012. }
\label{fig:polcalfits}
\end{figure*}

This procedure was used for two calibrators: 0854+201 and 1058+015. We do not assume stability
over long timescales in the polarization parameters of the calibrators (and in fact find evidence for significant variation
with time), so the polarization parameters for each source are derived in each epoch from the best datasets.
The flux scale in most of the datasets was calibrated using Mars, with brightness temperatures from the 
Caltech thermal model by Mark Gurwell, which includes seasonal variations. 
Datasets from November 2010 and May 2011 use 3C84 as the flux calibrator. Based on flux monitoring at 
CARMA\footnote{See CARMA Memo \#59 \citep{CARMAmemo59} for a description of flux monitoring at CARMA},
we assumed a flux of 11 Jy in November 2010 and 10 Jy in May 2011 (both at 107 GHz).

\begin{table}[t]
\centering
\begin{tabular}{| c | c | c | c | c | c |}
\hline
Calibrator & Epoch & $\nu_\mathrm{obs}$ & Stokes I & $p_\mathrm{qu}$ & Angle \\
 & & (GHz) & Flux (Jy) & & (deg) \\
\hline
0854+201 & Nov 2010 & 106.6 & 5.2 & 0.10 & 70 \\
& Sep 2011 & 98.1 & 4.0 & 0.10 & 67 \\
& Nov 2011 & 98.0 & 5.2 & 0.10 & 67 \\
& Feb 2012 & 98.0 & 5.2 & 0.08 & 54 \\
\hline
1058+015 & May 2011 & 107.0 & 4.6 & 0.08 & 19 \\
& Nov 2011 & 98.9 & 4.1 & 0.09 & 19 \\
& Feb 2012 & 98.9 & 3.6 & - & - \\
& Apr 2012 & 98.3 & 2.8 & 0.08 & 94 \\
\hline
\end{tabular}
\caption{Polarized calibrator parameters. The values given in the table are those
that were used in the reduction of the relevant datasets. In some cases, these are
not the same as the best fit values. See text for discussion.}
\label{tab:polcalvals}
\end{table}

The fits in the different epochs are shown in Figure \ref{fig:polcalfits}, with 0854+201 on the left and 1058+015 on the right. 
Flux versus parallactic angle is plotted, with data points for each dataset marked with a different symbol (indicated in the legend) 
and the best fit line (Equation \ref{equ:polflux}) in solid black. The dotted black line shows the best fit Stokes I flux. 
The observing frequency and best-fit parameters are given in the upper right corner of each panel.
The polarization parameters used for each calibrator, in each epoch, are given in Table \ref{tab:polcalvals}.

While the fit for 0854+201 in November 2010 is not well constrained due to limited parallactic angle coverage, 
it agrees with subsequent epochs, so we use it. The fit for 0854+201 in September 2011 is not very good due
to noisy data, but the good fit from November 2011 (scaled to the average flux in September 2011; black dashed line) 
appears to match the September 2011 data well. Therefore, we use the November 2011 polarization fraction and angle
for September 2011, with an average Stokes I flux of 4.0 Jy.

The data for 1058+015 from November 2011 to April 2012 suggest a radical change in polarization, from 9\% polarized
at 19 degrees to 8\% polarized at 94 degrees over the course of 6 months. The data from February 2012 show no indication
of polarization (in fact, the best fit is a negative polarization fraction). 
This disappearance and (rotated) reappearance of the linear polarization component of 1058+015 
is coupled with a significant decrease in flux.

\section{$uv$-Spectra}
\label{sec:uvspectra}

\begin{figure*}[b]
\centering
\includegraphics[width=\linewidth]{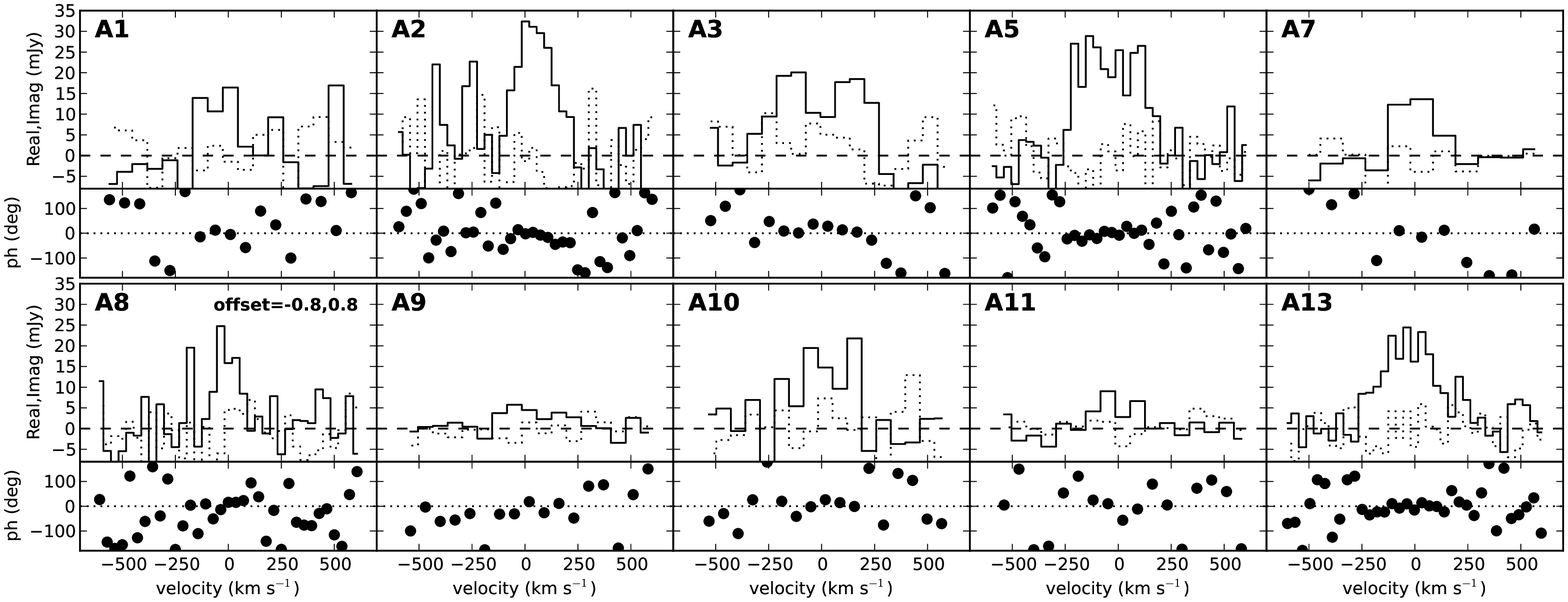}
\includegraphics[width=\linewidth]{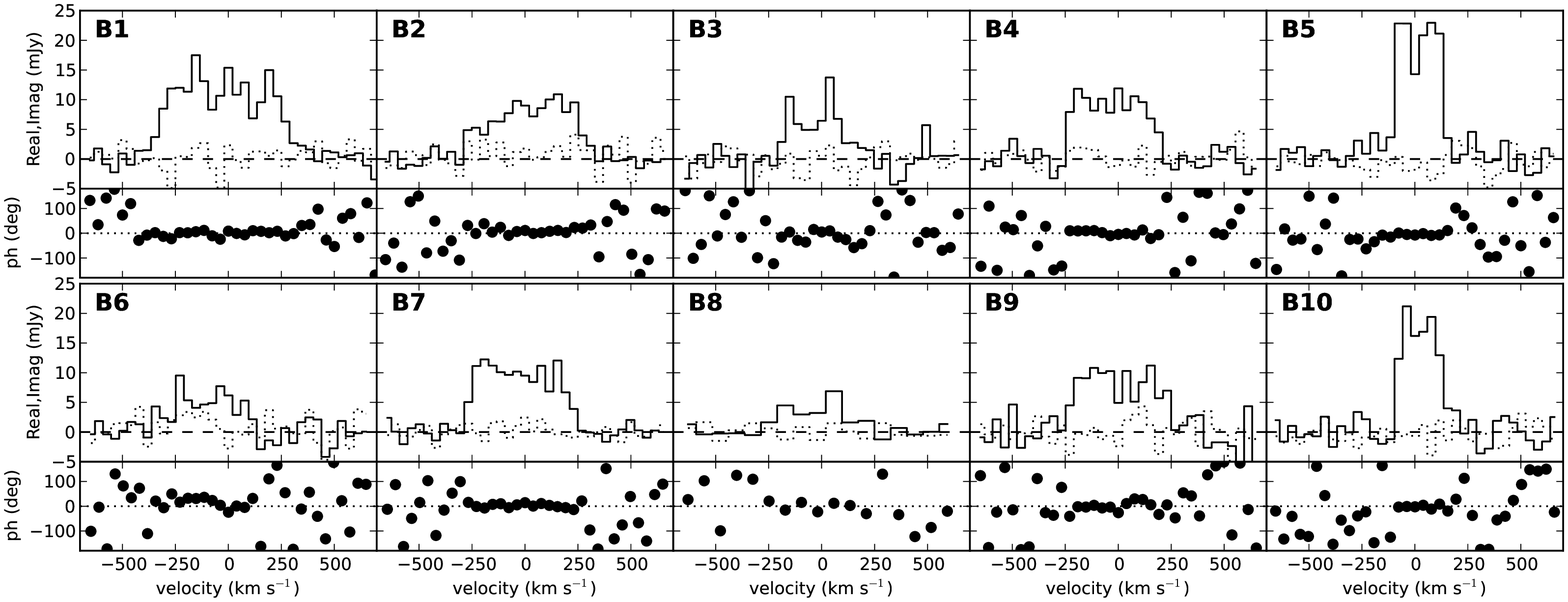}
\caption{$uv$-spectra of bin A and B galaxies. 
The top panel shows the vector-averaged {\it Real} (solid) and {\it Imaginary} (dotted) amplitudes (Real,Imag in mJy beam$^{-1}$) 
and the bottom panel shows phase (ph, degrees) of the calibrated $uv$ data versus velocity (\kms).
For galaxy A8, the spectrum is calculated at a $-0.8\arcsec,0.8\arcsec$ offset from the SDSS position (indicated
in the upper right corner of the A8 panel). }
\label{fig:uvspectra}
\end{figure*}

\begin{figure*}[t]
\centering
\includegraphics[height=2.8in]{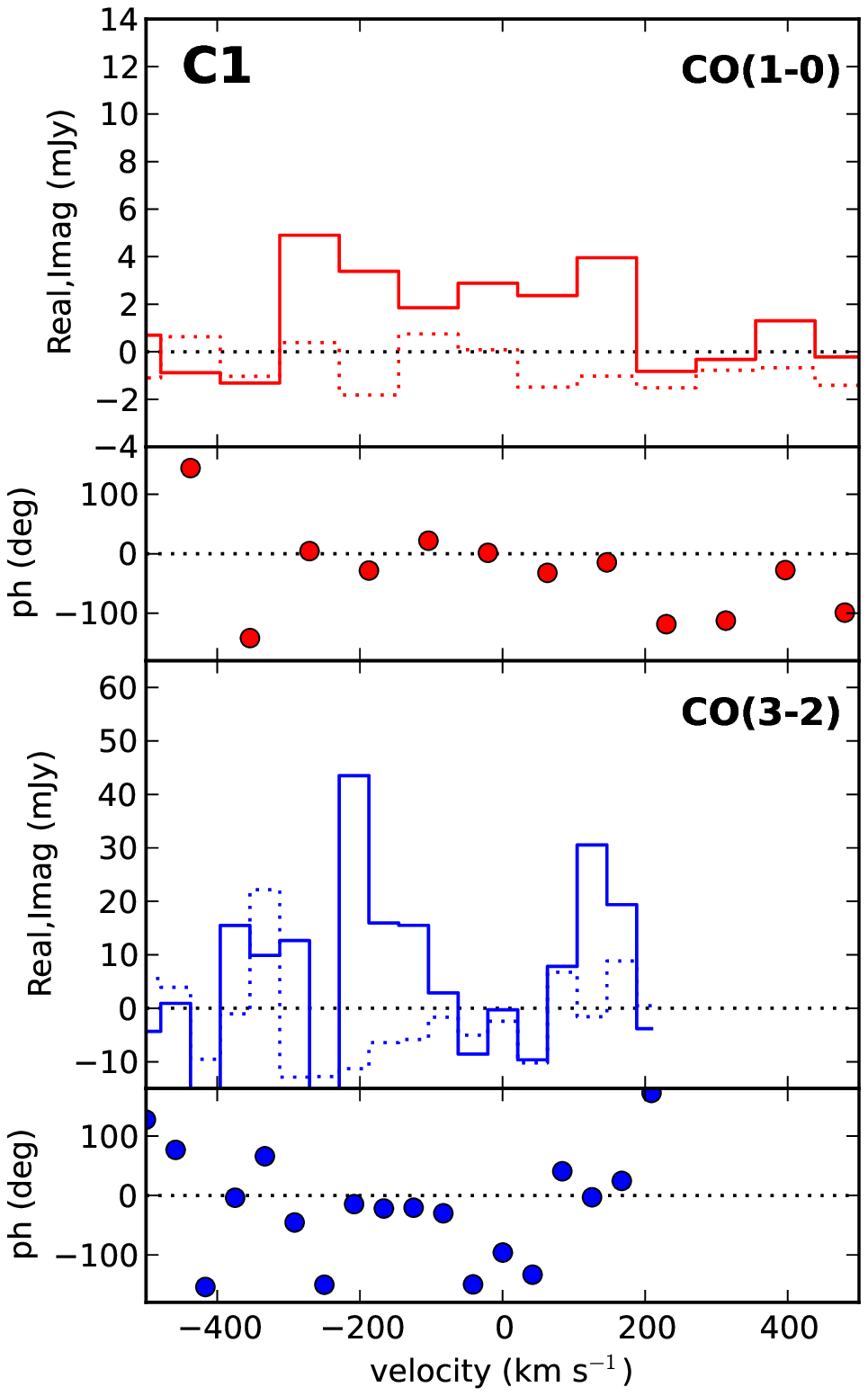}
\includegraphics[height=2.8in]{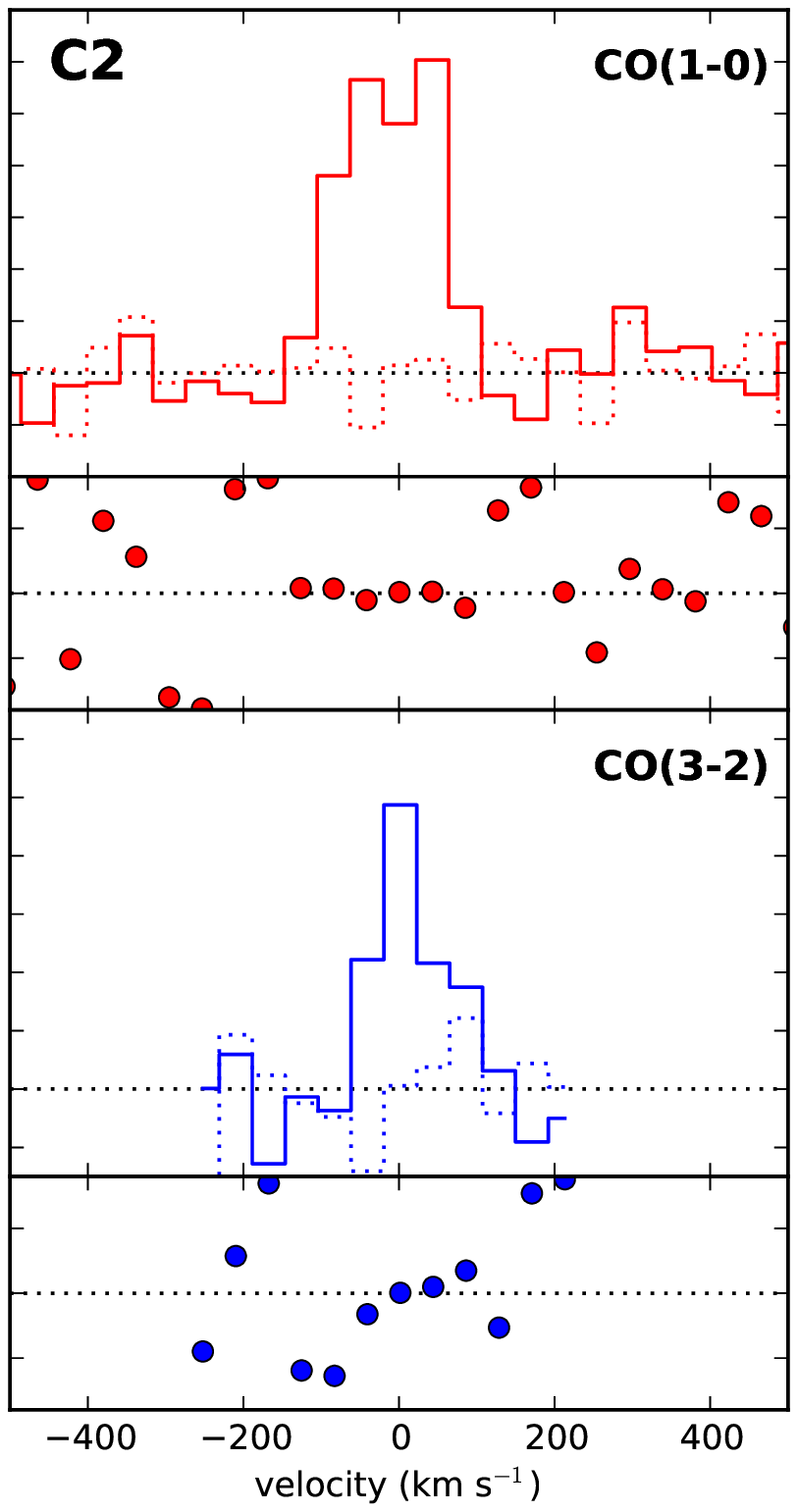}
\includegraphics[height=2.8in]{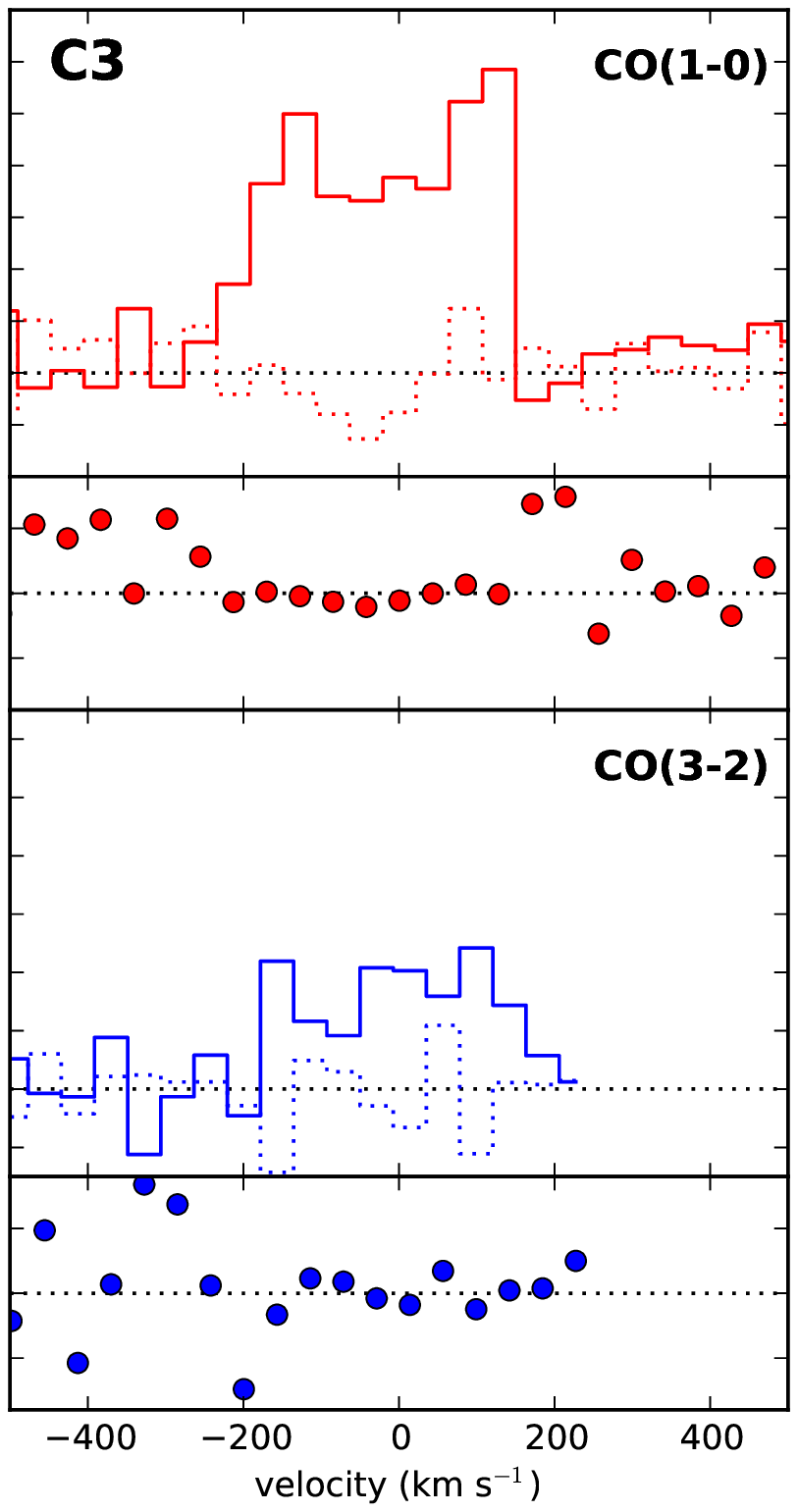}
\includegraphics[height=2.8in]{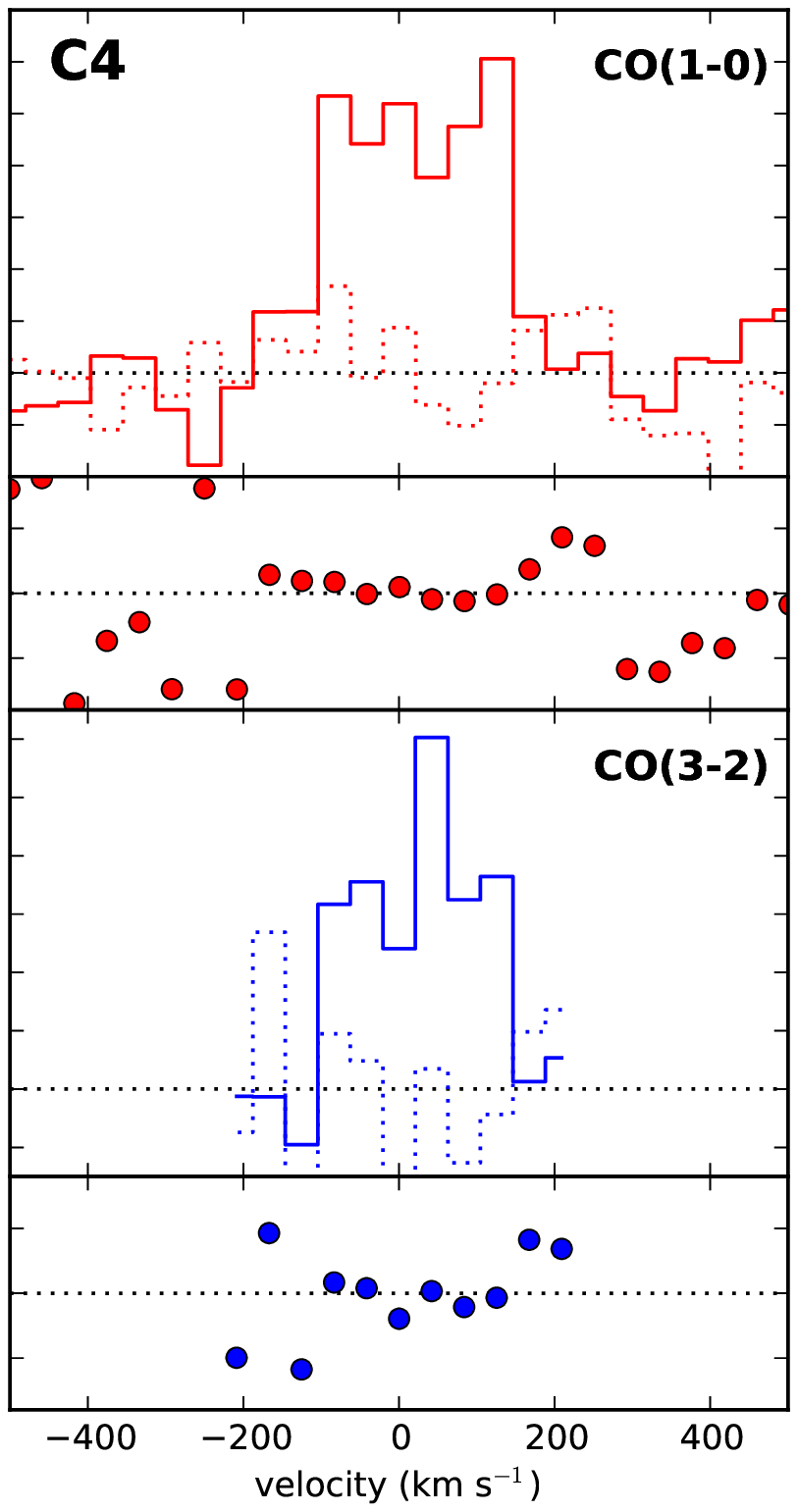}
\caption{$uv$-spectra of bin C galaxies. For each source, 
the top row (red) shows the CO\jone line and the bottom row (blue) shows the CO\jthree line. 
For each line, for each source, the top panel shows the vector-averaged {\it Real} (solid) and {\it Imaginary} (dotted) 
amplitudes (Real,Imag in mJy beam$^{-1}$) 
and the bottom panel shows phase (ph, degrees) of the calibrated $uv$ data versus velocity (\kms).
With the higher resolution of the CO\jthree
data, we observe the peak of the CO\jthree emission to be offset ($<1.5\arcsec \sim 6$ kpc)
from the centers of galaxies C1 and C2. 
In these cases, we calculate the $uv$ spectra at the peak of the CO\jthree emission.
For the CO\jone line of galaxy C1, we average 2 channels together
in order to increase the signal to noise ratio. }
\label{fig:binCuvspectra}
\end{figure*}

We show the $uv$-spectra for all detected galaxies in Figures \ref{fig:uvspectra} and \ref{fig:binCuvspectra}.
For each galaxy (indicated in the top left), the top (bottom) panel shows the 
vector-averaged {\it Real} and {\it Imaginary} amplitudes (phase) of the calibrated $uv$ data versus velocity.
In Figure \ref{fig:binCuvspectra}, the CO\jone line is shown in the top panels and
the CO\jthree line in the bottom panels. 
For galaxy A8 and the CO\jthree line in galaxies C1 and C2, the spectrum is calculated at a position 
offset from the SDSS position in order to coincide with the observed peak of the CO emission.
For a compact source at the center of the field of view, the
{\it Real} part shows the signal without a noise bias, and the {\it Imaginary} part shows the noise.
In all cases (except CO\jthree in galaxy C1), 
we see coherent emission (larger {\it Real} amplitudes coincident with noise-like {\it Imaginary}
amplitudes and phases of $\approx 0^\circ$) over multiple velocity channels, indicative of a detection. 
We do not detect the CO\jthree line in galaxy C1.

\section{Extraordinary CO Emission in Galaxies A3 and A8}
\label{sec:A3andA8}

The integrated CO emission maps of galaxies A3 ($z=0.06$) and A8 ($z=0.10$) show evidence for disturbed morphologies.
In both galaxies, we detect CO emission coincident with the optical emission, which shows a velocity gradient misaligned with
respect to the optical major axis. In galaxy A8 we detect CO emission outside of the optical emission. Our observations
are suggestive of CO emission outside the optical emission of A3 as well, but are not conclusive.
In this Appendix, we investigate these emission components using $uv$-spectra calculated
at various positions in the field to distinguish true emission from noise. We discuss the more compelling case of galaxy A8 first, 
then A3.

\subsection{EGNoG A8}

\begin{figure}[t]
\centering
\includegraphics[width=\linewidth]{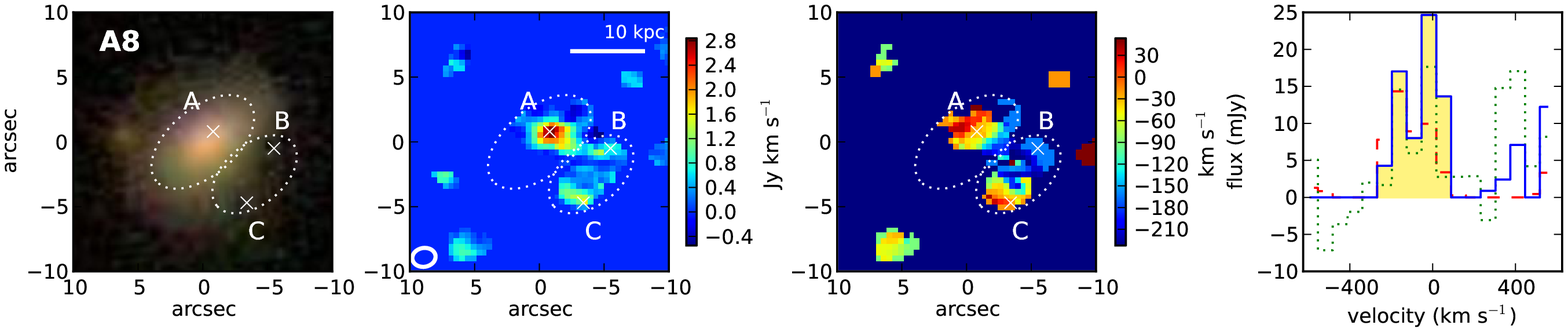}
\includegraphics[height=2in]{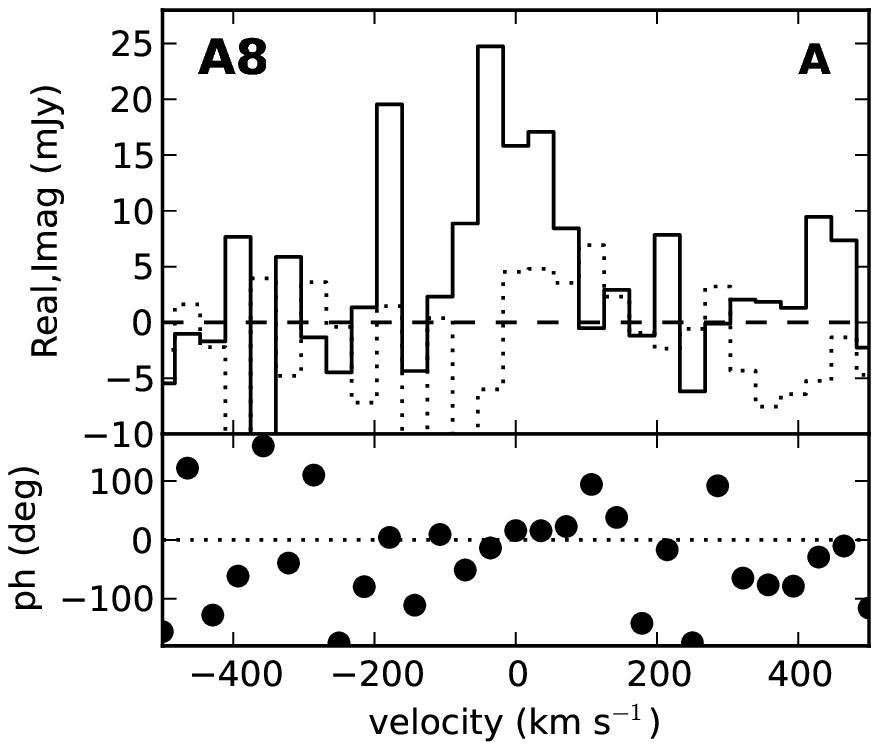}
\includegraphics[height=2in]{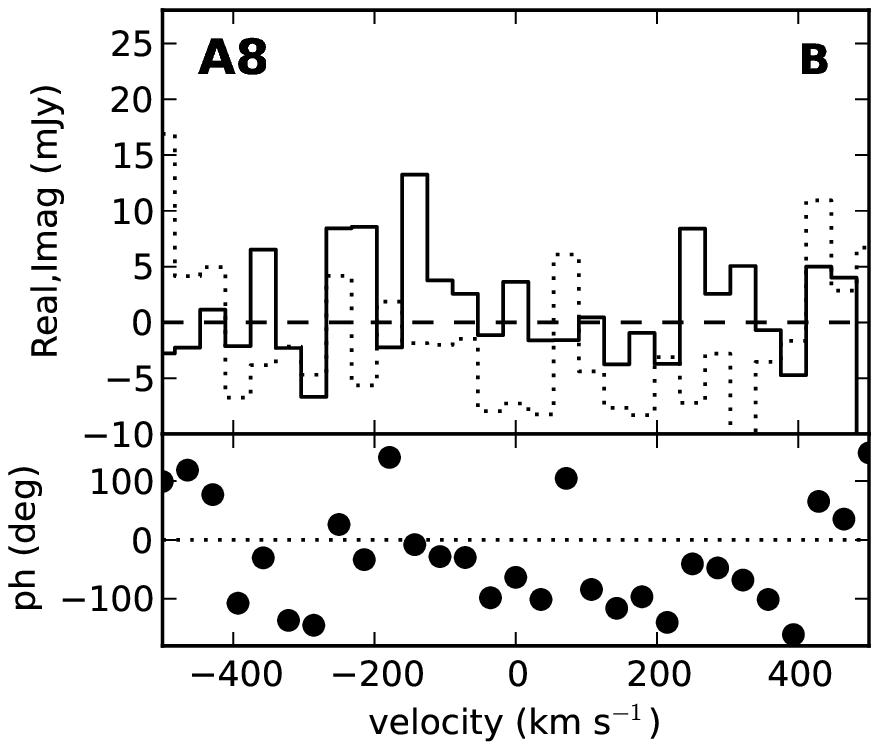}
\includegraphics[height=2in]{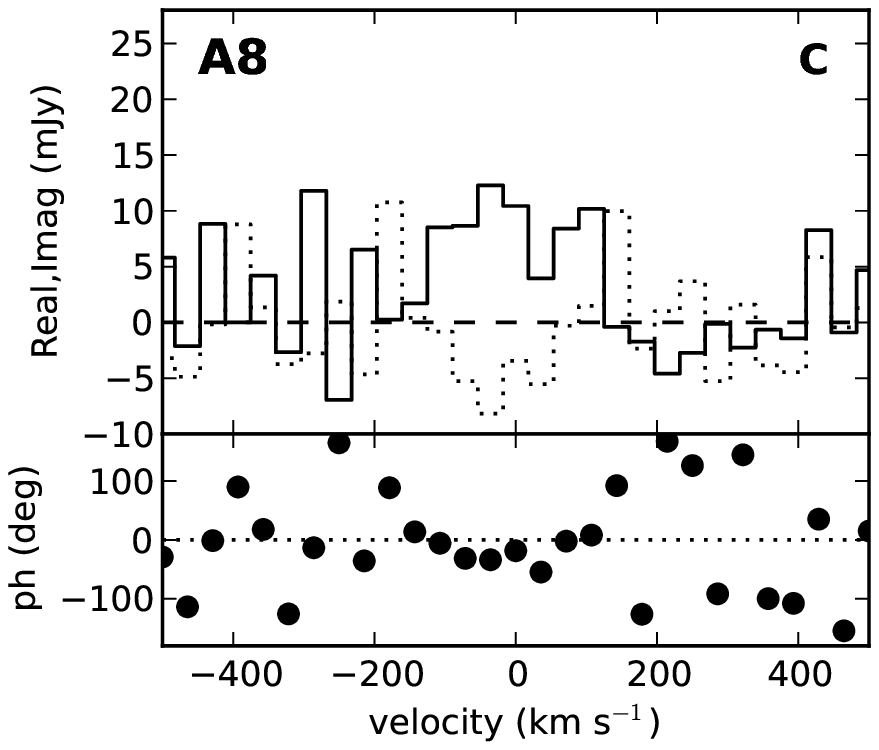}
\caption{Optical image, moment maps, integrated spectra and $uv$-spectra for EGNoG galaxy A8. The top four panels show
(from left to right) the optical image, the moment 0 map, the moment 1 map and 
the integrated spectrum (as in Figure \ref{fig:binAmaps1})
of the main part of the galaxy (solid blue shows spectrum with masking, dotted green without). 
The integrated spectrum of the external CO component is also shown by a red dashed line. 
The bottom three panels show the $uv$-spectra (amplitude and phase) versus velocity of 
each of the 3 points (A,B and C) marked in the moment maps in the top row.}
\label{fig:A8blob}
\end{figure}

In Figure \ref{fig:A8blob}, we show the optical image, moment maps, integrated spectrum (as in Figure \ref{fig:binAmaps1}), and
$uv$-spectra of galaxy A8. 
In the top right panel, the solid blue (green dotted) line shows the integrated spectrum with (without) masking 
of the main CO component, indicated by the central white dotted ellipse in the moment map panels.
The dashed red line shows the integrated spectrum of the emission component
outside the optical galaxy (the offset white dotted ellipse in the map panels).
The bottom three panels show the $uv$-spectra 
(amplitude and phase versus velocity) at 3 points, A, B and C, labeled on the map panels in the top row. 

Point A is centered on the CO emission coinciding with the optical emission. 
We only detect CO emission corresponding to the upper right (north-west) half of the
optical galaxy, at negative velocities (relative to the optical redshift from SDSS). 
The moment 1 map (intensity-weighted velocity) of this component
shows a velocity gradient suggestive of a rotating disk. 
However, this velocity gradient does not lie along the optical major axis, 
but appears to be rotated by 45 to 90 degrees. 

Points B and C are at two positions along the long axis of the CO component that lies outside the optical galaxy 
(following the velocity gradient). 
The optical image from the SDSS shows no optical counterpart to this component of the CO emission.
While the $uv$-spectrum at point B appears dominated by noise, point C shows phases of $\approx0^\circ$ 
over several velocity channels, indicative of real emission. 
This CO component shows a velocity gradient parallel to the major axis of the optical galaxy and is therefore not aligned with the 
velocity gradient of the CO disk component. However, the two components become roughly spatially coincident at negative velocities.

In summary, the molecular gas in galaxy A8 is disturbed: CO emission is observed coinciding with the north-west half the
optical galaxy but not in the south-east portion; the velocity gradient of the CO emission is not aligned with the optical major axis; 
and we observe significant CO emission outside the optical galaxy, with no optical counterpart.
The optical image does not indicate interaction. 
We note that there is another galaxy in the field (seen on the left side of the optical image), 
but this galaxy is in the SDSS catalog with a photometric redshift of 0.28. 
The CO flux of component outside the disk is 3.2 Jy \kms, with $v_\mathrm{center} = -110$ \kmssp and $\Delta V = 357$ \kms. 
This is very similar to the main CO component: $S_\mathrm{CO} = 4.5$ Jy \kms, 
with $v_\mathrm{center} = -67$ \kmssp and $\Delta V = 357$ \kms. 
If the emission outside of the optical disk traces dense molecular gas, 
the molecular gas mass of this component (assuming a Milky Way-like
conversion factor) would be $6.5\times10^9$ M$_\odot$, roughly 70\% the molecular mass of the component
coincident with the optical emission.

\subsection{EGNoG A3}

\begin{figure}[b]
\centering
\includegraphics[width=\linewidth]{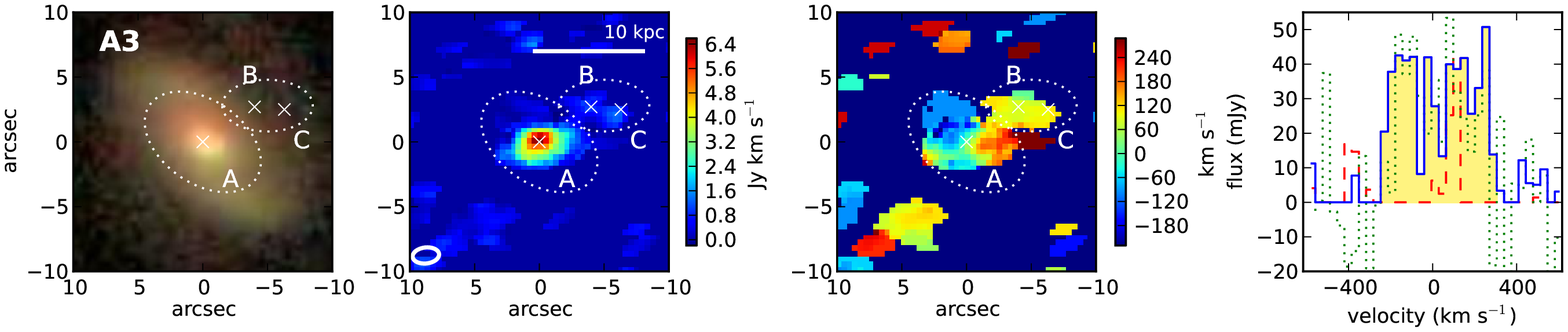}
\includegraphics[width=\linewidth]{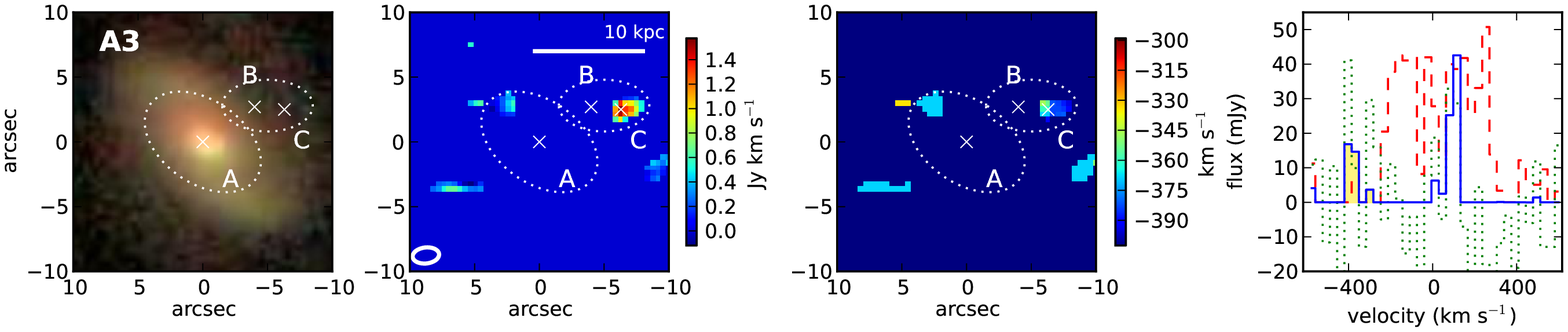}
\includegraphics[height=2in]{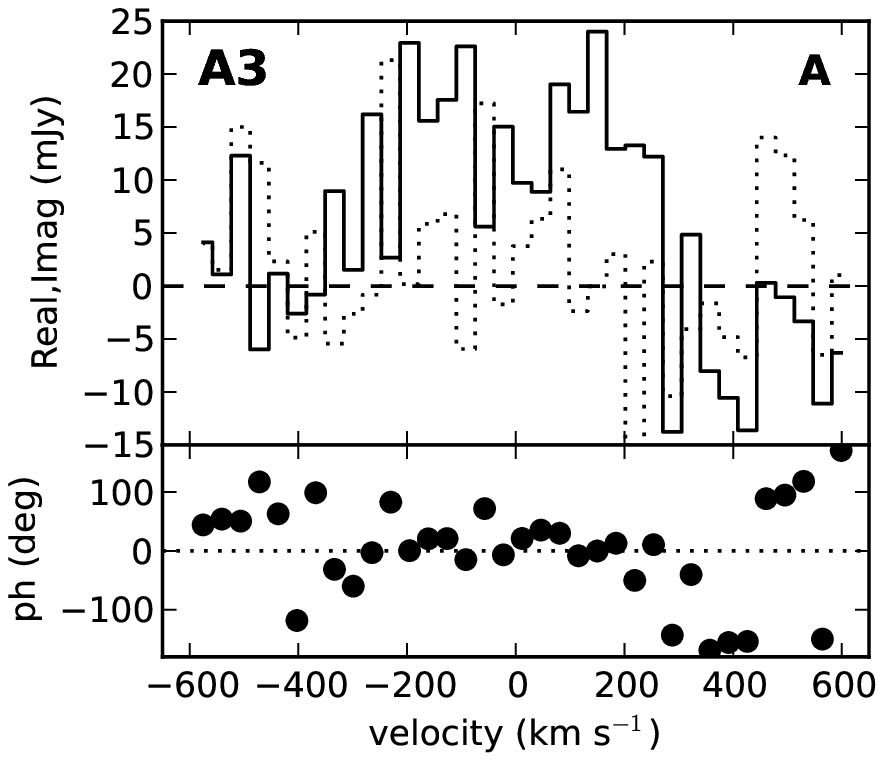}
\includegraphics[height=2in]{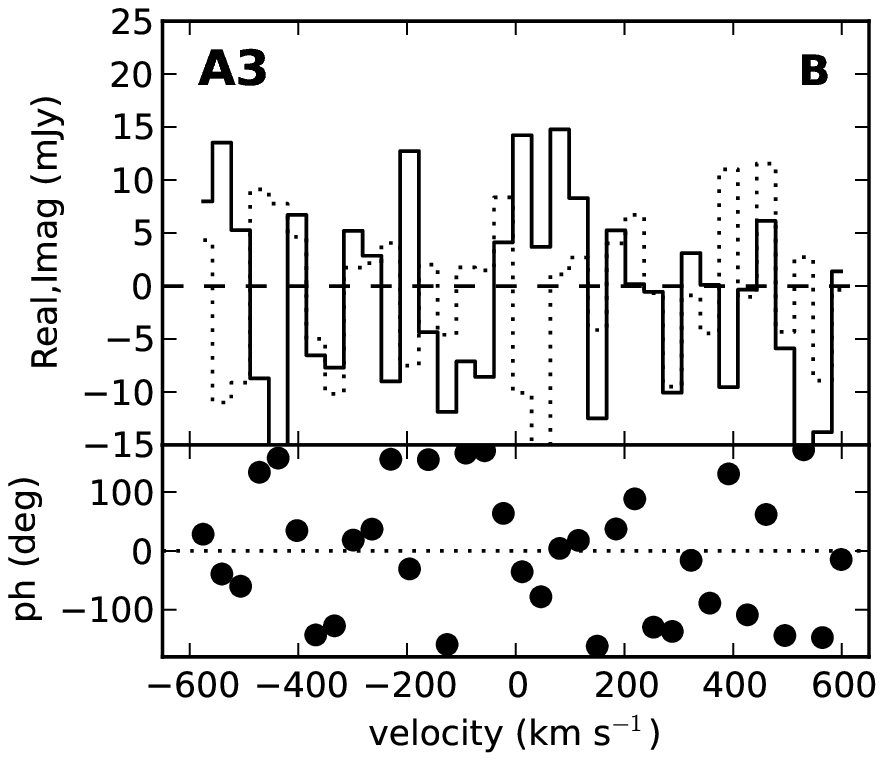}
\includegraphics[height=2in]{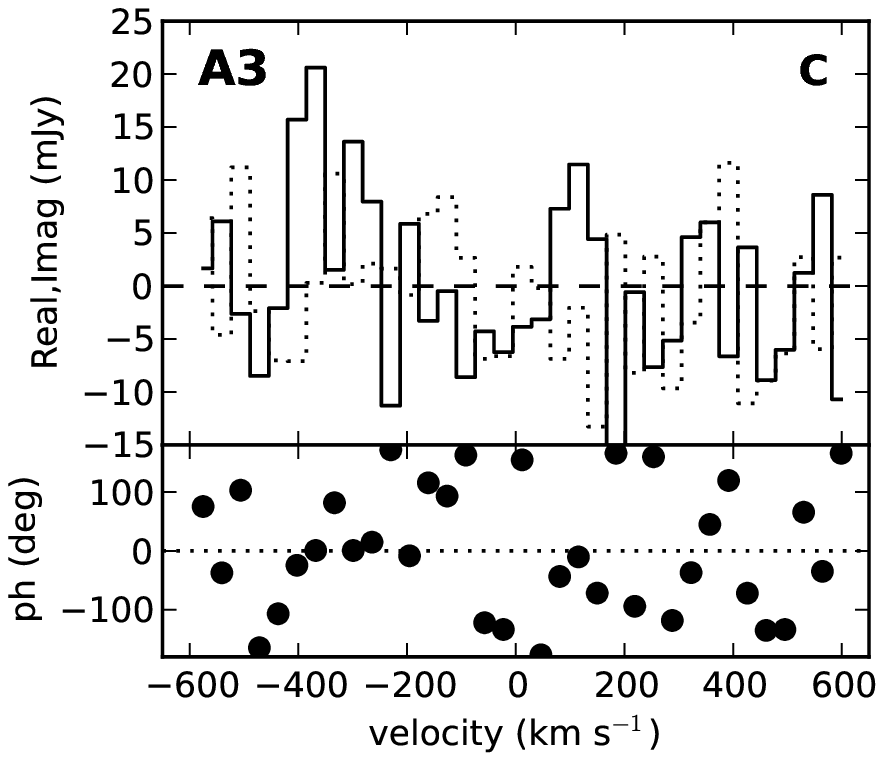}
\caption{Optical image, moment maps, integrated spectra and $uv$-spectra for EGNoG galaxy A3. The top four panels show
(from left to right) the optical image, the moment 0 map, the moment 1 map and the integrated spectrum (as in Figure \ref{fig:binAmaps1})
of the main part of the galaxy. The integrated spectrum of the external CO component is shown by a red dashed line. 
The middle four panels are the same as the top four, but for the negative velocity component of the external CO emission.
The red dashed line in the far right panel of the middle row shows the spectrum of the main part of the galaxy for reference.  
The bottom three panels show the $uv$-spectra (amplitude and phase versus velocity) 
at each of the 3 points (A,B and C) marked in the moment maps in the top and middle rows.}
\label{fig:A3blob}
\end{figure}

In Figure \ref{fig:A3blob}, we show the optical image, moment maps, integrated spectrum (as in Figure \ref{fig:binAmaps1}), and
$uv$-spectra of galaxy A3. The top row shows moment maps and integrated spectrum
calculated in the velocity channels of the CO emission corresponding to the optical emission of the galaxy. 
In the top right panel, the solid blue (green dotted) line shows the integrated spectrum with (without) masking 
of the main CO component, indicated by the central white dotted ellipse in the moment map panels.
The dashed red line shows the integrated spectrum of the emission component
outside the optical galaxy (the offset white dotted ellipse in the map panels).
The middle row is the same as the top row, but for the velocity channels of the negative velocity component
of the emission lying outside of the optical galaxy. 
In the middle right panel, the solid blue line shows the integrated spectrum of the emission
component outside the optical galaxy and the dashed red line shows the emission corresponding to the main CO component. 
The bottom three panels show the $uv$-spectra 
(amplitude and phase versus velocity) at 3 points, A, B and C, labeled on the map panels in the top row. 

Point A is centered on the CO emission coinciding with the optical galaxy. 
The moment 1 map of this component shows a velocity gradient suggestive of a rotating disk, 
but the gradient is misaligned with respect to the optical major axis (rotated 45 to 90 degrees from the major axis, as in galaxy A8). 

Points B and C are at two positions in the emission component that lie outside the optical galaxy: 
point B marks an area that is bright at positive velocities
(93 \kms) and point C marks an area that is bright at negative velocities ($-377$ \kms).
The optical image from the SDSS shows no optical counterpart to either component of the CO emission.
The $uv$-spectra at points B and C (bottom row, middle and right panels) are suggestive of real emission but do not show a
strong signal. The integrated flux from the positive (negative) velocity component is 
2.6 Jy \kmssp (1.2 Jy \kms), with $v_\mathrm{center} = 93$ (-377) \kmssp and $\Delta V = 138$ (138) \kms. 
If the emission is real and tracing dense molecular gas, the positive velocity component has a molecular gas mass
of $1.5\times10^9$ M$_\odot$ and the negative velocity component has $7.0\times10^8$ M$_\odot$. 
These masses are both small compared to the molecular gas mass of the main CO component, $1.4\times10^{10}$ M$_\odot$.

The misalignment (with respect to the optical major axis) of the velocity gradient of the main component 
of the CO emission is suggestive of a disturbed morphology. 
Our observations also suggest there is CO emission outside of the optical galaxy, but we do not detect this emission 
at a significant level.

\end{appendix}

\end{document}